\documentclass[11pt]{article}
\usepackage[english]{babel}
\usepackage{afterpage}
\usepackage{amsmath,amssymb,graphicx,calc,capt-of,subcaption,ifthen,theorem,pifont,amsfonts}
\usepackage[numbers,sort&compress]{natbib}

\begin{document}

\setlength\arraycolsep{2pt}

\renewcommand{\theequation}{\arabic{section}.\arabic{equation}}
\setcounter{page}{1}

\begin{titlepage}

\begin{center}

\vskip 1.5 cm

{\huge \bf Covariant evolution of perturbations during reheating in two-field inflation}

\vskip 2.0cm

{\Large
Pablo Gonz\'alez$^{a,b}$, Gonzalo A. Palma$^{b}$ and Nelson Videla$^{c,b}$
}

\vskip 0.5cm

{\it $^{a}$Sede Esmeralda, \mbox{Universidad de Tarapac\'a},\\ Avda. Luis Emilio Recabarren 2477, Iquique, Chile.\\
$^{b}$Grupo de Cosmolog\'ia y Astrof\'isica Te\'orica, Departamento de F\'{i}sica, FCFM, \mbox{Universidad de Chile}, Blanco Encalada 2008, Santiago, Chile.\\
$^{c}$Instituto de F\'isica, \mbox{Pontificia Universidad Cat\'olica de Valpara\'so},\\ Avda. Universidad 330, Curauma, Valparaiso, Chile.}

\vskip 2.5cm

\end{center}

\begin{abstract}

We develop a covariant method for studying the effects of a reheating phase on the primordial adiabatic and isocurvature perturbations in two-field models of inflation. To model the decay of the scalar fields into radiation at the end of inflation, we introduce a prescription in which radiation is treated as an additional effective scalar field, requiring us to extend the two-field setup into a three-field system. In this prescription, the coupling between radiation and the scalars can be interpreted covariantly in terms geometrical quantities that parametrize the evolution of a background trajectory in a three-field space. In order to obtain concrete results, we consider two scenarios characterized for having unsuppressed isocurvature fluctuations at the end of inflation: (1) canonical two-field inflation with the product exponential potential, which sources a large negative amount of non-gaussianity and, (2) two-field inflation with an ultra-light field, a model in which the isocurvature mode becomes approximately massless, and its interaction with the curvature perturbation persists during the entire period of inflation. In both cases we discuss how their predictions are modified by the coupling of the scalar fields to the radiation fluid.

\end{abstract}

\end{titlepage}

\newpage

\setcounter{equation}{0}
\section{Introduction}

Single field slow-roll inflation represents the most successful framework to describe the dynamics of the very early universe. It solves the classical shortcomings of the hot big-bang scenario: the horizon, flatness, and monopole problems~\cite{Starobinsky:1980te, Guth:1980zm, Sato:1980yn, Linde:1981mu, Linde:1983gd, Albrecht:1982wi}. It also gives us a mechanism to explain the origin of the large-scale structure (LSS) of the universe and the cosmic microwave background (CMB) anisotropies~\cite{Ade:2015xua,Ade:2013zuv,Hinshaw:2012aka}. Within this paradigm, the primordial curvature perturbations are the result of quantum fluctuations of the inflaton scalar field during the inflationary expansion~\cite{Starobinsky:1979ty, Mukhanov:1981xt, Hawking:1982cz, Guth:1982ec, Starobinsky:1982ee, Bardeen:1983qw}.

One of the salient points of single field slow-roll inflation is that the evolution of curvature perturbations happens adiabatically. While the inflationary background evolves slowly, curvature perturbations freeze, recording in their amplitude information about the background at the time of horizon crossing. The subsequent evolution of modes is insensitive to the details of the universe's background evolution, and so the measurements of correlation functions in the CMB provides direct information about the epoch of inflation when the fluctuations where produced.

This simple picture might break down in more realistic versions of inflation. More fundamental theories, where the standard model interactions are unified with the gravitational interaction, such as supergravity and string theory, require the existence of a large number of scalar fields. In these models, many of the scalar fields have a geometrical role, and so they are not necessarily massive at high energies. It is therefore not well understood how these scalar fields (isocurvature fields) could have had a role during inflation, where the typical energy available to fluctuations during horizon crossing is of the order of the grand unification theory scale.

One possibility is that isocurvature degrees of freedom did not affect the evolution of curvature perturbations during inflation after all. In that case, regardless of how complicated the ultraviolet theory underlying inflation is, one can understand the evolution of curvature perturbations in terms of an effective field theory (EFT) describing the evolution of a canonical scalar field. Another possibility is that isocurvature fields did not have a relevant kinematic role during inflation, but their couplings to the inflaton still affected the evolution of curvature perturbations. In such cases, the evolution of curvature perturbations continues to be well described by an effective field theory of a single scalar degree of freedom, but this time the theory contains self-interactions incorporating departures from the canonical picture~\cite{Alishahiha:2004eh, Cheung:2007st, Tolley:2009fg, Achucarro:2010da}. An example of this, is the appearance of the sound-speed $c_s$ of curvature perturbations that is able to generate potentially large values of equilateral and orthogonal non-Gaussianity~\cite{Seery:2005wm, Chen:2006nt}.

Yet another possibility is that isocurvature degrees of freedom were both coupled to curvature perturbations and kinematically relevant at the time of horizon crossing~\cite{Langlois:1999dw, Gordon:2000hv, GrootNibbelink:2000vx, GrootNibbelink:2001qt, Amendola:2001ni, Bartolo:2001rt, Wands:2002bn, Tsujikawa:2002qx, Enqvist:2004bk, Lyth:2005fi, Seery:2005gb, Rigopoulos:2005ae, Alabidi:2005qi, Byrnes:2006fr, Choi:2007su, Battefeld:2007en, Byrnes:2008wi, Byrnes:2009qy, Battefeld:2009ym, Chen:2009we, Chen:2009zp, Cremonini:2010sv, Cremonini:2010ua, Elliston:2011et, Mulryne:2011ni, McAllister:2012am, Byrnes:2012sc, Achucarro:2016fby, Bjorkmo:2017nzd}. In this case, the dynamics of curvature perturbations cannot be reduced to a single field EFT, and the computation of correlation functions has to take into account the full multi-field nature of inflation during horizon crossing. The wide range of potentially observable effects in this third possibility (multi-field inflation) is far from having been studied exhaustively.\footnote{For a comprehensive review on multi-field inflation, see Refs.~\cite{Gong:2016qmq,Langlois:2010xc}.} A crucial difference between single-field and multi-field inflation is the presence of isocurvature perturbations. Because of these modes, the slow-roll solution is no longer an attractor, leading to non-vanishing non-adiabatic pressure perturbations and thus possible super-horizon evolution of the primordial curvature perturbation \cite{Lalak:2007vi,GrootNibbelink:2001qt,Gordon:2000hv}.

Irrespective of the number of fields that drove inflation, at the end of the inflationary stage the universe is typically found to be in a highly non-thermal state. The key feature of inflation, allowing it to homogenize the universe, also means that it leaves the universe at an effectively zero temperature. Hence, a successful theory of inflation must also explain how the universe was heated to the high temperatures required by the standard hot big-bang picture. This is achieved at the end of inflation by reheating \cite{Kofman:1997yn,Kofman:1994rk}, a process in which the inflaton field starts oscillating around the minimum of its potential. Reheating sets the post-inflationary conditions of the universe before it enters the radiation era. Once the universe enters the radiation era, its evolution is described by the physics of the hot big bang model.

Essentially, all elementary particles are produced during reheating. This makes reheating one of the most fundamental phases of the very early universe. At the end of inflation all the energy stored in the inflaton field is transferred to the thermal energy of the particles produced by the decay of the inflaton. These particles interact and eventually thermalize into equilibrium at a reheating temperature $T_{\rm rh}$ which is rather model-dependent~\cite{Martin:2010kz}. Once all the energy of the inflaton is transferred to the particles, reheating ends and the Universe starts its radiation era. The details of reheating are sensitive to the inflationary background, choice of parameters, and initial conditions. In many inflationary models, reheating consists of distinct stages with different features, starting with a stage called preheating~\cite{Greene:1997fu}, where parametric resonance of the inflaton field leads to particle production, and then moving to a second stage of thermalization. For a treatment of preheating after multi-field inflation, see Refs.\cite{Krajewski:2018moi,DeCross:2016fdz,Kaiser:2015usz,DeCross:2015uza,Amin:2014eta}.

A fundamental issue related to reheating is whether it can affect the evolution of super-horizon primordial perturbations. Our models of inflation are tested with the help of CMB observations, hence, if reheating modifies how the spectra of perturbations depend on the background during horizon crossing, we might be unable to constrain the role of fundamental theories during inflation. In the case of single field inflation, the conditions for the conservation of curvature perturbations over superhorizon scales are well understood~\cite{Bardeen:1983qw, Mukhanov:1990me}. Satisfying these conditions allows one to constrain certain aspects of reheating with CMB observations~\cite{Martin:2014nya}.

On the other hand, if additional isocurvature fields (also known as entropy perturbations) remain unsuppressed during reheating the situation may change drastically. This is possible if the the effective mass of the entropy modes remain light compared to Hubble rate $H$ during the whole duration of inflation~\cite{Gordon:2000hv, Lalak:2007vi}. For instance, it was shown in Ref.~\cite{Bassett:1998wg} that parametric resonance during reheating may lead to an exponential amplification of super-horizon fluctuations. This has been confirmed in Refs.~\cite{Bassett:1999cg, Finelli:2000ya, Tsujikawa:2002qx}. However, these works did not take into account the possible coupling of the perturbations (curvature and isocurvature) to radiation and pressureless matter components. This issue was considered in Ref.~\cite{Choi:2008et}, for a model of multi-field chaotic inflation. There it was found that, as long as the isocurvature modes remain light, the total curvature perturbation may experience large variations due to the non-adiabatic components. For latter studies on reheating after multi-field inflation, see Refs.~\cite{Hotinli:2017vhx,Schimmrigk:2017jwa}.

It has been pointed out~\cite{Turzynski:2014tza} that multi-field scenarios should a priori be considered as non-predictive unless one demands the so called adiabatic limit, in which isocurvature perturbations decay before the end of inflation. The models for which the adiabatic limit is not reached are characterized by having isocurvature modes that remain nearly massless during the whole period of inflation. However, there are well motivated examples of models where the adiabatic limit is not reached, even at the end of inflation. One example is given by the product exponential potential $V(\phi_1,\phi_2)=V_0\,\phi_1^2\,e^{-\lambda \phi_2^2}$~\cite{Elliston:2011dr,Byrnes:2008zy}. Another example is offered by models with ultra-light entropy modes, where the isocurvature mode remains nearly massless thanks to a non-trivial symmetry of the two-field system~\cite{Achucarro:2016fby}.

The purpose of this article is to study the behavior of perturbations during reheating in models of inflation where the curvature perturbation interacts continuously with a non-adiabatic mode that remains approximately massless from horizon crossing until reheating~\cite{Amendola:2001ni,Wands:2002bn, Kobayashi:2010fm, Enqvist:2013qba, Vennin:2015egh, Achucarro:2016fby}. We are particularly interested in addressing the possibility of having the fluctuations coupled to an effective radiation component. We will show that a two-field system plus a radiation component may be effectively treated as a system with three scalar fields. In this way, we may use the covariant formalism in field-space, introduced in~\cite{GrootNibbelink:2001qt}, to derive the complete set of background and perturbed equations which describe the universe starting from two-field inflation up to reheating phase.

We have organized this article as follows. In Section~\ref{Sec: Two Scalar Fields plus Radiation}, we derive the background equations of motion for a two-field system coupled to a radiation fluid. There we show that the equations of motion may be effectively understood in terms of a three-field system. Then, in Section~\ref{Sec: Perturbations} we deduce the equations of motion for gauge invariant perturbations, such as the total curvature perturbation, the curvature perturbation associated with each individual component, and also the equations governing the evolution of the isocurvature modes. In Section~\ref{Sec: Analysis and Results}, we present the numerical results based on two inflation models: the product exponential potential and a model of inflation with ultra-light fields. We summarize our findings and present our conclusions in Section~\ref{Sec:conlusions}.

\setcounter{equation}{0}
\section{Two scalar fields plus radiation}  \label{Sec: Two Scalar Fields plus Radiation}

Here we introduce a covariant prescription to study two-field systems coupled to a radiation fluid. We will first consider the background equations of motion, and then move on to consider the dynamics for the perturbations (next section).

\subsection{Basic background equations} \label{SubSec: Background equations}

Let us start by considering the most general action describing a two-field system with two space-time derivatives~\cite{Langlois:2010xc} coupled to a radiation field (setting the reduced Planck mass $M_{\rm Pl}^2 = 1$ for simplicity):
\begin{eqnarray}
\label{action}
S &=& \int d^{4}x\sqrt{-g}\left[\frac{1}{2}R-\frac{1}{2}g^{\mu \nu}\gamma_{ab}\partial_{\mu}\phi^{a}\partial_{\nu}\phi^{b}-V(\phi)+\mathcal{L}_{\textup{rad}}+\mathcal{L}_{\textup{int}}\right].
\end{eqnarray}
Here $g_{\mu \nu}$ and $R$ are the usual metric and Ricci scalar respectively describing the gravitational sector. On the other hand $\gamma_{ab}$ is the $\sigma$-model metric characterizing the two-dimensional target space spanned by the fields $\phi^1$ and $\phi^2$ (with an inverse given by $\gamma^{ab}$), and $V$ is the scalar field potential of the model. The term $\mathcal{L}_{\textup{rad}}$ describes a radiation fluid that interacts with the fields $\phi^a$ through an interaction term $\mathcal{L}_{\textup{int}}$. These contributions will be parameterized through effective terms that will be introduced in what follows. The equations of motion for the scalar fields are found to be
\begin{eqnarray}
\label{KG1}
\Box \phi^a+\Gamma^{a}_{bc}g^{\mu \nu}\partial_{\mu}\phi^b\partial_{\nu}\phi^c = V^a+J^a,
\end{eqnarray}
where $\Box\phi^a=g^{\alpha \beta}\nabla_{\alpha}\nabla_{\beta}\phi^a=g^{\alpha \beta} (\partial_{\alpha}\partial_{\beta}\phi^a-\Gamma^{\gamma}_{\alpha \beta}\partial_{\gamma}\phi^a )$, and $V^a\equiv \gamma^{ab}V_b$ with $V_b=\partial_b V$. In the previous expressions, both $\Gamma^{\mu}_{\alpha \beta}$ and $\Gamma^a_{bc}$ correspond to the usual Christoffel symbols for the gravitational and scalar sectors respectively. For instance, in the case of the target space one has $\Gamma^a_{bc} = \frac{1}{2}\gamma^{ad}\left(\partial_c\gamma_{db}+\partial_b\gamma_{cd}-\partial_d\gamma_{bc}\right)$. Notice that we have introduced a source term $J^a$ that appears as a consequence of $\mathcal{L}_{\textup{int}}$. Now, if we take Eq.~(\ref{KG1}) at background level, the Klein-Gordon equation is reduced to
\begin{eqnarray}
\label{KG2}
\frac{D\dot{\phi}^a_0}{dt} + 3H\dot{\phi}^a_0 + V^a = - J_0^a,
\end{eqnarray}
with $\frac{D X^a}{dt} = \dot{X}^a + \Gamma^a_{bc}\dot{\phi}_0^b X^c$. The 0-label reminds us that we are dealing with background quantities. Note that while $V$ depends only on $\phi^a_0$, the source term $J^a_0$ may depend on both $\phi^a_0$ and the background radiation component.

Given that the source $J_0^a$ is due to the interaction between the scalar fields and radiation, we need to introduce a background equation of motion describing the evolution of the radiation energy density $\rho_R^0$ with a source term. This is given by:
\begin{eqnarray}
\dot{\rho}^0_R + 4H\rho_R^0 &=& Q.  \label{rad}
\end{eqnarray}
Here, $Q$ describes the energy transfer between the scalar fields and $\rho_R$. Notice that we have demanded that the radiation pressure satisfies $p_R^0 = \frac{1}{3}\rho_R^0$. In order to find the dependence between $J^a_0$ and $Q$ one may contract Eq.~(\ref{KG2}) by $\dot \phi^a_0$ to obtain:
\begin{eqnarray}
\dot{\varphi}_0\ddot{\varphi}_0 + 3H\dot{\varphi}_0^2 + \dot{\phi}_0^a\partial_a V &=& - J^0_a \dot{\phi}^a_0, \label{infl}
\end{eqnarray}
where we have defined $\dot{\varphi}^2_0 \equiv \gamma_{ab}\dot{\phi}^a_0 \dot{\phi}^b_0$. Now, in order to have energy-momentum conservation $\nabla_{\mu}T^{\mu \nu} = 0$, we require
\begin{eqnarray}
Q = J^0_a \dot{\phi}^a_0.
\end{eqnarray}
One way to think about $Q$ is that it corresponds to the zeroth component of a vector $Q^{\nu}=(Q,0)$ that defines the source for the radiation and scalar field contributions to the energy momentum tensor as~\cite{Malik:2008im}:
\begin{eqnarray}
\label{conserv rules 1}
\nabla_{\mu}T^{\mu \nu}_R &=& Q^{\nu}, \\
\label{conserv rules 2}
\nabla_{\mu}T^{\mu \nu}_{\phi} &=& - Q^{\nu}.
\end{eqnarray}
Next, it is direct to verify that the Friedmann equation is given by:
\begin{eqnarray}
\label{H2}
3H^2 = \rho^0 = \frac{1}{2}\dot{\varphi}^2_0 + V + \rho_R^0.
\end{eqnarray}
This equation, together with Eqs.~(\ref{rad}) and (\ref{infl}) implies the following additional equation for $\dot H$:
\begin{eqnarray}
\label{dotH}
\dot{H} = - \frac{\rho^0 + p^0}{2} = - \frac{1}{2}\left(\dot{\varphi}^2_0+\frac{4}{3}\rho_R^0\right).
\end{eqnarray}
Eqs.~(\ref{KG1}), (\ref{rad}) and (\ref{H2}) are the background equations of motion describing the system. The dependence of $J_0^a$ on the scalar fields and the radiation energy density is model dependent, and needs to be supplied.

\subsection{Standard parametrization of multi-field background trajectories} \label{SubSec: Standard parametrization of the background trajectory}

The impact of the inflationary background on the evolution of perturbations may be studied by the introduction of dimensionless parameters with clear geometrical interpretations. In particular, given that we are dealing with a multi-field trajectory, we will have two classes of useful dimensionless parameters: The first class corresponds to the set of slow-roll parameters, that give us information about the gradual evolution of the quasi-de Sitter background. The second class consists of quantities that parametrize the multi-field nature of the trajectory. In ref.~\cite{GrootNibbelink:2001qt} the following set of parameters were introduced in order to deal with general multi-field inflationary models:
\begin{eqnarray}
\label{epsilon def v1}
\varepsilon &\equiv& -\frac{\dot{H}}{H^2}, \\
\label{eta def v1}
\eta^a &\equiv& - \frac{1}{H\dot{\varphi}_0}\frac{D\dot{\phi}^a_0}{dt}. \label{def-eta-a}
\end{eqnarray}
The function $\varepsilon$ corresponds to the usual first slow-roll parameter that describes the rate at which the quasi-de Sitter spacetime change in time. The quantity $\eta^a$, on the other hand, contains mixed information about the evolution of the quasi-de Sitter spacetime and the multi-field inflationary trajectory. To clarify the meaning of $\eta^a$, it is convenient to introduce a basis of vectors aligned with the background inflationary trajectory. In the case of the present two-field system described by Eq.~(\ref{KG2}) one may introduce the following two unit vectors, $T^a$ and $N^a$, defined as:
\begin{eqnarray}
T^a &=& \frac{\dot \phi_0^a}{\dot \varphi_0}, \\
N^a &=& -\frac{1}{\Omega}  \frac{D T^a}{dt},
\end{eqnarray}
where $\Omega$ is a factor that keeps $N^{a}$ normalized, and that may be interpreted as the rate of turn of the inflationary path whenever it bends. Now, these two vectors may be used to project Eq.~(\ref{KG2}) along the inflationary trajectory, and orthogonal to it. One way of doing this is to directly compute a covariant time derivative of $T^a$ with the help of the equations of motion (\ref{KG2}). One finds:
\begin{eqnarray}
\label{DTa0}
\frac{DT^a}{dt} &=& - \left(\frac{\ddot{\varphi}_0}{\dot{\varphi}_0}+3H\right) T^a - \frac{1}{\dot{\varphi}_0}\left(V^a + J_0^a\right).
\end{eqnarray}
Then, projecting this equation along $T^a$, and using the fact that $T_a DT^a / dt = 0$, one finds:
\begin{eqnarray}
\ddot{\varphi}_0+3H \dot \varphi_0 +  V_T = - Q / \dot \varphi_0,
\end{eqnarray}
where $V_T= V^aT_a$. On the other hand, the projection of Eq.~(\ref{DTa0}) along $N^a$ gives
\begin{eqnarray}
\Omega =  \frac{1}{\dot{\varphi}_0}\left(V_N + J_0^a N_a \right),
\end{eqnarray}
where we have defined $V_N= V^aN_a$. Coming back to Eq.~(\ref{def-eta-a}), we may decompose $\eta^a$ along the two directions in the following way:
\begin{eqnarray}
\eta^a =  T^a \eta_{||} + N^a \eta_{\perp} .
\end{eqnarray}
Then, it is direct to find that the components $\eta_{||}$ and $\eta_{\perp}$ are given  by
\begin{eqnarray}
\eta_{||}  &=& - \frac{\ddot \varphi_0}{H \dot \varphi_0} = 3 H + \frac{V_T}{\dot \varphi_0} + \frac{Q}{\dot \varphi_0^2} , \\
\eta_{\perp} &=& \frac{\Omega}{H } = \frac{1}{H \dot{\varphi}_0}\left(V_N + J_0^a N_a \right) .
\end{eqnarray}
Now, it may be appreciated explicitly that this parametrization, based on Eq.~(\ref{def-eta-a}), becomes singular when $\dot \phi_0 \to 0$, which can happen at the end of inflation. In the following subsection we offer a simple extension of this parametrization allowing the essentially the same geometrical interpretation, but that has the benefit of avoiding any singular behavior at $\dot \phi_0 \to 0$.

\subsection{Radiation as an effective scalar} \label{SubSec: Radiation as an effective scalar}

We will now reconsider the previous set of background equations of motion by accommodating the radiation energy density (hereby described as a perfect fluid) as an extra scalar~\cite{Ray:1972}. The idea is to extend our previous two-dimensional field space to a three-dimensional one by introducing a new field $\phi^{(3)}_0$ satisfying a Klein-Gordon equation
\begin{eqnarray}
\label{KGphi3}
\frac{D\dot{\phi}^{(3)}_0}{dt} + 3H\dot{\phi}^{(3)}_0 + \mathcal{V}^{(3)} &=& - J_0^{(3)},
\end{eqnarray}
where $\mathcal{V}^{(3)}$ represents the derivative with respect to $\phi^{(3)}_0$ of an extended potential $\mathcal V$ that will be defined below [see Eq.~(\ref{def-V-mathcal})]. In order to achieve this, an effective field metric and an effective potential with a radiation component must be defined in such a way that they are consistent with Eqs.~ (\ref{KG2})-(\ref{dotH}). The aim is to accommodate this equation as part of the set of field equations describing multi-field inflation. With this in mind, we extend the set of two fields $\phi^a$ to the following set of scalar fields
\begin{eqnarray}
\phi^A_0 \equiv \left(
   \begin{array}{c}
     \phi^a_0 \\
    \phi^{(3)}_0 \\
   \end{array}
\right),
\end{eqnarray}
where $\phi^a_0$ stands for the original two fields. In addition, we extend the $\sigma$-model metric $\gamma_{ab}$ of the two-field target space to
\begin{eqnarray}
\mathbf{q}_{AB} =
\left(
\begin{array}{cc}
   \gamma_{ab} &    0   \\
         0     & \mathbf{q}_{(33)} \\
\end{array}
\right),
\end{eqnarray}
where $\mathbf{q}_{(33)}$ is an arbitrary function of $\phi^{(3)}$. having defined these extended objects, we may achieve Eq.~(\ref{KGphi3}), by identifying the radiation density with the kinetic energy of the new scalar field $\phi^{(3)}$ in the following way:
\begin{eqnarray}
\label{Rad Cond}
 \frac{2}{3}\rho_R^0 = \frac{1}{2} \mathbf{q}_{(33)}\left(\dot{\phi}^{(3)}_0\right)^2.
\end{eqnarray}
Then, the continuity equation (\ref{rad}) is equivalent to (\ref{KGphi3}) as long as the effective potential $\mathcal V$ and the source term $J_0^{(3)}$ are given by
\begin{eqnarray}
\mathcal{V} = V + \frac{1}{3}\rho_R^0 , \label{def-V-mathcal} \\
J_0^{(3)} = - \frac{Q}{\mathbf{q}_{(33)} \dot{\phi}^{(3)}_0}.
\end{eqnarray}
In the previous expressions, $\rho_R^0$ may be thought of as an explicit function of $\phi^{(3)}$. But given that $\phi^{(3)}$ is an auxiliary field, the specific dependence of $\rho_R^0$ on $\phi^{(3)}$ is in fact irrelevant. Instead, what matters is that the radiation density is determined by the scalar field $\phi^{(3)}$ in such a way that Eq.~(\ref{Rad Cond}) is satisfied. The auxiliary field $\phi^{(3)}$ satisfies Eq.~(\ref{KGphi3}), and it is equivalent to (\ref{rad}) multiply by $\mathbf{q}_{(33)} \dot \phi_0^{(3)}$.

Now, the full system consisting of two scalar fields plus radiation may be formally described through a covariant Klein-Gordon equation for three scalar fields given by:
\begin{eqnarray}
\label{KGfinal}
\frac{D\dot{\phi}^A_0}{dt} + 3H\dot{\phi}^A_0 + \mathcal{V}^A &=& - \mathcal{J}_0^A, \\
\label{H2final}
3H^2 &=& \frac{1}{2}\dot{\phi}^2_0 + \mathcal{V}, \\
\label{dotHfinal}
\dot{H} &=& - \frac{1}{2}\dot{\phi}^2_0,
\end{eqnarray}
where we have defined
\begin{eqnarray}
\dot{\phi}^2_0 \equiv \mathbf{q}_{AB}\dot{\phi}^A_0 \dot{\phi}^B_0 = \dot{\varphi}^2_0 + \frac{4}{3}\rho_R^0. \label{phi-varphi-rel}
\end{eqnarray}
This time, the covariant time derivative acts on a given vector $X^A$ as
\begin{eqnarray}
\frac{D X^A}{dt} = \dot{X}^A + \Gamma^A_{BC}\dot{\phi}_0^B X^C,
\end{eqnarray}
where the Christoffel symbols are now given by $\Gamma^A_{BC} = \frac{1}{2}\mathbf{q}^{AD}\left(\partial_C\mathbf{q}_{DB}+\partial_B\mathbf{q}_{CD}-\partial_D\mathbf{q}_{BC}\right)$. In addition, the source term $\mathcal{J}_0^A$ of Eq.~(\ref{KGfinal}) is given by
\begin{eqnarray}
\mathcal{J}_0^A = \left(  J_0^a \, , \,  -\frac{J^0_b\dot{\phi}^b_0}{\mathbf{q}_{(33)}\dot{\phi}_0^{(3)}} \right).
\end{eqnarray}
In all of the previous expressions, the capital latin indices go from 1 to 3. Then, Eq.~(\ref{KGfinal}) with $A=1,2$ correspond to (\ref{KG2}), $A=3$ to (\ref{rad}) and the Friedmann equations are given by (\ref{H2final}) and (\ref{dotHfinal}).

Following ref.~\cite{GrootNibbelink:2001qt}, we introduce a set of orthogonal unit vectors to parameterize our new three-dimensional field-space. First, we write the tangent vector to the inflationary trajectory as $\mathcal{T}^A = \dot{\phi}^A_0 / \dot{\phi}_0$. Then, by introducing it in Eq.~(\ref{KGfinal}), we obtain:
\begin{eqnarray}
\label{DTA0}
\frac{D\mathcal{T}^A}{dt} &=& - \left(\frac{\ddot{\phi}_0}{\dot{\phi}_0}+3H\right)\mathcal{T}^A - \frac{1}{\dot{\phi}_0}\left(\mathcal{V}^A + \mathcal{J}_0^A\right).
\end{eqnarray}
Now, by projecting this equation along $\mathcal{T}^A$, we obtain the following Klein-Gordon equation for the field $\phi_0$:
\begin{eqnarray}
\label{phimod}
\ddot{\phi}_0 + 3H\dot{\phi}_0 + \mathcal{V}_\mathcal{T} = 0,
\end{eqnarray}
where $\mathcal{V}_\mathcal{T} \equiv \mathcal{T}^A\partial_AV$. Notice that, in order to derive the previous equation, we used that $\mathcal{J}_\mathcal{T} = \mathcal{T}_A\mathcal{J}_0^A = 0$ and $\mathcal{T}_A\frac{D\mathcal{T}^A}{dt} = 0$. Next, we define two additional vectors, $\mathcal{N}^A$ and $\mathcal{B}^A$, to complete an orthogonal basis around the inflationary trajectory, such that $\mathcal{T}^A\mathcal{N}_A = \mathcal{T}^A\mathcal{B}_A = \mathcal{N}^A\mathcal{B}_A = 0$ and $\mathcal{T}^A\mathcal{T}_A = \mathcal{N}^A\mathcal{N}_A = \mathcal{B}^A\mathcal{B}_A = 1$ (for instance, see Ref.~\cite{Cespedes:2013rda}). The normal vector $\mathcal{N}^A$ is defined in such a way that it remains parallel to the time variation of $\mathcal{T}^A$ (that is $\mathcal{N}^A \propto D_t \mathcal{T}^A$).  This definition implies
\begin{eqnarray}
\label{DTA}
\frac{D\mathcal{T}^A}{dt} &=& - \frac{\left(\mathcal{V}_\mathcal{N} + \mathcal{J}_\mathcal{N}\right)}{\dot{\phi}_0}\mathcal{N}^A.
\end{eqnarray}
Then, it is direct to verify that the combination $\mathcal{V}^A + \mathcal{J}_0^A $ decomposes only along the subspace spanned by $\mathcal{T}^A$ and $\mathcal{N}^A$:
\begin{eqnarray}
\mathcal{V}^A + \mathcal{J}_0^A = \mathcal{V}_\mathcal{T}\mathcal{T}^A + \left(\mathcal{V}_\mathcal{N} + \mathcal{J}_\mathcal{N}\right)\mathcal{N}^A,
\end{eqnarray}
where $\mathcal{V}_\mathcal{N} = \mathcal{V}^A\mathcal{N}_A$ and $\mathcal{J}_\mathcal{N} = \mathcal{J}_0^A \mathcal{N}_A$. This, in turn, implies that the bi-normal vector $\mathcal{B}^A$ satisfies $\mathcal{V}_\mathcal{B} + \mathcal{J}_\mathcal{B} = 0$. The time variations of $\mathcal{N}^A$ and $\mathcal{B}^A$ are found to satisfy
\begin{eqnarray}
\label{DNA}
\frac{D\mathcal{N}^A}{dt} &=& \frac{\left(\mathcal{V}_\mathcal{N} + \mathcal{J}_\mathcal{N}\right)}{\dot{\phi}_0}\mathcal{T}^A - \frac{D\mathcal{B}^B}{dt}\mathcal{N}_B\mathcal{B}^A, \\
\label{DBA}
\frac{D\mathcal{B}^A}{dt} &=& \frac{D\mathcal{B}^B}{dt}\mathcal{N}_B\mathcal{N}^A.
\end{eqnarray}
We present more details about these orthogonal vectors like a base of the space fields in Appendix \ref{app:FieldsSpace}, where we introduce a special parametrization to solve the equations of motion. We can now introduce the slow-roll parameters $\varepsilon$ and $\eta^A$, defined as:
\begin{eqnarray}
\label{epsilon def}
\varepsilon &\equiv& -\frac{\dot{H}}{H^2} = \frac{\dot{\phi}^2_0}{2H^2}, \\
\label{eta def}
\eta^A &\equiv& - \frac{1}{H\dot{\phi}_0}\frac{D\dot{\phi}^A_0}{dt} = \left(3+\frac{\mathcal{V}_\mathcal{T}}{H\dot{\phi}_0}\right)\mathcal{T}^A + \left(\frac{\mathcal{V}_\mathcal{N} + \mathcal{J}_\mathcal{N}}{H\dot{\phi}_0}\right)\mathcal{N}^A,
\end{eqnarray}
where we have used Eq.~(\ref{KGfinal}). It is interesting to mention that Eq.~(\ref{eta def}) corresponds to the three-dimensional extension of $\eta^a$
of Ref.~\cite{Cespedes:2012hu}. We can decompose $\eta^A$ along the normal and tangent directions by introducing two independent parameters $\eta^A = \eta_{\parallel}\mathcal{T}^A + \eta_{\perp}\mathcal{N}^A$, with:
\begin{eqnarray}
\label{etpar}
\eta_{\parallel} &=& - \frac{\ddot{\phi}_0}{H\dot{\phi}_0} = \varepsilon - \frac{\dot{\varepsilon}}{2H\varepsilon}, \\
\label{etpr}
\eta_{\perp} &=& \frac{\mathcal{V}_\mathcal{N} + \mathcal{J}_\mathcal{N}}{H\dot{\phi}_0}.
\end{eqnarray}
We note that $\eta_{\parallel}$ may be recognized as the usual $\eta$ slow-roll parameter in single field inflation. On the other hand $\eta_{\perp}$ informs us about the rate with which $\mathcal{T}^A$ rotates, and therefore it parameterizes the rate of turn of the trajectory followed by the scalar fields.

Let us briefly emphasize an important aspect of this new parametrization. Notice that $\dot \phi_0^2$ and $\dot \varphi_0^2$ are related by Eq.~(\ref{phi-varphi-rel}). This implies that the new geometrical parameters appearing in (\ref{eta def}) will not be singular as $\dot \varphi_0 \to 0$. As we shall see, after inflation ends, and we enter into the radiation era, one has $\dot{\phi}_0 \rightarrow 2H$. In conclusion, this parametrization allows us to deal with well-defined inflationary parameters using an effective potential $\mathcal V$ in addition to multi-field interactions. The relevant equations relating geometrical dimensionless parameters and the potential $\mathcal V$, are given by
\begin{eqnarray}
\label{V}
\mathcal{V} &=& H^2(3-\varepsilon), \\
\label{VT}
\mathcal{V}_\mathcal{T} &=& - H\dot{\phi}_0\left(3-\eta_{\parallel}\right), \\
\label{VN}
\mathcal{V}_\mathcal{N} + \mathcal{J}_\mathcal{N} &=& H\dot{\phi}_0\eta_{\perp}, \\
\label{VB}
\mathcal{V}_\mathcal{B} + \mathcal{J}_\mathcal{B} &=& 0,
\end{eqnarray}
and the equations of motion:
\begin{eqnarray}
\label{bg eq 1}
\dot{\phi}_0^2 &=& 2H^2\varepsilon, \\
\label{bg eq 2}
\frac{D\mathcal{T}^A}{dt} &=& - H\eta_{\perp}\mathcal{N}^A,  \\
\label{bg eq 3}
\frac{D\mathcal{N}^A}{dt} &=& H\eta_{\perp}\mathcal{T}^A - HC\mathcal{B}^A,  \\
\label{bg eq 4}
\frac{D\mathcal{B}^A}{dt} &=& HC\mathcal{N}^A,
\end{eqnarray}
with $C \equiv H^{-1}\frac{D\mathcal{B}^B}{dt}\mathcal{N}_B$. Additionally, the initial conditions will be fixed in such a way that the radiation density is null in the beginning of inflation. This means that $\mathcal{T}^{(3)},\mathcal{N}^{(3)} \rightarrow 0$ and we can choose:
\begin{eqnarray}
\mathcal{B}^A = \left( 0 \, , \, 0 \, , \, \frac{1}{\sqrt{\mathbf{q}_{(33)}}}   \right).
\end{eqnarray}
More details about this are provided in Appendix \ref{app:FieldsSpace}. Also, we describe the interaction contribution, given by $\mathcal{J}_\mathcal{N}$ and $\mathcal{J}_\mathcal{B}$, in Appendix \ref{app:Interaction}. Finally, in order to deal with the perturbative equations in the next section, we introduce the following parameters:
\begin{eqnarray}
\label{xis}
\xi_{\parallel} &\equiv& - \frac{\dot{\eta}_{\parallel}}{H\eta_{\parallel}}, \\
\xi_{\perp} &\equiv& - \frac{\dot{\eta}_{\perp}}{H\eta_{\perp}}.
\end{eqnarray}

\setcounter{equation}{0}
\section{Perturbations} \label{Sec: Perturbations}

In this section we consider the dynamics of scalar perturbations, parameterizing departures from the homogeneous and isotropic background. This may be done by defining perturbations $\delta \phi^a$, $\delta g_{\mu \nu}$, $\delta \Gamma^{\gamma}_{\mu \nu}$ and $\delta \Gamma^{a}_{bc}$ as
\begin{eqnarray}
\phi^a\left(t,{\bf{x}}\right) &=& \phi^a_0\left(t\right) + \delta \phi^a\left(t,\bf{x}\right), \\
g_{\mu \nu}\left(t,\bf{x}\right) &=& g^0_{\mu \nu}\left(t\right) + \delta g_{\mu \nu}\left(t,\bf{x}\right), \\
\Gamma^{\gamma}_{\mu \nu}\left(t,\bf{x}\right) &=& \Gamma^{\gamma0}_{\mu \nu}\left(t\right) + \delta \Gamma^{\gamma}_{\mu \nu}\left(t,\bf{x}\right), \\
\gamma_{ab}\left(\phi\right) &=& \gamma^0_{ab}\left(\phi_0\right) + \delta \gamma_{ab}\left(\phi_0,\delta \phi\right), \\
\Gamma^a_{bc}\left(\phi\right) &=& \Gamma^{a0}_{bc}\left(\phi_0\right) + \delta \Gamma^{a}_{bc}\left(\phi_0,\delta \phi\right),
\end{eqnarray}
where the $0$-label denotes background quantities. The Greek and Latin indices correspond to the coordinate space and field space respectively. So, $\Gamma^{\gamma}_{\mu \nu}$ are the Christoffel symbols associated to $g_{\mu \nu}$ and $\Gamma^a_{bc}$ is related to $\gamma_{ab}$.

We must consider perturbations to the homogeneous background space-time and the energy-momentum tensor of the universe. The most general first-order perturbation to a spatially flat FLRW metric is \cite{Malik:2008im}:
\begin{eqnarray}
\label{generalds}
ds^2 = - \left(1+2\Phi\right)dt^2 + 2aB_idx^idt + a^2\left(\left(1-2\Psi\right)\delta_{ij}+2E_{ij}\right)dx^idx^j,
\end{eqnarray}
where $\Phi$ is the lapse, $B_i$ is the shift, $\Psi$ is the spatial curvature perturbation, and $E_{ij}$ is the symmetric shear tensor. The symmetries of the FLRW background space-time allow linear perturbations to be decomposed into independent scalar, vector and tensor components. This reduces the linearized Einstein equations to a set of uncoupled ordinary differential equations. So, the metric in Eq.~(\ref{generalds}) can be rewritten as:
\begin{eqnarray}
\label{descds}
ds^2 &=& - (1+2\Phi)dt^2 + 2a\partial_iBdx^idt + a^2\left(\left(1-2\Psi\right)\delta_{ij}+2\partial_i\partial_jE\right)dx^idx^j \nonumber \\
&&  + 2aG_idx^idt + 2a^2\left(\partial_iC_{j}+\partial_jC_{i}\right)dx^idx^j  + h_{ij}dx^idx^j,
\end{eqnarray}
where we have introduced the scalars $B$ and $E$, the vectors $G_i$ and $C_i$ and the pure tensor field $h_{ij}$ to decompose the shift and the shear. For this decomposition to be unique, the fields $G_i$, $C_i$ and $h_{ij}$ must satisfy the constraints $\partial_iG_i = \partial_iC_i = \partial_ih_{ij} = \partial_jh_{ij} = h_{ii} = 0$. Note that two of the four scalar perturbations ($\Phi$, $B$, $\Psi$ and $E$) can be eliminated by the specific gauge choice. In the same way, one of the vector perturbations ($G_i$ and $C_i$) can be removed, but the tensor perturbation ($h_{ij}$) is a gauge invariant. In this paper, we will rewrite the perturbation equations in a gauge-invariant form, because these become particularly simple as we will see soon.

Then focusing our attention to the scalar degrees of freedom, the metric is found to be given as:
\begin{equation}
\label{g}
g_{\mu \nu} = \left[\begin{array}{cc}
                    -\left(1+2\Phi\right) & a\partial_i B \\
                    a\partial_i B & a^2\left(\left(1-2\Psi\right)\delta_{ij}+2\partial_i \partial_j E\right)
                  \end{array}
\right],
\end{equation}
whereas the inverse metric, up to linear order, is given by
\begin{equation}
\label{g inv}
g^{\mu \nu} = \left[\begin{array}{cc}
                    -\left(1-2\Phi\right) & a^{-1}\partial^i B \\
                    a^{-1}\partial^i B & a^{-2}\left(\left(1+2\Psi\right)\delta^{ij}-2\partial^i \partial^j E\right)
                  \end{array}
\right].
\end{equation}
On the other hand the Christoffel symbols become:
\begin{eqnarray}
\label{Christ}
\Gamma^0_{00} &=& \dot{\Phi},\,\,\,\,\,\,\,\,\,\,\Gamma^0_{0i}=\Gamma^0_{i0}=\partial_{i}\left(\Phi+aHB\right),  \\
\Gamma^0_{ij} &=& \delta_{ij} a^2\left(H-2H\left(\Psi+\Phi\right)-\dot{\Psi}\right) +a^2\partial_i \partial_j\left(2HE+\left(\dot{E}-\frac{B}{a}\right)\right),  \\
\Gamma^{i}_{00} &=& a^{-2}\partial_{i}\left(\Phi+\dot{\left(aB\right)}\right), \\
\Gamma^{i}_{0j} &=& \Gamma^{i}_{j0}=\delta_{ij}\left(H-\Psi\right)+\partial_i \partial_j\dot{E},  \\
\Gamma^{i}_{jk} &=& -aH\delta_{jk}\partial_i B-\delta_{ij}\partial_k\Psi-\delta_{ik}\partial_j\Psi + \delta_{jk}\partial_i\Psi+\partial_i \partial_j \partial_k E.
\end{eqnarray}
From these, we may compute the perturbed Einstein tensor, whose components are given by:
\begin{eqnarray}
\delta G^0_0 &=& 6H\left(\dot{\Psi}+H\Phi\right)-2\frac{\partial^2}{a^2}\left[H(a^2\dot{E}-aB)+\Psi\right], \\
\delta G^i_j &=& \left[2\left(3H^2+2\dot{H}\right)\Phi+2H\left(\dot{\Phi}+3\dot{\Psi}\right)+2\ddot{\Psi}-\partial^2\left(\frac{d}{dt}\left(\dot{E}-\frac{B}{a}\right)+3H\left(\dot{E}-\frac{B}{a}\right)\right.\right.\nonumber\\
&&\left.\left.+\frac{\left(\Psi-\Phi\right)}{a^2}\right)\right]\delta^i_j+\delta^{ik}\partial_k \partial_j\left[\frac{d}{dt}\left(\dot{E}-\frac{B}{a}\right)+3H\left(\dot{E}-\frac{B}{a}\right)+\frac{\left(\Psi-\Phi\right)}{a^2}\right], \\
\delta G^i_0 &=& \frac{2}{a^2}\delta^{ij}\partial_j\left[\dot{\Psi}+H\Phi-\dot{H}B\right], \\
\delta G^0_i &=& -\frac{2}{a^2}\partial_i\left[\dot{\Psi}+H\Phi\right].
\end{eqnarray}
In order to construct the perturbed field equations, we need to consider, besides the gravitational fields, the effects that induce the matter perturbations. The components of the perturbed total energy-momentum tensor can be expressed as \cite{Malik:2008im}:
\begin{eqnarray}
\delta T^0_0 &=& - \rho^1, \\
\delta T^0_i  &=& q^1_i, \\
\delta T^i_0  &=& \frac{\delta^{ij}}{a^2}\left[\left(\rho^0+p^0\right)\partial_j B-q^1_j\right], \\
\delta T^i_j &=& p^1\delta^i_j+\Sigma^i_j,
\end{eqnarray}
where $q^1_{i} = - \left(\rho^0+p^0\right)u^1_{i}$ is the 3-momentum density, and $u^1_{i}$ is the velocity perturbation of the matter content. $\Sigma^i_j$ is a anisotropic stress tensor. To relate the metric and energy-momentum perturbations, we must use the perturbed Einstein field equations at linear order given by:
\begin{eqnarray}
2\frac{\partial^2}{a^2}\left[H\left(a^2\dot{E}-aB\right)+\Psi\right]-6H\left(\dot{\Psi}+H\Phi\right) &=& \rho^1, \\
\left[2\left(3H^2+2\dot{H}\right)\Phi+2H\left(\dot{\Phi}+3\dot{\Psi}\right)+2\ddot{\Psi}-\partial^2\left(\frac{d}{dt}\left(\dot{E}-\frac{B}{a}\right)\right.\right. \nonumber \\
\left.\left. +3H\left(\dot{E}-\frac{B}{a}\right)+\frac{\left(\Psi-\Phi\right)}{a^2}\right)\right]\delta^i_j \nonumber \\
+\delta^{ik}\partial_k \partial_j\left[\frac{d}{dt}\left(\dot{E}-\frac{B}{a}\right)+3H\left(\dot{E}-\frac{B}{a}\right)+\frac{\left(\Psi-\Phi\right)}{a^2}\right] &=& p^{1}\delta^i_j+\Sigma^i_j, \\
- 2\partial_i\left(H\Phi+\dot{\Psi}\right) &=& q^1_i.
\end{eqnarray}
In general, $\Sigma^i_j$ for $i\neq j$ contains a non diagonal component called viscous pressure. On the other hand, the velocity perturbation can be decomposed into a scalar and vector components, where the first one can be written as $u^1_{i}=\partial_i u^1$. Then, the perturbed field equations, in the absence of anisotropic stress, reduce to:
\begin{eqnarray}
\label{pert eq 1}
\frac{d}{dt}\left(\dot{E}-\frac{B}{a}\right)+3H\left(\dot{E}-\frac{B}{a}\right)+\frac{\left(\Psi-\Phi\right)}{a^2} &=& 0, \\
\label{pert eq 2}
2\frac{\partial^2}{a^2}\left[a^2H\left(\dot{E}-\frac{B}{a}\right)+\Psi\right]-6H\left(\dot{\Psi}+H\Phi\right)-\rho^1 &=& 0, \\
\label{pert eq 3}
2\left(3H^2+2\dot{H}\right)\Phi+2H\left(\dot{\Phi}+3\dot{\Psi}\right)+2\ddot{\Psi} - p^1 &=& 0, \\
\label{pert eq 4}
\left(\dot{\Psi}+H\Phi\right) + \dot{H}u^1 &=& 0,
\end{eqnarray}
where we used that $\rho^0+p^0 = -2\dot{H}$. To extract physical results it is useful to define gauge-invariant combinations of the scalar metric perturbations. Two relevant quantities are known as the Bardeen potentials, defined as \cite{Malik:2008im}:
\begin{eqnarray}
\label{bard1}
\Phi_\mathbf{B} &\equiv& \Phi-\frac{d}{dt}\left[a^2\left(\dot{E}-\frac{B}{a}\right)\right], \\
\label{bard2}
\Psi_\mathbf{B} &\equiv& \Psi+a^2H\left(\dot{E}-\frac{B}{a}\right).
\end{eqnarray}
On the other hand, the matter perturbations are also gauge-dependent. However, it is possible to construct a set of gauge-invariant quantities given by:
\begin{eqnarray}
\label{drhoinv}
\rho^{1I} &=& \rho^1 + \frac{\dot{\rho}^0}{H}\Psi, \\
\label{dpinv}
p^{1I} &=& p^1 + \frac{\dot{p}^0}{H}\Psi, \\
\label{Rinv}
\mathcal{R} &=& \Psi + H u^1.
\end{eqnarray}
These correspond to gauge invariant expressions for the energy density, pressure, and the total curvature perturbation, respectively. In terms of these gauge-invariant quantities, the perturbed Einstein equations may be rewritten as:
\begin{eqnarray}
\label{Einst EQ 1}
\Phi_\mathbf{B} - \Psi_\mathbf{B} &=& 0, \\
\label{Einst EQ 2}
2\frac{\partial^2}{a^2}\Phi_\mathbf{B} - 6H^2\varepsilon\mathcal{R} - \rho^{1I} &=& 0, \\
\label{Einst EQ 3}
2H\varepsilon\dot{\mathcal{R}} + 2H^2\varepsilon\left(3-2\eta_{\parallel}\right)\mathcal{R} - p^{1I} &=& 0, \\
\label{Einst EQ 4}
\dot{\Phi}_\mathbf{B} - H\varepsilon\mathcal{R} + H\left(1+\varepsilon\right)\Phi_\mathbf{B} &=& 0,
\end{eqnarray}
for which we used the inflationary parameters given by Eqs.~(\ref{epsilon def})-(\ref{etpr}). In this work, we will consider that the total matter content of the universe is given by a two-fields system interacting with radiation. However, in the last section, radiation was represented by an additional scalar field. This fact has to be taken into account to define the total density, pressure and scalar velocity, presented in Appendix \ref{app:FluidComp} where $\left(\delta\phi^{(3)}\right)$ is introduced in the system. We will give more detail about this later.

Now, we will perturb the Klein-Gordon equation (\ref{KG1}) at first order. This yields:
\begin{eqnarray}
&& \left(g^{\alpha \beta}_0+\delta g^{\alpha \beta}\right)\left[\partial_{\alpha}\partial_{\beta}\left(\phi^a_0+\delta \phi^a\right)-\left(\Gamma^{\gamma0}_{\alpha \beta}+\delta \Gamma^{\gamma}_{\alpha \beta}\right)\partial_{\gamma}\left(\phi^a_0+\delta \phi^a\right)\right] \\
&& +\left(\Gamma^{a0}_{bc}+\delta\Gamma^{a}_{bc}\right)\left(g^{\mu \nu}_0+\delta g^{\mu \nu}\right)\partial_{\mu}\left(\phi^b_0+\delta \phi^b\right)\partial_{\nu}\left(\phi^c_0+\delta \phi^c\right) = V^a+\partial_bV^a\delta \phi^b+J_0^a+\left(\delta J^a\right). \nonumber
\end{eqnarray}
Expanding, we note that the zeroth order term corresponds to the Klein Gordon equation. Now, the first order term is:
\begin{eqnarray}
\ddot{\left(\delta \phi^a\right)}-\frac{\partial^2}{a^2}\left(\delta \phi^a\right)+3H\dot{\left(\delta \phi^a\right)}+2\Gamma^{a0}_{bc}\dot{\phi}^b_0\dot{\left(\delta \phi^c\right)}
+\dot{\phi}^b_0 \dot{\phi}^c_0\left(\delta \Gamma^{a}_{bc}\right)+\partial_b\left(\gamma^{ac}\partial_cV\right)\left(\delta \phi^b\right) &&  \nonumber \\
 + 2\left(V^{a}+J_0^{a}\right)\Phi - \dot{\phi}^a_0\left(\dot{\Phi} + 3\dot{\Psi} - \partial^2\left(\dot{E}-\frac{B}{a}\right)\right) + \left(\delta J^a\right) &=& 0,
\end{eqnarray}
for which we used Eqs.~(\ref{g})-(\ref{Christ}). To write this equation in a covariant form, we have:
\begin{eqnarray}
\partial_b\left(\gamma^{ac}\partial_cV\right) &=& V^a_{\,\,\,\,\,b} - \Gamma^{a0}_{bc}V^c \textrm{, } \quad \left(\delta \Gamma^{a}_{bc}\right) = \partial_d\Gamma^{a0}_{bc}(\delta \phi^d), \\
\left(\delta J^a\right) &=& \left(\Delta J^a\right) - \Gamma^{a0}_{bc}J_0^c\left(\delta\phi^b\right) \textrm{, } \quad \dot{\left(\delta \phi^a\right)} = \frac{D\left(\delta\phi^a\right)}{dt} - \Gamma^{a0}_{bc}\dot{\phi}_0^b\left(\delta\phi^c\right), \\
\ddot{\left(\delta \phi^a\right)} &=& \frac{D^2\left(\delta\phi^a\right)}{dt^2} - 2\Gamma^{a0}_{bc}\dot{\phi}_0^b\frac{D\left(\delta\phi^c\right)}{dt} - \Gamma^{a0}_{bc}\frac{D\dot{\phi}^b_0}{dt}\left(\delta\phi^c\right) \nonumber \\
&& - \dot{\phi}_0^b\dot{\phi}_0^c\left(\partial_c\Gamma^{a0}_{bd} - \Gamma^{a0}_{be}\Gamma^{e0}_{dc} - \Gamma^{a0}_{de}\Gamma^{e0}_{bc}\right)\left(\delta\phi^d\right).
\end{eqnarray}
Here, $V^a_{\,\,\,\,\,b} = D_bD^a V$, where $D_a$ represents the covariant derivative in the field space, and $\left(\Delta J^a\right)$ is a vector in the field space. So, by using Eq.~(\ref{KG2}) and the expressions from above, the perturbative Klein-Gordon's equation becomes:
\begin{eqnarray}
\label{pert eq 5 0}
\frac{D^2}{dt^2}\left(\delta\phi^a\right) - \frac{\partial^2}{a^2}\left(\delta \phi^a\right) + 3H\frac{D}{dt}\left(\delta\phi^a\right) + V^a_{\,\,\,\,\,b}\left(\delta \phi^b\right) - \dot{\phi}_0^b\dot{\phi}_0^c\mathbb{R}^a_{\,\,bcd}\left(\delta\phi^d\right) && \nonumber \\
+ 2\left(V^a+J_0^a\right)\Phi - \dot{\phi}^a_0\left(\dot{\Phi} + 3\dot{\Psi} - \partial^2\left(\dot{E}-\frac{B}{a}\right)\right) + \left(\Delta J^a\right) &=& 0,
\end{eqnarray}
where $\mathbb{R}^a_{\,\,bcd} = \partial_c\Gamma^{a0}_{bd} - \partial_d\Gamma^{a0}_{bc} + \Gamma^{a0}_{ce}\Gamma^{e0}_{bd} - \Gamma^{a0}_{de}\Gamma^{e0}_{bc}$ is the Riemann tensor in the 2D-field space. Now, just as we did with the background, we now extend Eq.~(\ref{pert eq 5 0}) to a three-field version, in which one of the perturbations $\left(\delta\phi^{(3)}\right)$ corresponds to an auxiliary field identified with the radiation perturbation. In other words, we define a fluctuation $\delta\phi^{(3)}$ in such a way that the following set of equations are valid:
\begin{eqnarray}
\label{pert eq 5}
\frac{D^2}{dt^2}\left(\delta\phi^A\right) - \frac{\partial^2}{a^2}\left(\delta \phi^A\right) + 3H\frac{D}{dt}\left(\delta\phi^A\right) + \left(\mathcal{V}^A_{\,\,\,\,\,B} - \dot{\phi}_0^2\mathbb{R}^A_{\,\,\mathcal{T}\mathcal{T}B}\right)\left(\delta\phi^B\right) && \nonumber\\
+ 2\left(\mathcal{V}^A+\mathcal{J}_0^A\right)\Phi - \dot{\phi}_0\mathcal{T}^A\left(\dot{\Phi} + 3\dot{\Psi} - \partial^2\left(\dot{E}-\frac{B}{a}\right)\right) + \left(\Delta \mathcal{J}^A\right) &=& 0,
\end{eqnarray}
where $\mathbb{R}^A_{\,\,\mathcal{T}\mathcal{T}B} \equiv \mathcal{T}^C\mathcal{T}^D\mathbb{R}^A_{\,\,CDB}$ corresponds to the 3D Riemann tensor with the metric $\mathbf{q}_{AB}$, mentioned in Section \ref{SubSec: Background equations}, and $\left(\Delta \mathcal{J}^A\right)$ is the effective perturbative interaction. Recall that $\mathcal{V} = V(\phi_0^{(1)},\phi_0^{(2)}) + \frac{1}{3}\rho^0_R(\phi_0^{(3)})$, and so $\mathcal{V}^{A=1,2}_{\,\,\,\,\,(3)} = 0$. In addition, the only non-zero component of the Riemann tensor is $\mathbb{R}_{(1212)}$. Therefore, it is easy to see that Eq.~(\ref{pert eq 5}) with $A=1,2$ reduces to (\ref{pert eq 5 0}). In Appendix \ref{app:FluidComp}, we show that the third component of Eq.~(\ref{pert eq 5}) is equivalent to a perturbative equation of a perfect radiation fluid.

To continue, we will rewrite (\ref{pert eq 5}) in terms of gauge invariant quantities. For this, we define:
\begin{eqnarray}
\label{dphi}
\left(\delta\phi^A\right) &=& \frac{\dot{\phi}_0}{H}\left[ \left(\mathcal{R} - \Psi\right)\mathcal{T}^A + \mathcal{S}_\mathcal{N}\mathcal{N}^A + \mathcal{S}_\mathcal{B}\mathcal{B}^A\right], \\
\label{J1I}
\left(\Delta \mathcal{J}^A\right) &=& \left(\Delta \mathcal{J}^A\right)^I - \frac{1}{H}\frac{D}{dt}\left(\mathcal{J}_0^A\right)\Psi,
\end{eqnarray}
where the total curvature $\mathcal{R}$, the isocurvature components $\mathcal{S}_\mathcal{N}$ and $\mathcal{S}_\mathcal{B}$ and $\left(\Delta \mathcal{J}^A\right)^I$ are gauge invariant. Now, from Eq.~(\ref{dphi}), we have:
\begin{eqnarray}
\label{DQ a}
\frac{D}{dt}\left(\delta\phi^A\right) &=& f_\mathcal{T} \mathcal{T}^A + f_\mathcal{N} \mathcal{N}^A + f_\mathcal{B} \mathcal{B}^A, \\
\label{DDQ a}
\frac{D^2}{dt^2}\left(\delta\phi^A\right) &=& \left(\dot{f}_\mathcal{T} + H\eta_{\perp}f_\mathcal{N}\right)\mathcal{T}^A + \left(\dot{f}_\mathcal{N} - H\eta_{\perp}f_\mathcal{T} + HCf_\mathcal{B}\right)\mathcal{N}^A + \left(\dot{f}_\mathcal{B} - HCf_\mathcal{N}\right)\mathcal{B}^A,
\end{eqnarray}
with $C=\frac{\mathcal{N}_B}{H}\frac{D}{dt}\left(\mathcal{B}^B\right)$, and:
\begin{eqnarray}
\label{fT}
f_\mathcal{T} \equiv \mathcal{T}_A\frac{D}{dt}\left(\delta\phi^A\right) &=& \frac{\dot{\phi}_0}{H}\left(\dot{\mathcal{R}}-\dot{\Psi}\right) + \dot{\phi}_0\left(\varepsilon-\eta_{\parallel}\right)\left(\mathcal{R}-\Psi\right) + \dot{\phi}_0\eta_{\perp}\mathcal{S}_\mathcal{N}, \\
\label{fN}
f_\mathcal{N} \equiv \mathcal{N}_A\frac{D}{dt}\left(\delta\phi^A\right) &=& \frac{\dot{\phi}_0}{H}\dot{\mathcal{S}}_\mathcal{N} + \dot{\phi}_0\left(\varepsilon-\eta_{\parallel}\right)\mathcal{S}_\mathcal{N} - \dot{\phi}_0\eta_{\perp}\left(\mathcal{R}-\Psi\right) + \dot{\phi}_0C\mathcal{S}_\mathcal{B}, \\
\label{fB}
f_\mathcal{B} \equiv \mathcal{B}_A\frac{D}{dt}\left(\delta\phi^A\right) &=& \frac{\dot{\phi}_0}{H}\dot{\mathcal{S}}_\mathcal{B} + \dot{\phi}_0\left(\varepsilon-\eta_{\parallel}\right)\mathcal{S}_\mathcal{B} - \dot{\phi}_0C\mathcal{S}_\mathcal{N}.
\end{eqnarray}
In addition, from Appendix \ref{app:Interaction}, we have:
\begin{eqnarray}
\left(\Delta \mathcal{J}_\mathcal{T}\right)^I &=& \eta_{\perp}\mathcal{J}_\mathcal{N}\mathcal{R} + \dot{\phi}_0\tau_\mathcal{N}\dot{\mathcal{S}}_\mathcal{N} + \left(\frac{\dot{\mathcal{J}}_\mathcal{N}}{H}+C\mathcal{J}_\mathcal{B}+\dot{\phi}_0\left(\dot{\tau}_\mathcal{N}+H\left(3-2\eta_{\parallel}\right)\tau_\mathcal{N}\right)\right)\mathcal{S}_\mathcal{N} \nonumber \\
&& + \dot{\phi}_0\tau_\mathcal{B}\dot{\mathcal{S}}_\mathcal{B} + \left(\frac{\dot{\mathcal{J}}_\mathcal{B}}{H}-C\mathcal{J}_\mathcal{N}+\dot{\phi}_0\left(\dot{\tau}_\mathcal{B}+H\left(3-2\eta_{\parallel}\right)\tau_\mathcal{B}\right)\right)\mathcal{S}_\mathcal{B},
\\
\left(\Delta \mathcal{J}_\mathcal{N}\right)^I &=& - \dot{\phi}_0\tau_\mathcal{N}\dot{\mathcal{R}} + \dot{\phi}_0j_0\dot{\mathcal{S}}_\mathcal{B} + \left(\frac{\dot{\mathcal{J}}_\mathcal{N}}{H}+C\mathcal{J}_\mathcal{B}+\dot{\phi}_0\left(\dot{\tau}_\mathcal{N}+H\left(3-2\varepsilon\right)\tau_\mathcal{N}\right)+H\dot{\phi}_0C\tau_\mathcal{B}\right)\mathcal{R} \nonumber \\
&& + H\dot{\phi}_0\Lambda_{\mathcal{N}\mathcal{N}}\mathcal{S}_\mathcal{N} + H\dot{\phi}_0\left(\Lambda_{\mathcal{N}\mathcal{B}}+\left(\varepsilon-\eta_{\parallel}\right)j_0\right)\mathcal{S}_\mathcal{B} + H\dot{\phi}_0\kappa_\mathcal{N}\Phi_{\mathbf{B}}, \\
\left(\Delta \mathcal{J}_\mathcal{B}\right)^I &=& - \dot{\phi}_0\tau_\mathcal{B}\dot{\mathcal{R}} - \dot{\phi}_0j_0\dot{\mathcal{S}}_\mathcal{N} \nonumber \\
&& + \left(\frac{\dot{\mathcal{J}}_\mathcal{B}}{H}-C\mathcal{J}_\mathcal{N}+\dot{\phi}_0\left(\dot{\tau}_\mathcal{B}+H\left(3-2\varepsilon\right)\tau_\mathcal{B}\right)+H\dot{\phi}_0\left(\eta_{\perp}j0-C\tau_\mathcal{N}\right)\right)\mathcal{R} \nonumber \\
&& + H\dot{\phi}_0\left(\Lambda_{\mathcal{N}\mathcal{B}}-\left(\varepsilon-\eta_{\parallel}\right)j_0-\eta_{\perp}\tau_\mathcal{B}\right)\mathcal{S}_\mathcal{N} + H\dot{\phi}_0\Lambda_{\mathcal{B}\mathcal{B}}\mathcal{S}_\mathcal{B} + H\dot{\phi}_0\kappa_\mathcal{B}\Phi_{\mathbf{B}}.
\end{eqnarray}
On the other hand
\begin{eqnarray}
\mathcal{V}^A_{\,\,\,\,\,B} &=& \mathcal{V}_{\mathcal{T}\mathcal{T}}\mathcal{T}^A\mathcal{T}_B + \mathcal{V}_{\mathcal{N}\mathcal{N}}\mathcal{N}^A\mathcal{N}_B + \mathcal{V}_{\mathcal{B}\mathcal{B}}\mathcal{B}^A\mathcal{B}_B + \mathcal{V}_{\mathcal{T}\mathcal{N}}\left(\mathcal{T}^A\mathcal{N}_B + \mathcal{N}^A\mathcal{T}_B\right) \nonumber \\
&& + \mathcal{V}_{\mathcal{T}\mathcal{B}}\left(\mathcal{T}^A\mathcal{B}_B + \mathcal{B}^A\mathcal{T}_B\right) + \mathcal{V}_{\mathcal{N}\mathcal{B}}\left(\mathcal{N}^A\mathcal{B}_B + \mathcal{B}^A\mathcal{N}_B\right),
\end{eqnarray}
where the projections are defined as:
\begin{eqnarray}
\mathcal{V}_{\mathcal{T}\mathcal{T}} = \mathcal{T}^A\mathcal{T}^BD_AD_B \mathcal{V} \textrm{, }  & \quad  \mathcal{V}_{\mathcal{T}\mathcal{N}} = \mathcal{N}^A\mathcal{T}^BD_AD_B \mathcal{V} \textrm{, } &  \quad \mathcal{V}_{\mathcal{T}\mathcal{B}} = \mathcal{B}^A\mathcal{T}^AD_AD_B \mathcal{V}, \\
\mathcal{V}_{\mathcal{N}\mathcal{N}} = \mathcal{N}^A\mathcal{N}^BD_AD_B \mathcal{V} \textrm{, } & \quad \mathcal{V}_{\mathcal{N}\mathcal{B}} = \mathcal{N}^A\mathcal{B}^BD_AD_B \mathcal{V} \textrm{, }& \quad \mathcal{V}_{\mathcal{B}\mathcal{B}} = \mathcal{B}^A\mathcal{B}^BD_AD_B \mathcal{V},
\end{eqnarray}
where $D_A$ is the covariant derivative in the 3D field space. From Eqs.~(\ref{epsilon def})-(\ref{xis}), we know that:
\begin{eqnarray}
\mathcal{V}_{\mathcal{T}\mathcal{T}} &=& \frac{\dot{\mathcal{V}}_\mathcal{T} + H\eta_{\perp}\mathcal{V}_\mathcal{N}}{\dot{\phi}_0} = H^2\left(\left(3-\eta_{\parallel}\right)\left(\varepsilon+\eta_{\parallel}\right) - \eta_{\parallel}\xi_{\parallel} + \eta_{\perp}^2\right) - \frac{H\eta_{\perp}\mathcal{J}_\mathcal{N}}{\dot{\phi}_0}, \\
\mathcal{V}_{\mathcal{T}\mathcal{N}} &=& \frac{\dot{\mathcal{V}}_\mathcal{N} - H\eta_{\perp}\mathcal{V}_\mathcal{T} + HC\mathcal{V}_\mathcal{B}}{\dot{\phi}_0} = H^2\eta_{\perp}\left(3 - \varepsilon - 2\eta_{\parallel} - \xi_{\perp}\right) - \frac{\dot{\mathcal{J}}_\mathcal{N} + HC\mathcal{J}_\mathcal{B}}{\dot{\phi}_0}, \\
\mathcal{V}_{\mathcal{T}\mathcal{B}} &=& \frac{\dot{\mathcal{V}}_\mathcal{B} - HC\mathcal{V}_\mathcal{N}}{\dot{\phi}_0} = - H^2C\eta_{\perp} - \frac{\dot{\mathcal{J}}_\mathcal{B} - HC\mathcal{J}_\mathcal{N}}{\dot{\phi}_0}.
\end{eqnarray}
Finally, we can rewrite the perturbative Einstein equations (\ref{Einst EQ 1})-(\ref{Einst EQ 4}) using Appendix \ref{app:FluidComp}:
\begin{eqnarray}
\label{Einst EQ 1 Final}
\Psi_\mathbf{B} &=& \Phi_\mathbf{B}, \\
\label{Einst EQ 2 Final}
\frac{\partial^2}{a^2}\Phi_\mathbf{B} &=& H^2\varepsilon\left(\frac{\dot{\mathcal{R}}}{H} + \left(2\eta_{\perp}+\tau_\mathcal{N}\right)\mathcal{S}_\mathcal{N} + \tau_\mathcal{B}\mathcal{S}_\mathcal{B}\right), \\
\label{Einst EQ 3 Final}
\dot{\Phi}_\mathbf{B} + H\left(1+\varepsilon\right)\Phi_\mathbf{B} &=& H\varepsilon\mathcal{R}.
\end{eqnarray}
Then, with all these expressions and using Eqs.~(\ref{bard1})-(\ref{bard2}), the components of the perturbative Klein-Gordon equation (\ref{pert eq 5}), become:
\begin{eqnarray}
\label{KG EQ 1 Final}
\ddot{\mathcal{R}} - \frac{\partial^2}{a^2}\mathcal{R} + H\left(3+2\varepsilon-2\eta_{\parallel}\right)\dot{\mathcal{R}} && \\
+ H\left(2\eta_{\perp}+\tau_\mathcal{N}\right)\dot{\mathcal{S}}_\mathcal{N} + H^2\left(2\eta_{\perp}\left(3+\varepsilon-2\eta_{\parallel}-\xi_{\perp}\right)+\frac{\dot{\tau}_\mathcal{N}}{H}+\left(3+\varepsilon-2\eta_{\parallel}\right)\tau_\mathcal{N}\right)\mathcal{S}_\mathcal{N} \nonumber \\
+ H\tau_\mathcal{B}\dot{\mathcal{S}}_\mathcal{B} + H^2\left(\frac{\dot{\tau}_\mathcal{B}}{H}+\left(3+\varepsilon-2\eta_{\parallel}\right)\tau_\mathcal{B}\right)\mathcal{S}_\mathcal{B} &=& 0, \nonumber
\label{KG EQ 2 Final}
\end{eqnarray}
\begin{eqnarray}
\ddot{\mathcal{S}}_\mathcal{N} - \frac{\partial^2}{a^2}\mathcal{S}_\mathcal{N} + H\left(3+2\varepsilon-2\eta_{\parallel}\right)\dot{\mathcal{S}}_\mathcal{N} + H\left(2C+j_0\right)\dot{\mathcal{S}}_\mathcal{B} && \\
+ H^2\left(\frac{\mathcal{V}_{\mathcal{N}\mathcal{N}}}{H^2}+\varepsilon\left(3+2\varepsilon-2\mathbb{R}_{\mathcal{N}\mathcal{T}\mathcal{T}\mathcal{N}}\right)-\eta_{\perp}^2-C^2-\eta_{\parallel}\left(3+3\varepsilon-\eta_{\parallel}-\xi_{\parallel}\right)+\Lambda_{\mathcal{N}\mathcal{N}}\right)\mathcal{S}_\mathcal{N} && \nonumber \\
+ H^2\left(\frac{\mathcal{V}_{\mathcal{N}\mathcal{B}}}{H^2}-2\varepsilon\mathbb{R}_{\mathcal{N}\mathcal{T}\mathcal{T}\mathcal{B}}+C\left(3+\varepsilon-2\eta_{\parallel}\right)+\frac{\dot{C}}{H}+\Lambda_{\mathcal{N}\mathcal{B}}+\left(\varepsilon-\eta_{\parallel}\right)j_0\right)\mathcal{S}_\mathcal{B} && \nonumber \\
- H\left(2\eta_{\perp}+\tau_\mathcal{N}\right)\dot{\mathcal{R}} + H^2\left(\frac{\dot{\tau}_\mathcal{N}}{H}+\left(3-2\varepsilon\right)\tau_\mathcal{N}+C\tau_\mathcal{B}\right)\mathcal{R} + H^2\kappa_\mathcal{N}\Phi_{\mathbf{B}} &=& 0 , \nonumber
\end{eqnarray}
\begin{eqnarray}
\label{KG EQ 3 Final}
\ddot{\mathcal{S}}_\mathcal{B} - \frac{\partial^2}{a^2}\mathcal{S}_\mathcal{B} + H\left(3+2\varepsilon-2\eta_{\parallel}\right)\dot{\mathcal{S}}_\mathcal{B} - H\left(2C+j_0\right)\dot{\mathcal{S}}_\mathcal{N} && \\
+ H^2\left(\frac{\mathcal{V}_{\mathcal{B}\mathcal{B}}}{H^2}+\varepsilon\left(3+2\varepsilon-2\mathbb{R}_{\mathcal{B}\mathcal{T}\mathcal{T}\mathcal{B}}\right)-C^2-\eta_{\parallel}\left(3+3\varepsilon-\eta_{\parallel}-\xi_{\parallel}\right)+\Lambda_{\mathcal{B}\mathcal{B}}\right)\mathcal{S}_\mathcal{B} && \nonumber \\
+ H^2\left(\frac{\mathcal{V}_{\mathcal{N}\mathcal{B}}}{H^2}-2\varepsilon\mathbb{R}_{\mathcal{N}\mathcal{T}\mathcal{T}\mathcal{B}}-C\left(3+\varepsilon-2\eta_{\parallel}\right)-\frac{\dot{C}}{H}+\Lambda_{\mathcal{N}\mathcal{B}}-\left(\varepsilon-\eta_{\parallel}\right)j_0-\eta_{\perp}\tau_\mathcal{B}\right)\mathcal{S}_\mathcal{N} && \nonumber \\
- H\tau_\mathcal{B}\dot{\mathcal{R}} + H^2\left(\frac{\dot{\tau}_\mathcal{B}}{H}+\left(3-2\varepsilon\right)\tau_\mathcal{B}+\eta_{\perp}j0-C\tau_\mathcal{N}\right)\mathcal{R} + H^2\kappa_\mathcal{B}\Phi_{\mathbf{B}} &=& 0. \nonumber
\end{eqnarray}
We note that the background interaction, given by $\mathcal{J}_{\mathcal{N}}$ and $\mathcal{J}_{\mathcal{B}}$, does not appear explicitly in the perturbative equations. It is represented by the inflationary parameters in the coefficients of Eqs.~(\ref{Einst EQ 1 Final})-(\ref{KG EQ 3 Final}). However we have other parameters to represent the interaction in a perturbative level. In Appendix \ref{app:FluidComp} we show that only two parameters, $\tau_\mathcal{N}$ and $\tau_\mathcal{B}$, represent the interaction in the total fluid system. Actually, they are the only contribution by interaction in Eq.~(\ref{KG EQ 1 Final}) and produce curvature terms in (\ref{KG EQ 2 Final})-(\ref{KG EQ 3 Final}). This means that the other six parameters represent the internal effects of the interaction and they are related to $\tau_\mathcal{N}$ and $\tau_\mathcal{B}$. Besides, it is clear that $j_0$, $\Lambda_{\mathcal{N}\mathcal{N}}$, $\Lambda_{\mathcal{N}\mathcal{B}}$ and $\Lambda_{\mathcal{B}\mathcal{B}}$ are interaction parameters related to the isocurvature components, therefore they are important, for instance, to understand how a particular interaction affects the isocurvature evolution in the reheating era. On the other side, $\kappa_\mathcal{N}$ and $\kappa_\mathcal{B}$ are produced by a possible dependence on the metric in the lagrangian interaction. These kind of terms could be uncommon, but they can not be discarded.

All these parameters are completely arbitrary and could give us interesting properties. However, in this paper, we will focus in the effects of the background interaction parameters represented by $\mathcal{J}_{\mathcal{N}}$ and $\mathcal{J}_{\mathcal{B}}$. So, to simplify the calculation, we will fix them to zero, but they will be considered in future works.

Finally, it will be useful to express the cosmic time derivatives as derivatives with respect to the number of $e$-folds, $N$, according to the following relations:
\begin{eqnarray}
\label{tn}
\frac{d}{dt} &=& H\frac{d}{dN},  \qquad ()^\prime\equiv \frac{d}{dN}.
\end{eqnarray}
Additionally, we introduce the following useful dimensionless parameters:
\begin{eqnarray}
\label{Qk}
&\mathbf{Q} = \frac{k}{aH}, & \\
\label{vs}
v_{\mathcal{N}\mathcal{N}} = \frac{\mathcal{V}_{\mathcal{N}\mathcal{N}}}{H^2} \textrm{, } & \quad v_{\mathcal{N}\mathcal{B}} = \frac{\mathcal{V}_{\mathcal{N}\mathcal{B}}}{H^2} \textrm{, }& \quad v_{\mathcal{B}\mathcal{B}} = \frac{\mathcal{V}_{\mathcal{B}\mathcal{B}}}{H^2} \\
\label{qs}
q_\mathcal{N} = \frac{\mathcal{J}_\mathcal{N}}{H^2\sqrt{2\varepsilon}} \textrm{, } & \quad q_\mathcal{B} = \frac{\mathcal{J}_\mathcal{B}}{H^2\sqrt{2\varepsilon}} \textrm{. }&
\end{eqnarray}
Then, the background equations of motion acquire the following forms:
\begin{eqnarray}
\label{bg eq 1 N}
\mathcal{V} &=& H^2\left(3-\varepsilon\right), \\
\label{bg eq 2 N}
\phi_0'^2 &=& 2\varepsilon, \\
\label{bg eq 3 N}
\frac{D \mathcal{T}^A}{dN} &=& - \eta_{\perp}\mathcal{N}^A, \\
\label{bg eq 4 N}
\frac{D\mathcal{N}^A}{dN} &=& \eta_{\perp}\mathcal{T}^A - C\mathcal{B}^A, \\
\label{bg eq 5 N}
\frac{D\mathcal{B}^A}{dN} &=& C\mathcal{N}^A,
\end{eqnarray}
and the projections of the covariant derivative of $\mathcal{V}$ are found to be given by:
\begin{eqnarray}
\label{bg eq Ex N}
\mathcal{V}_\mathcal{T} = H^2\sqrt{2\varepsilon}\left(\eta_{\parallel}-3\right) \textrm{, } & \quad \mathcal{V}_\mathcal{N} = H^2\sqrt{2\varepsilon}\left(\eta_{\perp} - q_\mathcal{N}\right) \textrm{, }&  \quad \mathcal{V}_\mathcal{B} = - H^2\sqrt{2\varepsilon}q_\mathcal{B},
\end{eqnarray}
where we have introduced the covariant derivative with respect to $N$ as:
\begin{eqnarray}
\frac{D X^A}{dN} &=& X^{A \prime} + \Gamma^A_{BC}\phi_0^{B\prime}X^C, \\
\Gamma^A_{BC} &=& \frac{1}{2}\mathbf{q}^{AD}\left(\partial_C\mathbf{q}_{DB}+\partial_B\mathbf{q}_{CD}-\partial_D\mathbf{q}_{BC}\right).
\end{eqnarray}
In addition, the perturbative equations (\ref{Einst EQ 1 Final})-(\ref{KG EQ 3 Final}) become:
\begin{eqnarray}
\label{Pert EQ 1 Final}
\mathcal{R}'' + \mathbf{Q}^2\mathcal{R} + \left(3+\varepsilon-2\eta_{\parallel}\right)\mathcal{R}' + 2\eta_{\perp}\mathcal{S}_\mathcal{N}' + \eta_{\perp}\left(3+\varepsilon-2\eta_{\parallel}-\xi_{\perp}\right)\mathcal{S}_\mathcal{N} &=& 0, \\[10pt]
\label{Pert EQ 2 Final}
\mathcal{S}_\mathcal{N}'' + \mathbf{Q}^2\mathcal{S}_\mathcal{N} + \left(3+\varepsilon-2\eta_{\parallel}\right)\mathcal{S}_\mathcal{N}' && \nonumber \\
+ \left(v_{\mathcal{N}\mathcal{N}}-2\varepsilon\mathbb{R}_{\mathcal{N}\mathcal{T}\mathcal{T}\mathcal{N}}-\eta_{\perp}^2-C^2-\eta_{\parallel}\left(3+3\varepsilon-\eta_{\parallel}-\xi_{\parallel}\right)+3\varepsilon+2\varepsilon^2\right)\mathcal{S}_\mathcal{N} && \nonumber \\
+ 2C\mathcal{S}_\mathcal{B}' + \left(v_{\mathcal{N}\mathcal{B}}-2\varepsilon\mathbb{R}_{\mathcal{N}\mathcal{T}\mathcal{T}\mathcal{B}}+C\left(3+\varepsilon-2\eta_{\parallel}\right)+C'\right)\mathcal{S}_\mathcal{B} - 2\eta_{\perp}\mathcal{R}' &=& 0, \\[10pt]
\label{Pert EQ 3 Final}
\mathcal{S}_\mathcal{B}'' + \mathbf{Q}^2\mathcal{S}_\mathcal{B} + \left(3+\varepsilon-2\eta_{\parallel}\right)\mathcal{S}_\mathcal{B}' && \nonumber \\
+ \left(v_{\mathcal{B}\mathcal{B}}-2\varepsilon\mathbb{R}_{\mathcal{B}\mathcal{T}\mathcal{T}\mathcal{B}}-C^2-\eta_{\parallel}\left(3+3\varepsilon-\eta_{\parallel}-\xi_{\parallel}\right)+3\varepsilon+2\varepsilon^2\right)\mathcal{S}_\mathcal{B} && \nonumber \\
- 2C\mathcal{S}_\mathcal{N}' + \left(v_{\mathcal{N}\mathcal{B}}-2\varepsilon\mathbb{R}_{\mathcal{N}\mathcal{T}\mathcal{T}\mathcal{B}}-C\left(3+\varepsilon-2\eta_{\parallel}\right)-C'\right)\mathcal{S}_\mathcal{N} &=& 0, \\[10pt]
\label{Pert EQ 4 Final}
\mathbf{Q}^2\Phi_\mathbf{B} + \varepsilon\left(\mathcal{R}' + 2\eta_{\perp}\mathcal{S}_\mathcal{N}\right) &=& 0, \\[10pt]
\label{Pert EQ 5 Final}
\Phi_\mathbf{B}' - \varepsilon\mathcal{R} + \left(1+\varepsilon\right)\Phi_\mathbf{B} &=& 0,
\end{eqnarray}
(recall that $\Psi_\mathbf{B} = \Phi_\mathbf{B}$). On the other side, each single matter component has an intrinsic entropy perturbation associated. A single component perfect fluid, by definition, does not have any intrinsic entropy perturbation, whereas for a multi-scalar field system, the intrinsic entropy perturbation is defined as (see Ref.~\cite{Malik:2008im}):
\begin{eqnarray}
\label{S phi}
S_{\phi} &=& \frac{p_{nad}^1}{2H^2\left(3-\eta_{\parallel}\right)}, \\
p_{nad}^1 &\equiv& p^1 - \frac{\dot{p}^0}{\dot{\rho}^0}\rho^1 = p^{1I} - \left(\frac{2}{3}\eta_{\parallel}-1\right)\rho^{1I},
\end{eqnarray}
where $p_{nad}^1$ is the non-adiabatic pressure perturbation of our composite system. Using Eqs.~(\ref{drhoinv})-(\ref{dpinv}), it is easy to see that the intrinsic entropy is a gauge-invariant quantity. Using the density and pressure components presented in Appendix \ref{app:FluidComp}, the non-adiabatic pressure perturbation becomes:
\begin{eqnarray}
\label{dpNAD}
p_{nad}^1 &=& 2H^2\varepsilon\left(\mathcal{R}' + \left(3-2\eta_{\parallel}\right)\mathcal{R}\right) - \left(\frac{2}{3}\eta_{\parallel}-1\right)2H^2\varepsilon\left(\mathcal{R}' - 3\mathcal{R} + 2\eta_{\perp}\mathcal{S}_\mathcal{N}\right) \nonumber \\
&=& \frac{4H^2\varepsilon}{3}\left(\left(3-\eta_{\parallel}\right)\mathcal{R}' + \left(3-2\eta_{\parallel}\right)\eta_{\perp}\mathcal{S}_\mathcal{N}\right).
\end{eqnarray}
In Appendix \ref{app:FieldsSpace}, a new basis $\left(\mathcal{T}^A,\mathcal{N}_0^A,\mathcal{B}_0^A\right)$ is introduced, where $\mathcal{N}_0^A$ is such that $\mathcal{N}_0^{(3)} = 0$. The equations, in this case, are given by Eqs.~(\ref{EQ base0-1})-(\ref{EQ base0-3}). In that new basis, similarly to Eq.~(\ref{dphi}), we can write $(\delta \phi^A)$ as:
\begin{eqnarray}
\label{dphi 0}
\left(\delta\phi^A\right) &=& \frac{\dot{\phi}_0}{H}\left[\left(\mathcal{R} - \Psi\right)\mathcal{T}^A + \mathcal{S}_{\mathcal{N}0}\mathcal{N}_0^A + \mathcal{S}_{\mathcal{B}0}\mathcal{B}_0^A\right],
\end{eqnarray}
where $\mathcal{S}_{\mathcal{N}0}$ and $\mathcal{S}_{\mathcal{B}0}$ represent the isocurvature elements in this basis. On the other side, we can introduce the inflaton curvature $\zeta_{\phi} \equiv \left(\frac{\dot{\varphi}_0}{\dot{\phi}_0}\right)^2\mathcal{R}_{\phi}$ (which is defined on hypersurfaces orthogonal to comoving world-lines \cite{Malik:2008im}) such that:
\begin{eqnarray}
\label{dphi old}
\left(\delta\phi^a\right) &=& \frac{\dot{\varphi}_0}{H}\left(\left(\mathcal{R}_{\phi} - \Psi\right)\mathcal{T}_{2D}^a + F_{\phi}\mathcal{N}_{2D}^a\right).
\end{eqnarray}
Here, $\left(\mathcal{T}_{2D}^a,\mathcal{N}_{2D}^a\right)$ is the 2D basis, with $a=1,2$ considering just the two inflatons, $\phi_0^{(1)}$ and $\phi_0^{(2)}$, and $F_{\phi}$ is the inflaton isocurvature component. In this case, we have $\mathcal{T}_{2D}^a = \frac{\dot{\phi}_0^a}{\dot{\varphi}_0}$, so:
\begin{eqnarray}
\label{2D basis}
\mathcal{T}^a = \mathcal{T}_{2D}^a\cos(\alpha) \textrm{, } & \quad  \mathcal{N}_0^a = \mathcal{N}_{2D}^a \textrm{, }& \quad \mathcal{B}_0^a = - \mathcal{T}_{2D}^a\sin(\alpha),
\end{eqnarray}
with $\cos(\alpha) = \frac{\dot{\varphi}_0}{\dot{\phi}_0}$, $\mathcal{T}_{2D}^a\mathcal{T}^{2D}_a = \mathcal{N}_{2D}^a\mathcal{N}^{2D}_a = 1$, $\mathcal{T}_{2D}^a\mathcal{N}^{2D}_a = 0$ and we used Eq.~(\ref{Parameterize0}) from Appendix \ref{app:FieldsSpace}. Therefore, if we compare Eq.~(\ref{dphi 0}) with (\ref{dphi old}), we obtain:
\begin{eqnarray}
\label{zeta_phi}
\zeta_{\phi} &=& \cos^2(\alpha)\mathcal{R} - \sin(\alpha)\cos(\alpha)\mathcal{S}_{\mathcal{B}0}, \\
\label{F phi}
F_{\phi} &=& \frac{\mathcal{S}_{\mathcal{N}0}}{\cos(\alpha)}.
\end{eqnarray}
From here, we conclude that $\mathcal{S}_{\mathcal{B}0}$ represents the change of the contribution to the total curvature from inflaton perturbations to radiation. On the other side, $\mathcal{S}_{\mathcal{N}0}$ represents the inflaton's isocurvature. In fact, in the radiation epoch $\cos(\alpha) = 0$, so $\mathcal{S}_{\mathcal{N}0} = \zeta_{\phi} = 0$ as we expected it. Additionally, we can define de radiation curvature as:
\begin{eqnarray}
\label{zeta_R}
\zeta_R &\equiv& \mathcal{R} - \zeta_{\phi} = \sin^2(\alpha)\mathcal{R} + \sin(\alpha)\cos(\alpha)\mathcal{S}_{\mathcal{B}0}.
\end{eqnarray}
In the next section, we will use all these definitions to describe the evolution of perturbations in different cases.\\

\setcounter{equation}{0}
\section{Analysis and Results} \label{Sec: Analysis and Results}

In this section we set ourselves to study a few concrete examples of multi-field models coupled to radiation. Our analysis will benefit from the covariant formalism offered in the previous two sections to parametrize the evolution of both, background fields and perturbations. All of the models that we consider in the next subsections are characterized for having a diagonal metric in the extended field space:
\begin{equation}
\label{Metric q}
\mathbf{q}_{AB} =
\left(
\begin{array}{ccc}
   \mathbf{q}_{(11)} &        0          &        0          \\
          0          & \mathbf{q}_{(22)} &        0          \\
          0          &        0          & \mathbf{q}_{(33)} \\
\end{array}
\right).
\end{equation}
With this form of the field-metric, we may work with the basis of unit vectors given in Eq.~(\ref{Parameterize}), parametrized by the angles $\alpha$, $\beta$ and $\gamma$. Alternatively, we may work with the basis shown in (\ref{Parameterize0}) parametrized by the same angles. Using this parametrization back in Eqs.~(\ref{bg eq 1 N})-(\ref{bg eq Ex N}), the $\beta$-angle is found to be given by (see Appendix~\ref{app:FieldsSpace})
\begin{eqnarray}
\label{bg eq 1 angle}
\tan(\beta) &=& \frac{\sin(\alpha)\left(g\left(\theta\right) + \left(1+q_1\left(1-\frac{\cos^2(\alpha)}{3}\right)\right)\cos(\alpha)\right)}{f\left(\theta\right) + q_2} ,
\end{eqnarray}
where $f(\theta)$ and $g(\theta)$ are defined in Eqs.~(\ref{fyg1})-(\ref{fyg2}). On the other hand, the parameters $q_1$ and $q_2$ are defined to respect Eq.~(\ref{JA}). That is, they parametrize the projection of the source $\mathcal J_0^A$ on the directions $\mathcal N_0^A$ and $\mathcal B_0^A$ respectively. It may be seen that $q_1$ parameterizes the coupling of the two fields system and radiation, whereas the self-interaction of the two fields system is parameterized by $q_2$. To continue, the inflationary parameters describing the background dynamics acquire the form:
\begin{eqnarray}
\label{bg eq 2 angle}
\varepsilon &=& \frac{\phi_0'^2}{2} = \frac{3-\frac{V}{H^2}}{1+\frac{\sin^2(\alpha)}{2}}, \\
\label{bg eq 3 angle}
\eta_{\parallel} &=& 2 + \cos^2(\alpha)\left(1 + \frac{q_1\sin^2(\alpha)}{3}\right) + g(\theta)\cos(\alpha), \\
\label{bg eq 4 angle}
\eta_{\perp} &=& \sqrt{\left(f(\theta) + q_2\right)^2 + \sin^2(\alpha)\left(g(\theta) + \left(1+q_1\left(1-\frac{\cos^2(\alpha)}{3}\right)\right)\cos(\alpha)\right)^2}.
\end{eqnarray}
In this way, the equations of motion for the background are reduced to:
\begin{eqnarray}
\label{bg eq 5 angle}
\alpha' - \sin(\alpha)\left(g\left(\theta\right) + \left(1+q_1\left(1-\frac{\cos^2(\alpha)}{3}\right)\right)\cos(\alpha)\right) &=& 0, \\
\label{bg eq 6 angle}
\theta' - \frac{\left(f\left(\theta\right)+q_2\right)}{\cos(\alpha)} + \sqrt{\frac{\varepsilon}{2}}\cos(\alpha)\left(\frac{\cos(\theta)\partial_{(2)}\mathbf{q}_{(11)}}{\sqrt{\mathbf{q}_{(22)}}\mathbf{q}_{(11)}} + \frac{\sin(\theta)\partial_{(1)}\mathbf{q}_{(22)}}{\sqrt{\mathbf{q}_{(11)}}\mathbf{q}_{(22)}}\right) &=& 0, \\
\label{bg eq 7 angle}
\beta' - C + \left(f\left(\theta\right)+q_2\right)\tan(\alpha) &=& 0.
\end{eqnarray}
Equation~(\ref{bg eq 1 angle}) comes from $\mathcal{V}_\mathcal{B} + \mathcal{J}_\mathcal{B} = 0$, hence (\ref{bg eq 7 angle}) gives us an expression of $C$. In addition, from Eq.~(\ref{JA}) in Appendix \ref{app:Interaction}, we deduce:
\begin{eqnarray}
\label{qN}
q_\mathcal{N} &=& q_2\cos(\beta) + q_1\sin(\alpha)\cos(\alpha)\sin(\beta), \\
\label{qB}
q_\mathcal{B} &=& q_2\sin(\beta) - q_1\sin(\alpha)\cos(\alpha)\cos(\beta) .
\end{eqnarray}
In order to solve the perturbative equations, we will need $\mathbb{R}_{\mathcal{N}\mathcal{T}\mathcal{T}\mathcal{N}}$, $\mathbb{R}_{\mathcal{N}\mathcal{T}\mathcal{T}\mathcal{B}}$ and $\mathbb{R}_{\mathcal{B}\mathcal{T}\mathcal{T}\mathcal{B}}$. Given that the field metric is given by (\ref{Metric q}), the only non-zero component of the Riemann tensor is given by:
\begin{eqnarray}
\label{Riemann}
\mathbb{R}_{(1212)} = \frac{\mathbf{q}_{(11)}\mathbf{q}_{(22)}}{2}\mathbb{R},
\end{eqnarray}
with $\mathbb{R}$ the Ricci scalar given by:
\begin{eqnarray}
\label{Riemann}
\mathbb{R} &=& \frac{1}{2\mathbf{q}_{(11)}^2\mathbf{q}_{(22)}^2}\left(- 2\mathbf{q}_{(11)}\mathbf{q}_{(22)}\partial_{(2)}\partial_{(2)}\mathbf{q}_{(11)} - 2\mathbf{q}_{(11)}\mathbf{q}_{(22)}\partial_{(1)}\partial_{(1)}\mathbf{q}_{(22)} + \mathbf{q}_{(22)}\left(\partial_{(2)}\mathbf{q}_{(11)}\right)^2 \right. \nonumber \\
&& \left. + \mathbf{q}_{(22)}\left(\partial_{(1)}\mathbf{q}_{(11)}\right)\left(\partial_{(1)}\mathbf{q}_{(22)}\right) + \mathbf{q}_{(11)}\left(\partial_{(2)}\mathbf{q}_{(11)}\right)\left(\partial_{(2)}\mathbf{q}_{(22)}\right) + \mathbf{q}_{(11)}\left(\partial_{(1)}\mathbf{q}_{(22)}\right)^2\right).
\end{eqnarray}
Then, it follows that
\begin{eqnarray}
\label{RNTTN}
\mathbb{R}_{\mathcal{N}\mathcal{T}\mathcal{T}\mathcal{N}} &=& \left(\mathcal{N}^{(1)}\mathcal{T}^{(2)}\mathcal{T}^{(1)}\mathcal{N}^{(2)} - \mathcal{N}^{(1)}\mathcal{T}^{(2)}\mathcal{T}^{(2)}\mathcal{N}^{(1)} - \mathcal{N}^{(2)}\mathcal{T}^{(1)}\mathcal{T}^{(1)}\mathcal{N}^{(2)} \right. \nonumber \\
&& \left.+ \mathcal{N}^{(2)}\mathcal{T}^{(1)}\mathcal{T}^{(2)}\mathcal{N}^{(1)}\right)\mathbb{R}_{(1212)} \nonumber \\
&=& - \cos^2(\alpha)\cos^2(\beta)\frac{\mathbb{R}}{2}, \\
\label{RNTTB}
\mathbb{R}_{\mathcal{N}\mathcal{T}\mathcal{T}\mathcal{B}} &=& \left(\mathcal{N}^{(1)}\mathcal{T}^{(2)}\mathcal{T}^{(1)}\mathcal{B}^{(2)} - \mathcal{N}^{(1)}\mathcal{T}^{(2)}\mathcal{T}^{(2)}\mathcal{B}^{(1)} - \mathcal{N}^{(2)}\mathcal{T}^{(1)}\mathcal{T}^{(1)}\mathcal{B}^{(2)} \right. \nonumber \\
&& \left.+ \mathcal{N}^{(2)}\mathcal{T}^{(1)}\mathcal{T}^{(2)}\mathcal{B}^{(1)}\right)\mathbb{R}_{(1212)} \nonumber \\
&=& - \cos^2(\alpha)\sin(\beta)\cos(\beta)\frac{\mathbb{R}}{2}, \\
\label{RBTTB}
\mathbb{R}_{\mathcal{B}\mathcal{T}\mathcal{T}\mathcal{B}} &=& \left(\mathcal{N}^{(1)}\mathcal{T}^{(2)}\mathcal{T}^{(1)}\mathcal{N}^{(2)} - \mathcal{N}^{(1)}\mathcal{T}^{(2)}\mathcal{T}^{(2)}\mathcal{N}^{(1)} - \mathcal{N}^{(2)}\mathcal{T}^{(1)}\mathcal{T}^{(1)}\mathcal{N}^{(2)}  \right. \nonumber \\
&& \left.+ \mathcal{N}^{(2)}\mathcal{T}^{(1)}\mathcal{T}^{(2)}\mathcal{N}^{(1)}\right)\mathbb{R}_{(1212)} \nonumber \\
&=& - \cos^2(\alpha)\sin^2(\beta)\frac{\mathbb{R}}{2}.
\end{eqnarray}
It is important to recall that the main motivation for developing the covariant formalism in a three-dimensional field-space is that it provides us well-defined slow-roll parameters $\varepsilon$, $\eta_{\parallel}$, and $\eta_{\perp}$ during inflation as well as the transition from inflation up to radiation-dominated epoch. In addition, we stress that the present prescription necessarily introduces a coupling between the two-field system and a radiation fluid, parameterized by $q_1$,  making the smooth transition into reheating possible. On the other hand, $q_2$ parameterizes the self-interaction of the two-field system. Thus, in order to analyze the consequences of the interaction between the field fluctuations and the thermal bath (and to simplify our analysis) we will set $q_2=0$ from the beginning. As we shall see, the range of acceptable values for $q_1$ is somewhat restricted. If $q_1$ is too small, the transition from inflation to the radiation-dominated epoch becomes extremely oscillatory, and the thermalization of the universe is not efficient. On the other hand, if $q_1$ is too large, radiation will start to be produced too early during inflation, drastically modifying the usual analytical predictions describing the epoch at which the fluctuations crossed the horizon (we will come back to this issue later on).

In what follows, we apply the angular parametrization specified by $(\alpha, \beta, \theta)$ and solve the equations of motion for these angles, given by Eqs.~(\ref{bg eq 5 angle})-(\ref{bg eq 7 angle}). In this way, the slow-roll parameters $\varepsilon$, $\eta_\parallel$, and $\eta_{\perp}$ which describe the background evolution may be completely determined. Once the background evolution of our system is known, we solve the set of Eqs.~(\ref{Pert EQ 1 Final})-(\ref{Pert EQ 5 Final}) for the variables $\mathcal{R}$, $\Phi_\mathbf{B}$, $\mathcal{S}_\mathcal{N}$, and $\mathcal{S}_\mathcal{B}$ numerically. By imposing the standard Bunch-Davies initial conditions, we shall evolve these perturbations from the sub-Hubble to the super-Hubble scales, and analyze the effects of the transition from inflation to the radiation dominated epoch on the evolution of the large-scale perturbations.

\subsection{Product-Exponential (PE) Potential} \label{SubSec: Product Exponential-Potential}

The first model we consider is the product-exponential (PE) potential. This is a canonical two-field model with the product separable form:
\begin{equation}
\label{Vexp}
V(\phi_1,\phi_2)=V_0\,\phi_1^2\,e^{-\lambda \phi_2^2},
\end{equation}
where $V_0$ sets the energy scale of the potential and is of mass dimension two. Its value sets the scale of inflation and determines the amplitude of the primordial power spectrum. Inflation takes place at super-Planckian values for the $\phi_1$ field and the exponential factor is very much suppressed, i.e. $\lambda \phi_2^2\ll 1$. For this model $\phi_1$ is identified as the inflaton and $\phi_2$ as the sub-dominant field which sources the isocurvature perturbations. Additionally, the field-space metric is given by Eq.~(\ref{Metric q}) with $\mathbf{q}_{(11)}=\mathbf{q}_{(22)}=1$. This potential was first introduced in \cite{Byrnes:2008zy} to study the generation of non-Gaussianity.

\begin{figure}[ht]
  \centering
  \begin{subfigure}[b]{0.5\linewidth}
    \centering\includegraphics[width=240pt]{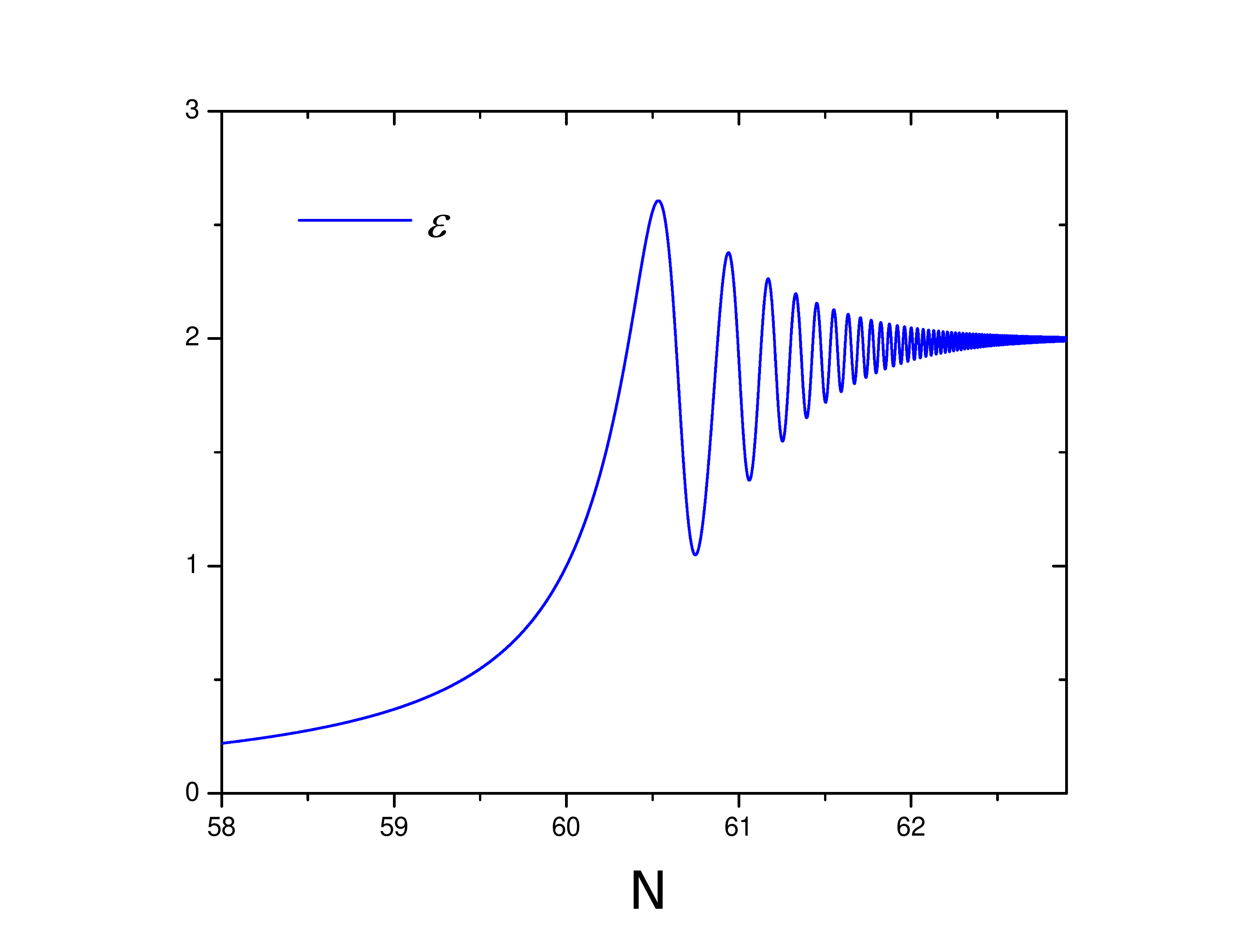}
    \vspace{0 cm}
    \caption{\label{fig:fig1a}}
  \end{subfigure}
    \hspace{-1 cm}
  \begin{subfigure}[b]{0.5\linewidth}
    \centering\includegraphics[width=240pt]{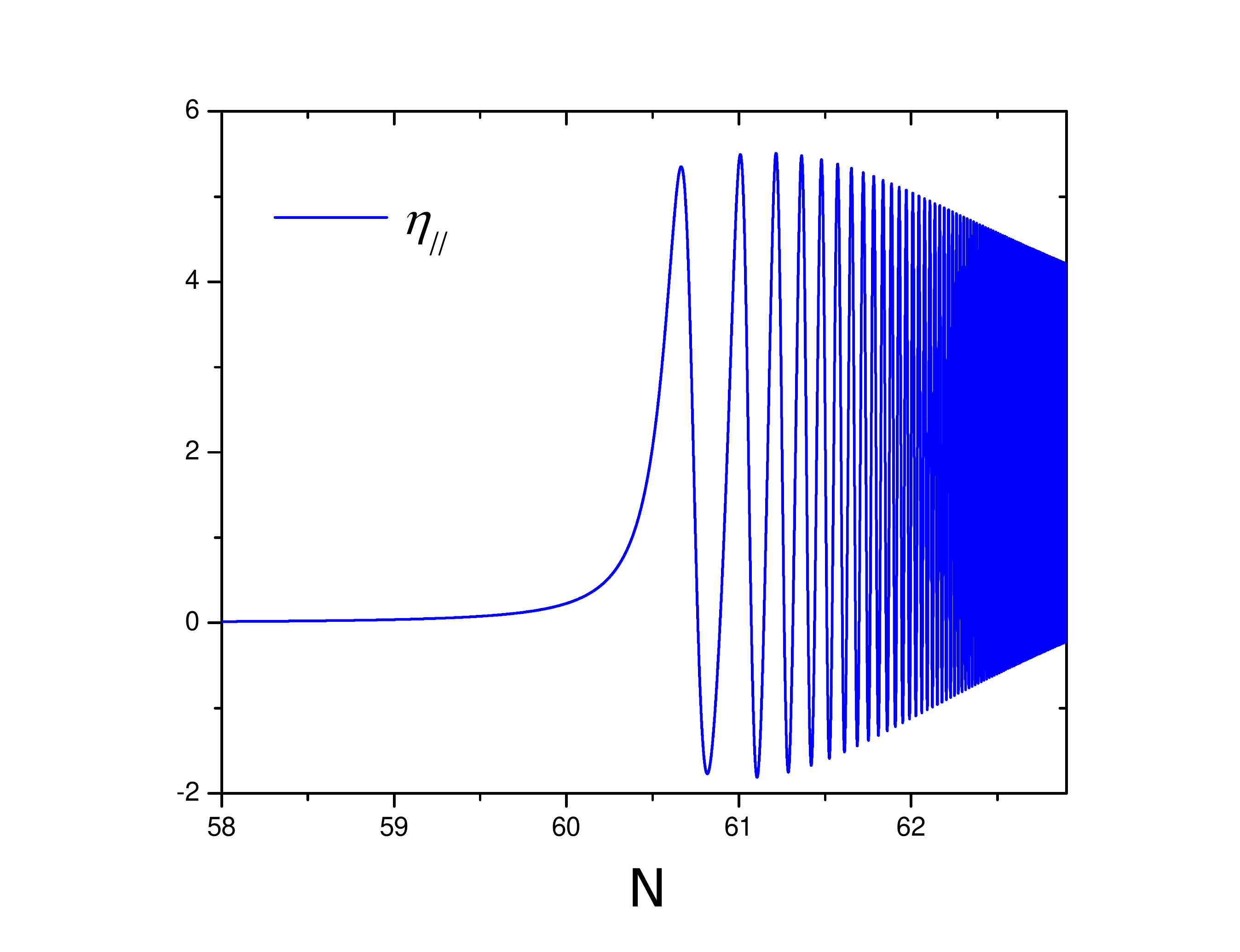}
     \vspace{0 cm}
    \caption{\label{fig:fig1b}}
  \end{subfigure}
  \begin{subfigure}[b]{0.5\linewidth}
    \centering\includegraphics[width=240pt]{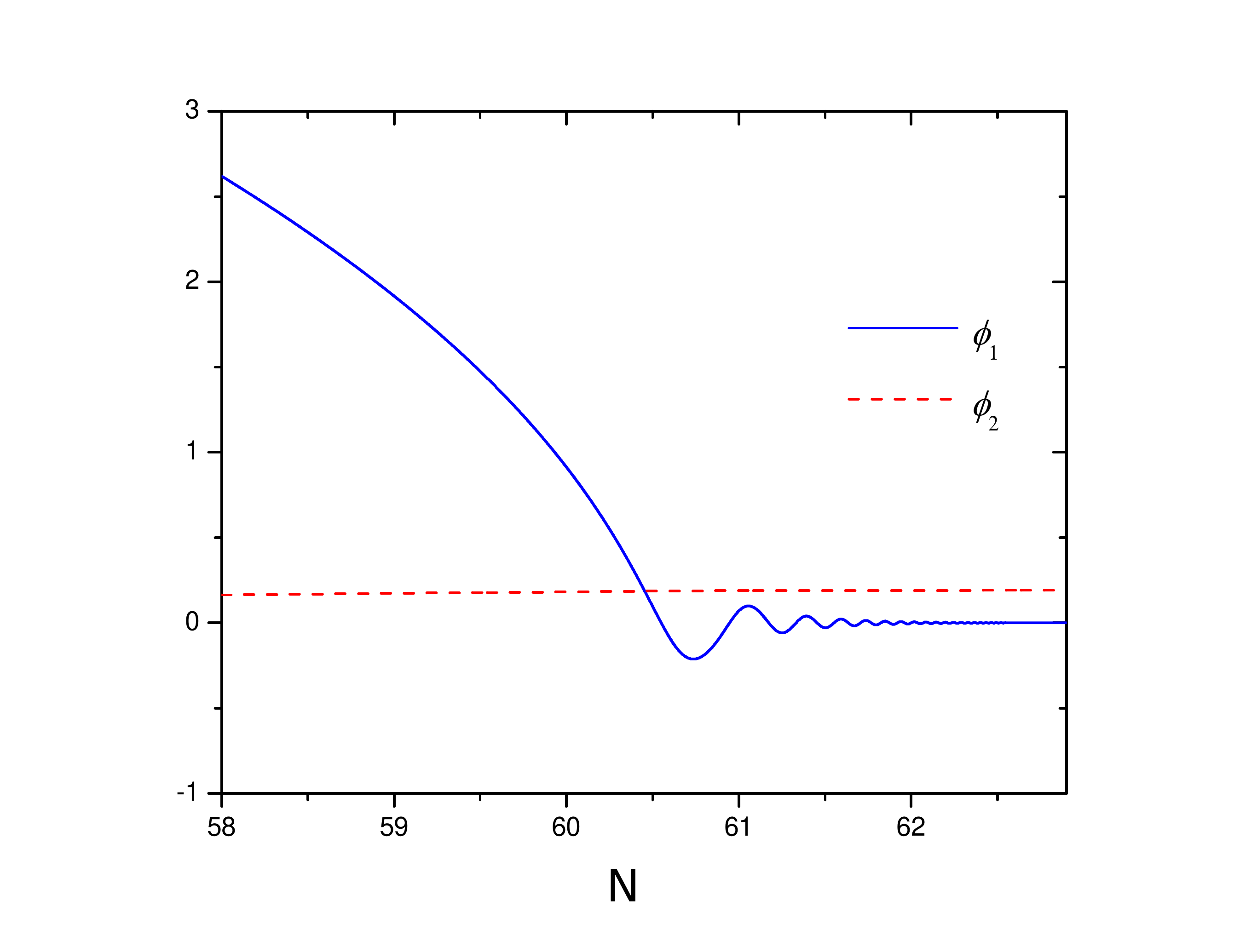}
    \vspace{0 cm}
    \caption{\label{fig:fig1c}}
  \end{subfigure}
    \hspace{-1 cm}
  \begin{subfigure}[b]{0.5\linewidth}
    \centering\includegraphics[width=240pt]{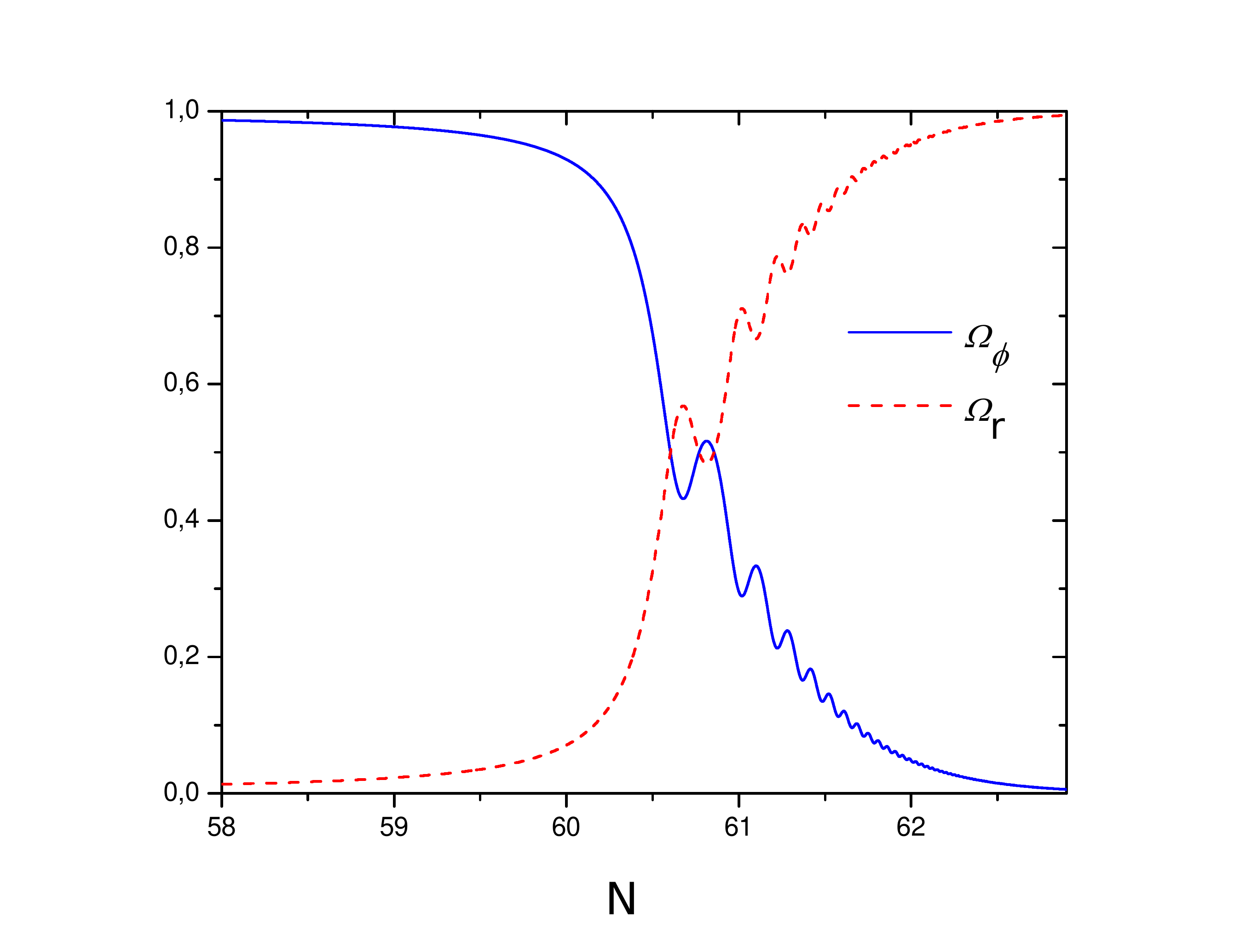}
     \vspace{0 cm}
    \caption{\label{fig:fig1d}}
  \end{subfigure}
  \caption{Transition from inflation to the radiation dominated era, through reheating, at background level for the product-exponential potential (\ref{Vexp}). Figures \subref{fig:fig1a} and \subref{fig:fig1b} show the evolution from the last stage of inflation to the radiation dominated epoch for the slow-roll parameters $\varepsilon$ and $\eta_{\parallel}$, respectively, which are plotted against the number of $e$-folds starting from the end of inflation. The fields $\phi_1$ (solid blue line) and $\phi_2$ (dashed red line) as function of the number of $e$-folds from the last $e$-folds of inflation to the end of reheating is depicted in Figure \subref{fig:fig1c}. Figure \subref{fig:fig1d} shows the evolution of fractional contributions of the energy density from the inflaton fields (solid blue line) and radiation (dashed red line).}
  \label{Fig1exp}
\end{figure}

\afterpage{\clearpage}

An interesting feature of this model is that the adiabatic regime is not reached during inflation, and so the curvature perturbation $\mathcal{R}$ and its power spectrum will continue to evolve after inflation has ended. In addition, the scale dependence of the curvature perturbation is not close to the maximum likelihood value from current observational data. In particular, at the end of inflation the spectral index of the curvature perturbations is around $n_{\mathcal{R}}\simeq 0.794$ \cite{Dias:2011xy,Huston:2011fr}. However, the effects of reheating can not be neglected for this model. Indeed, the coupling of this model to radiation makes it possible that the adiabatic limit be reached before the end of inflation. In addition, after reheating the value of $n_{\mathcal{R}}$ will be closer to the currently accepted value.

For the dimensionless parameter $q_1$, which parameterizes the coupling of the two-field system and radiation, we set the value $q_1=3.4$. Regarding the choice for the parameter values of the product exponential potential, firstly we consider $\lambda=0.03/M_p^2$, which is close to the value used in Refs.~\cite{Dias:2011xy,Huston:2011fr}. Now, by regarding that this model may reach the adiabatic limit at the end of reheating, $V_0$ should be chosen to match the Planck 2015 maximum likelihood value $\mathcal{P}_{\mathcal{R}}\simeq 2.2\times 10^{-9}$ for the pivot scale $k_0=0.002$ Mpc$^{-1}$ at that time. As it can be seen, the value $V_{0}=1.23\times 10^{-11}M_p^2$ sets the correct normalization of the power spectra at the end of reheating. For the results shown here, the initial values taken for the scalar fields $\phi_1$ and $\phi_2$ are $17$ $M_p$ and $0.0025$ $M_p$, respectively. Finally, the initial value for the dimensionless density parameter for radiation is set to $3.46\times 10^{-17}$.

Figure~\ref{Fig1exp} shows the transition from inflation to radiation-dominated epoch. In particular, Figure \ref{fig:fig1a} presents the evolution of the $\varepsilon$ slow-roll parameter from the last e-folds of inflation to a late evolution where $\varepsilon=2$, consistent with a radiation-dominated universe, after oscillating during a few e-folds. On the other hand, Figure \ref{fig:fig1b} presents the evolution of $\eta_{\parallel}$ during the last e-folds of inflation and the beginning of the radiation-dominated epoch. It is interesting to note that $\eta_{\parallel}$ has an oscillatory behavior with a larger amplitude than $\varepsilon$ about the value $2$, which is the expected value that this takes when the universe becomes radiation-dominated. As it is depicted in Figure \ref{fig:fig1c}, the $\phi_1$ field oscillates about its minimum and its kinetic energy is transferred to the radiation fluid. As radiation becomes the dominant component, Hubble damping slows the motion of $\phi_2$, which becomes a constant value. Finally, in Figure \ref{fig:fig1d}, we have plotted the fractional contributions (with respect to the total energy density) of the energy densities of the inflaton fields (solid blue line) and radiation (dashed red line). Notice that the value of the fractional contribution of the energy density from radiation at the end of inflation becomes $\Omega_R\simeq 0.07$, being negligible in comparison the contribution of the inflaton fields. On the other hand, we also confirm that the universe reheats properly in this model. We can see that as the model reheats, the proportion of energy density stored in the fields drops to zero and is converted into radiation. The numerical computation stops around $3$ $e$-folds of reheating, when $\Omega_R\simeq 0.994$. Then we assume that reheating ends at that time.

\begin{figure}[ht]
  \centering
  \begin{subfigure}[b]{0.5\linewidth}
    \centering\includegraphics[width=240pt]{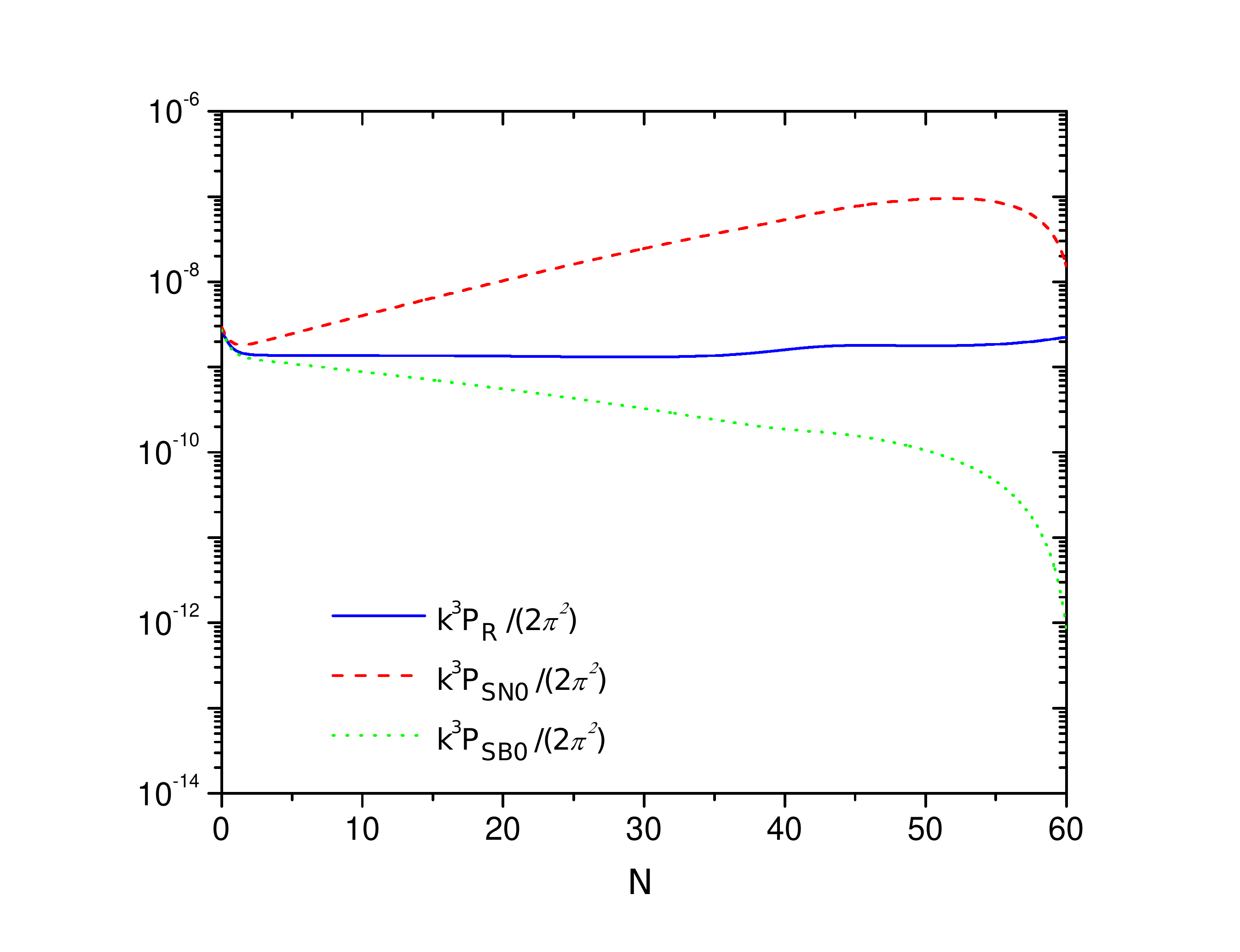}
    \vspace{0 cm}
    \caption{\label{fig:fig2a}}
  \end{subfigure}
  \hspace{-1 cm}
  \begin{subfigure}[b]{0.5\linewidth}
    \centering\includegraphics[width=240pt]{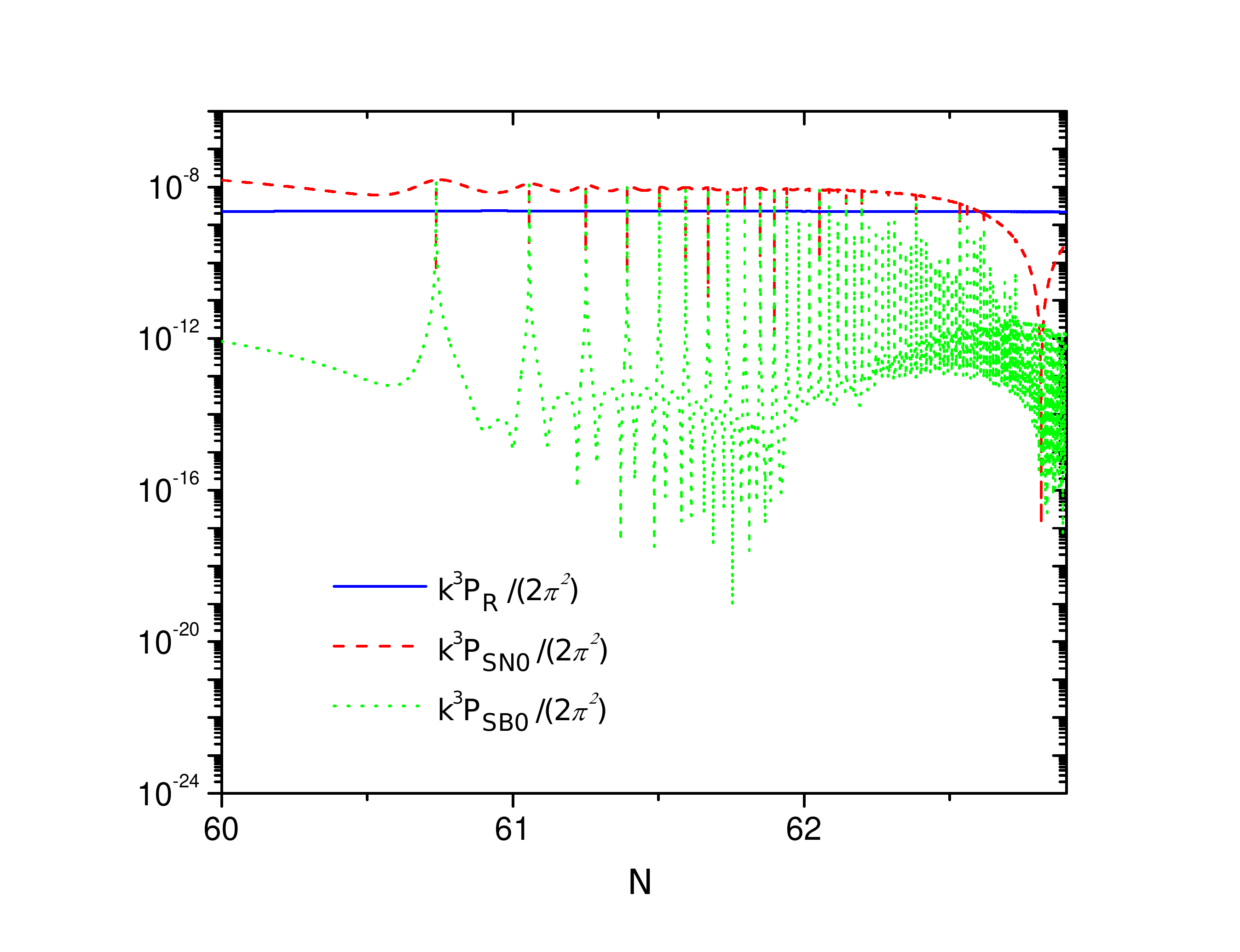}
     \vspace{0 cm}
    \caption{\label{fig:fig2b}}
\end{subfigure}
\begin{subfigure}[b]{0.5\linewidth}
    \centering\includegraphics[width=240pt]{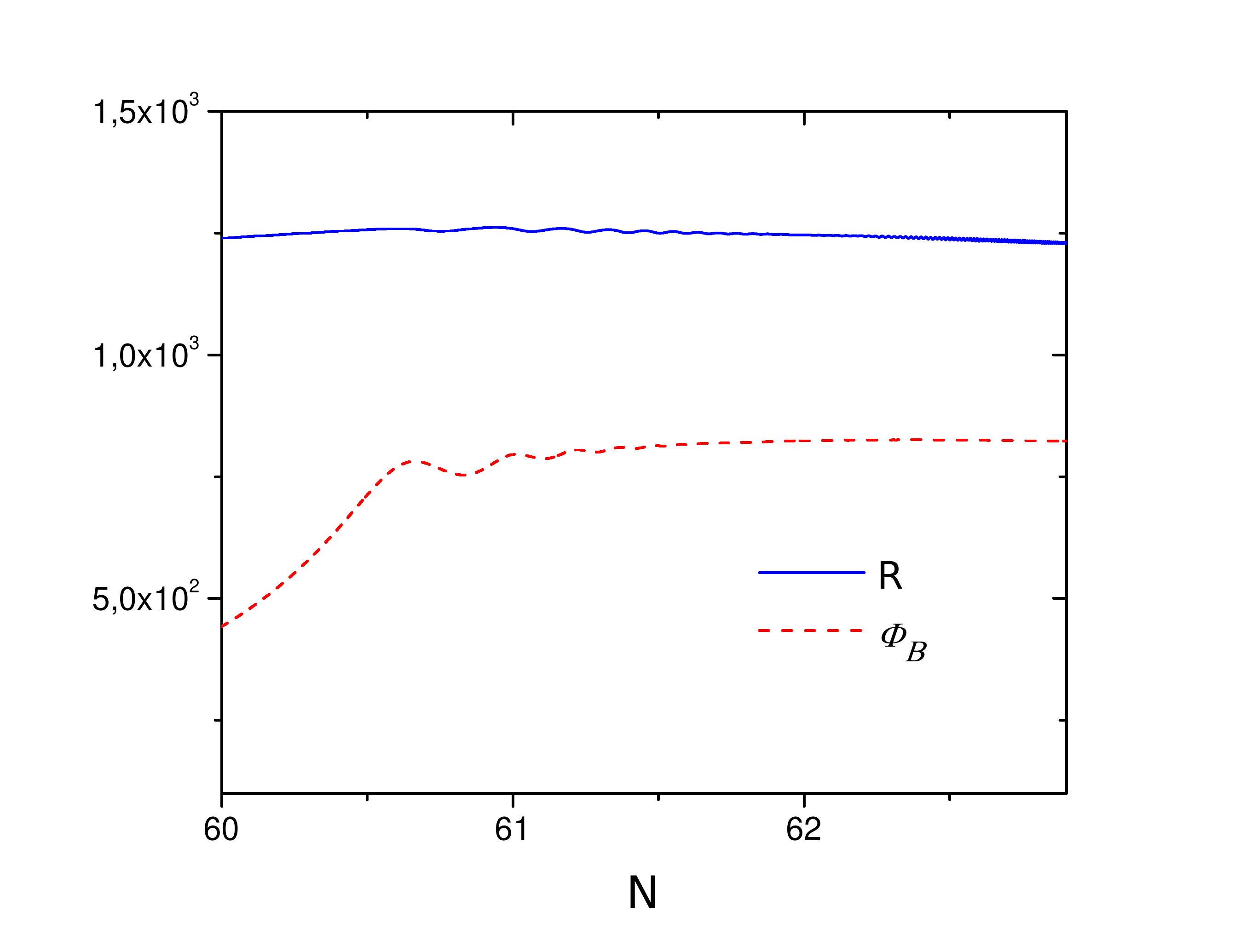}
     \vspace{0 cm}
    \caption{\label{fig:fig2c}}
\end{subfigure}
\hspace{-1 cm}
\begin{subfigure}[b]{0.5\linewidth}
    \centering\includegraphics[width=240pt]{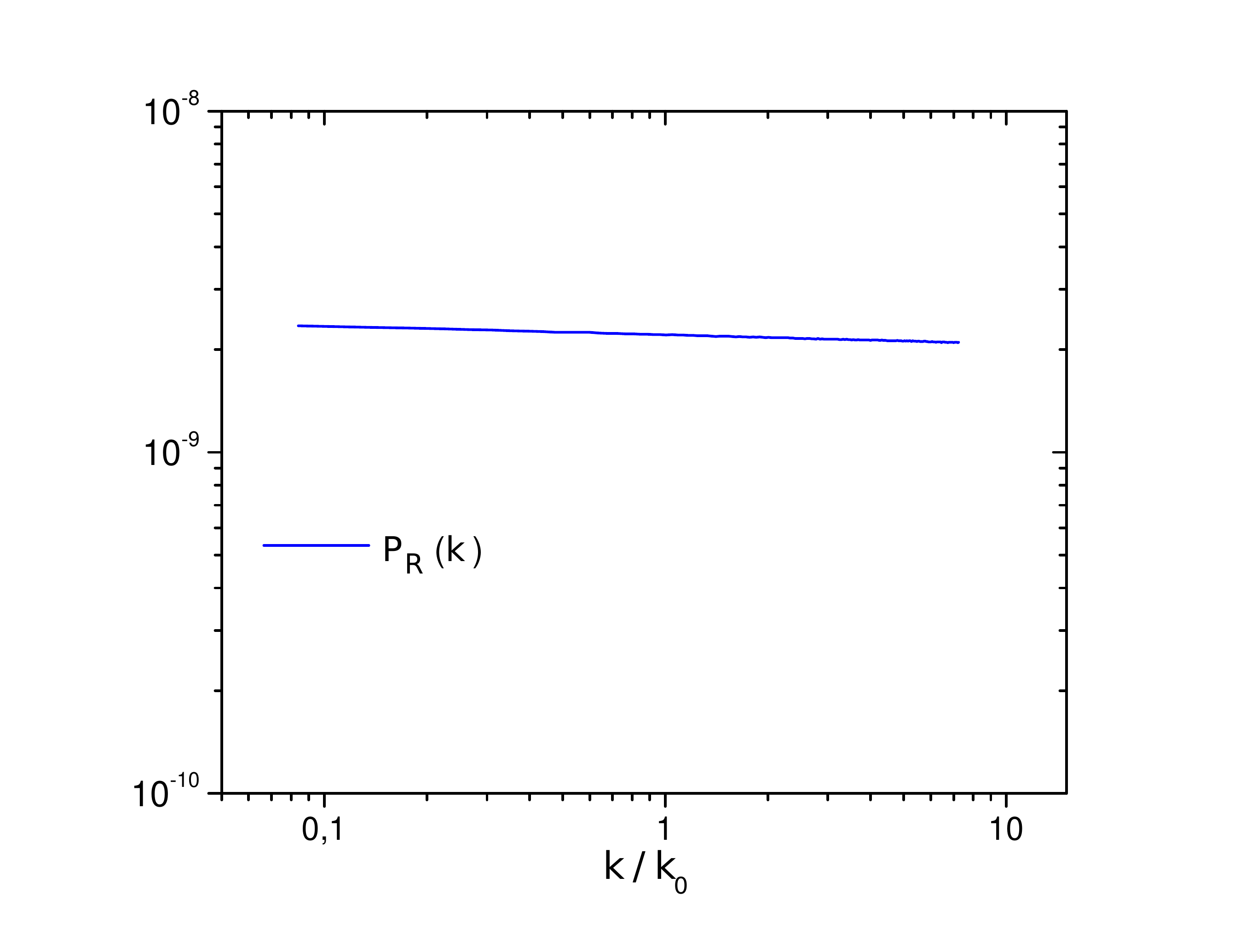}
     \vspace{0 cm}
    \caption{\label{fig:fig2d}}
\end{subfigure}
  \caption{Figures \ref{fig:fig2a} and \ref{fig:fig2b} show the comparison of the power spectra $\mathcal{P}_{\mathcal{R}}$ (solid blue line), $\mathcal{P}_{\mathcal{S}_{\mathcal{N}0}}$ (dashed red line), and $\mathcal{P}_{\mathcal{S}_{\mathcal{B}0}}$ (dotted green line) at the pivot scale $k_0=0.002$ Mpc$^{-1}$ during inflation and reheating, respectively, for the product exponential potential. In Figure \ref{fig:fig2c} we present the evolution of the amplitudes of the Bardeen potential $\Phi_\mathbf{B}$ and the total curvature perturbation during reheating, which also has been plotted as a function of the number of $e$-folds $N$. The power spectrum $\mathcal{P}_{\mathcal{R}}$ (solid blue line) in terms of the ratio $k/k_0$ at the end of reheating is plotted in Figure \ref{fig:fig2d}.}
   \label{Fig2exp}
\end{figure}

\afterpage{\clearpage}

Figure \ref{Fig2exp} shows the evolution of primordial perturbations during inflation as well as during reheating. In first place, Figures \ref{fig:fig2a} and \ref{fig:fig2b} show the comparison of the power spectra $\mathcal{P}_{\mathcal{R}}$ (solid blue line), $\mathcal{P}_{\mathcal{S}_{\mathcal{N}0}}$ (dashed red line), and $\mathcal{P}_{\mathcal{S}_{\mathcal{B}0}}$ (dotted green line) at the pivot scale $k_0=0.002$ Mpc$^{-1}$ during inflation and reheating, respectively. At this point we recall that, in the basis $\left(\mathcal{T}_0^A,\mathcal{N}_0^A,\mathcal{B}_0^A\right)$, $\mathcal{S}_{\mathcal{B}0}$ represents the interaction between the inflaton fields and radiation  at perturbative level, in the sense that it interchanges the contribution to the total curvature from one of them to another, as it can be inferred from Eq.~(\ref{zeta_R}). In addition, Eq.~(\ref{F phi}) shows that $\mathcal{S}_{\mathcal{N}0}$ represents the isocurvature of the inflaton fields. What is clear from Figure \ref{fig:fig2a} is that, during inflation, the isocurvature perturbation measured by $\mathcal{S}_{\mathcal{N}0}$ has a significantly larger amplitude than the curvature perturbation $\mathcal{R}$ which slowly increases. In particular, when inflation ends, the amplitude of the isocurvature $\mathcal{S}_{\mathcal{N}0}$ becomes $10^{-8}$, being one order of magnitude larger than the curvature power spectra. In addition, an interesting feature produced by the small, but non-vanishing coupling between the two-field system and the radiation fluid during inflation, is that the scalar spectral index at the end of inflation becomes $n_{\mathcal{R}}=0.974$. On the other hand, regarding the power spectra of the isocurvature $\mathcal{S}_{\mathcal{B}0}$, it has a lower value in comparison with the other ones during inflation. In addition, as it can be seen from Figure \ref{fig:fig2b}, the amplitude of the curvature perturbation remains almost a constant value after the end of inflation and also presents an oscillatory behavior with tiny amplitude. This behavior is more clearly shown in Figure \ref{fig:fig2b}, which displays the evolution of the magnitude of $\mathcal{R}$ in comparison to the Bardeen potential $\Phi_\mathbf{B}$. During reheating, the amplitude of $\mathcal{S}_{\mathcal{N}0}$ also presents oscillatory patterns, staying almost the same value reached at the end of inflation.  Approximately at 2.5 $e$-folds after inflation ends, the power spectra of $\mathcal{S}_{\mathcal{N}0}$ begins to be suppressed and then, at the end of reheating, the curvature and isocurvature power spectra take the values $2.2\times 10^{-9}$ and $3.2\times 10^{-10}$, respectively. Our results suggest that the adiabatic limit is only reached at the end of reheating. This is due that $\phi_1$ asymptotes to a constant as we approach to $\Omega_R\simeq 1$, then the trajectory in the field space does not evolve any longer and $\mathcal{R}$ approaches to a constant value. However, the numerical computation needs to be improved in order to go more $e$-folds further and obtain a more definitive result.

Regarding the post-inflationary evolution of both the Bardeen potential $\Phi_\mathbf{B}$ and the curvature perturbation $\mathcal{R}$, Figure \ref{fig:fig2c} shows that at the end of the reheating phase, $\Phi_\mathbf{B} \simeq 0.669 \mathcal{R}$, being in agreement with the condition $\Phi_\mathbf{B}=\frac{2}{3}\mathcal{R}$, which holds for a radiation-dominated phase~\cite{Ellis}. The comparison of the scale dependence of the power spectra at the end of reheating is displayed in Figure \ref{fig:fig2b} for the range $0.1\leq k/k_0 \leq 10$. This plot illustrate the near scale invariance of the power spectra. In particular, for the curvature power spectra (solid blue line), the scalar spectral index becomes $n_{\mathcal{R}}\simeq 0.973$, which deviates from the maximum likelihood value from Planck 2015.

\begin{figure}[ht]
  \centering
  \begin{subfigure}[b]{0.5\linewidth}
    \centering\includegraphics[width=240pt]{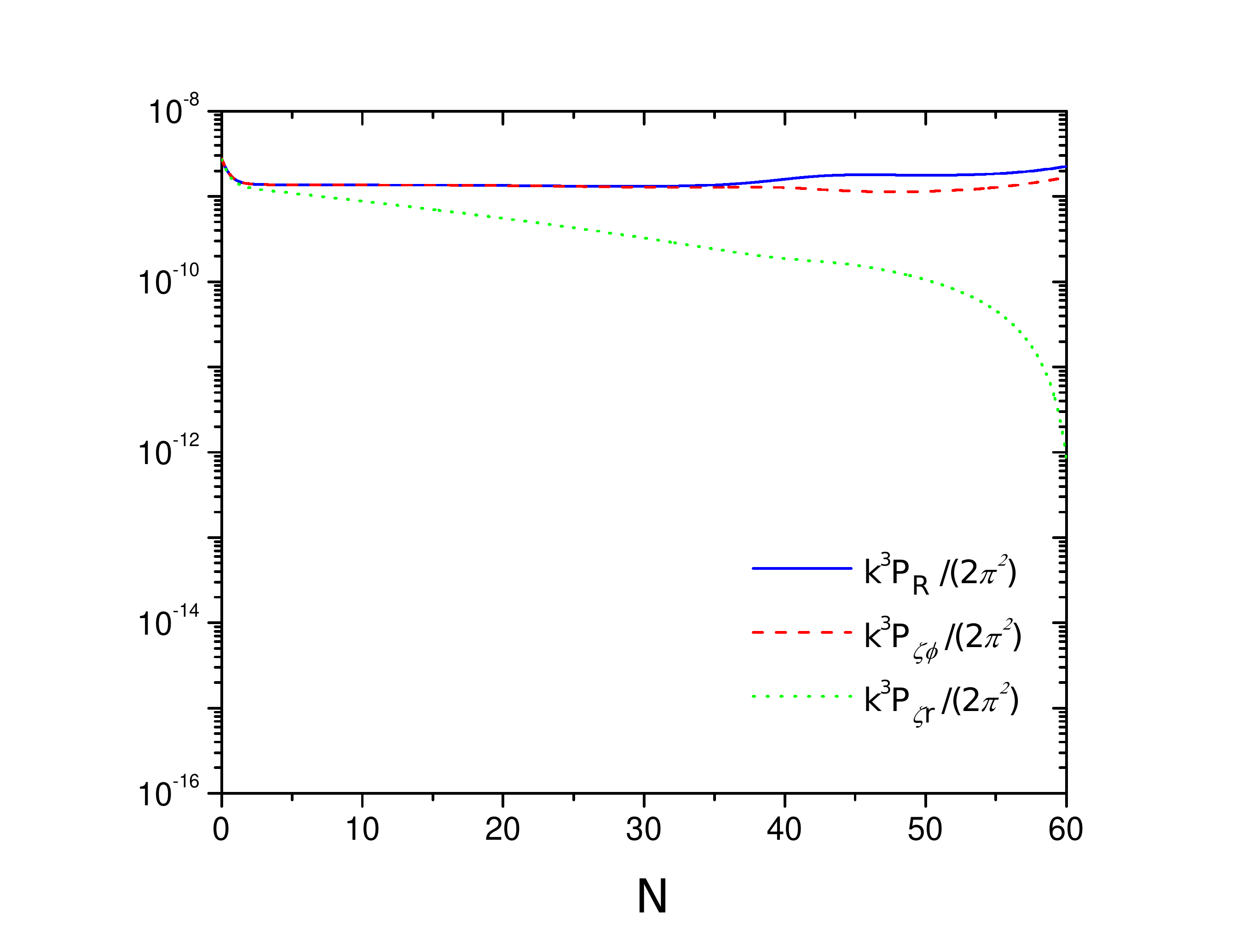}
    \vspace{0 cm}
    \caption{\label{fig:fig3a}}
  \end{subfigure}
  \begin{subfigure}[b]{0.5\linewidth}
    \centering\includegraphics[width=240pt]{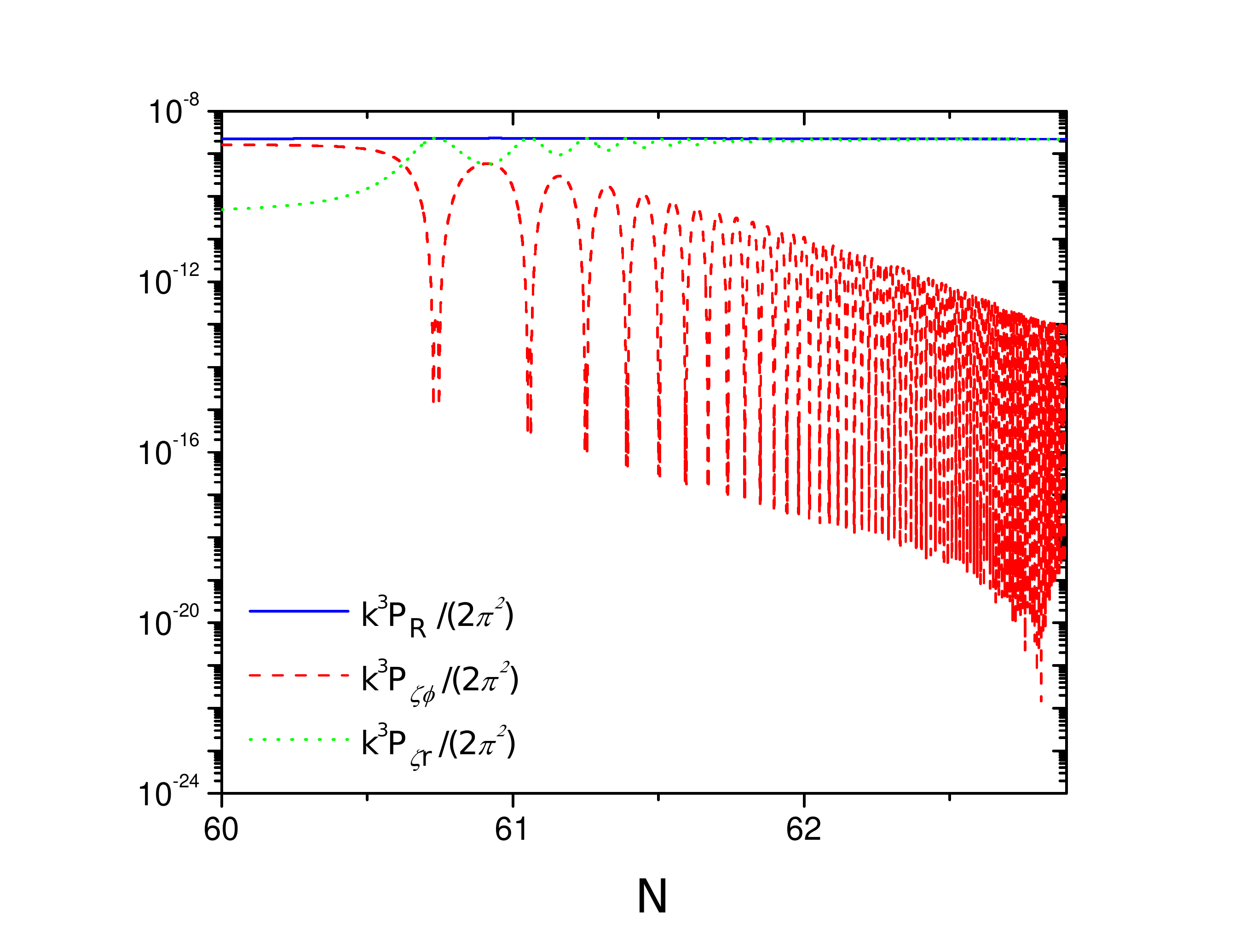}
     \vspace{0 cm}
    \caption{\label{fig:fig3b}}
  \end{subfigure}
  \caption{Figures~\ref{fig:fig3a} and~\ref{fig:fig3b} show the evolution of the curvature perturbations (total as solid blue line, dashed red and dotted green for the fields and fluid
 contribution, respectively) for the product exponential potential during inflation and reheating as functions of the number of $e$-folds, respectively.}
   \label{Fig3exp}
\end{figure}

\afterpage{\clearpage}

Alternatively, we may split the total curvature perturbation $\mathcal{R}$ as $ \mathcal{R} = \zeta_{\phi} + \zeta_R$, where $\zeta_{\phi}$ and $\zeta_{R}$ are the individual curvature perturbations of the two-field system and radiation defined on hypersurfaces orthogonal to comoving world lines, given by Eqs.~(\ref{zeta_phi}) and (\ref{zeta_R}), respectively. The evolution of power spectra of $\zeta_{\phi}$ (dashed red line) and $\zeta_R$ (dotted green line) during inflation as well as during reheating, in comparison to the power spectra of $\mathcal{R}$ (solid blue line), is depicted in Figure \ref{Fig3exp}. Since we have introduced a coupling between the inflaton fields and radiation fluid from the beginning of the numerical computation, Figure \ref{fig:fig3a} confirms that during the last 20 $e$-folds of inflation, $\zeta_{\phi}$  deviates from $\mathcal{R}$, however the contribution from the radiation fluid to curvature perturbation may be regarded negligible. Around 0.5 $e$-folds after the end of inflation, the power spectra of $\zeta_{\phi}$ and $\zeta_{R}$ become equal, but shortly after, the power spectra of $\zeta_{\phi}$ begins to increase with oscillatory behavior and becomes equal to the power spectra of curvature perturbation, i.e. $\mathcal{R}\simeq \zeta_{R}$, and hence the power spectra of $\zeta_{\phi}$ rapidly gets suppressed, as can be seen from Figure~\ref{fig:fig3b}. Our results show that during the first stages of reheating, although the energy stored in the fields has not been totally transferred to radiation fluid, the main contribution to total curvature comes from radiation fluid.

We have shown for this model that, at the end of reheating, the curvature perturbation becomes almost constant, achieving the adiabatic limit. Additionally, the initial conditions for the perturbations during radiation-dominated phase are set at the end of reheating phase. Finally, the tilt of the curvature power spectrum, measured by $n_{\mathcal{R}}$ is enhanced during reheating, reaching a value marginally consistent with the maximum likelihood from current Planck data.\\

\subsection{An Ultra-Light Field (ULF) Coupled to Inflation} \label{SubSec: An Ultra-Light Field Coupled to Inflation}

As it was mentioned in the introduction, the presence of light scalar field with masses much smaller than the Hubble expansion rate are known to produce potentially large levels of isocurvature perturbations, leading to super-Hubble evolution of curvature perturbations and possibly observable features in the CMB. Motivated by this, in Ref.~\cite{Achucarro:2016fby}, the authors studied the consequences of considering the extreme situation in which a non-adiabatic mode is approximately massless, and its interaction with the curvature perturbation persists during the whole period of inflation, from horizon crossing until reheating. The authors provided a concrete example in which an ultra-light field emerges, that appears within a well studied class of models consisting of a multi-field action with a non-canonical kinetic term~\cite{Lalak:2007vi}, typical of supergravity and string theory compactifications. For this class of models, the field-space metric is given by Eq.~(\ref{Metric q}) with $\mathbf{q}_{(11)}=e^{2\phi_2/R_0}$ and $\mathbf{q}_{(22)}=1$, which describes a two-dimensional hyperbolic manifold of curvature $-2/R^2_0$.

In order to obtain concrete results, the authors studied the dynamics of the system for a monomial potential of the form:
\begin{eqnarray}
\label{PGA}
V(\phi_1)=V_0\left(\phi_1/\phi_0\right)^n,
\end{eqnarray}
where $\phi_0$ is the value of the field $\phi_1$ at a given reference time $t_0$. For this non-canonical model with the monomial potential, the authors found that, in order to obtain $60$ $e$-folds of inflation and a value for the scalar spectral index $n_\mathcal{R}$ close to the maximum likelihood value from Planck 2015, the power $n$ must satisfy $n<4/5$, implying that the potential $V$ must be concave. For the particular case $n=1/2$, it is required that $R_0=2/3$ which implies that $n_s=0.967$. This ensures a huge enhancement to the curvature power spectrum at super-horizon scales. Then, the presence of such a light field implies non-vanishing isocurvature modes and a super-Hubble evolution of the curvature perturbation $\mathcal{R}$. This makes it really interesting to study the post-inflationary evolution of this class of model to see how reheating modifies the evolution of the primordial observables.

A first point to note here is that if the power $n$ of the monomial potential is an odd number, the potential of Eq.~ (\ref{PGA}) is not suitable to properly finish  inflation. However, by introducing the modified potential given by:
\begin{eqnarray}
\label{Vmod}
V_1(\phi_1)=V_0\left[\left(1+\left(\frac{\phi_1}{\tilde{\phi}_{0}}\right)^2\right)^{\frac{n}{2}}-1\right],
\end{eqnarray}
it is possible to address the end of inflation. For this modified potential, $\tilde{\phi}_0$ denotes the value of the scalar field $\phi_1$ for which the potential changes its concavity. When $\phi_1/\tilde{\phi}_0\gg 1$, this potential behaves as the monomial one and also presents a minimum at $\phi_1=0$. On the other hand, by adding an axion-like potential \cite{Palma:2017lww}:
\begin{eqnarray}
V_2(\phi_1)=\Lambda^4\left[1-\cos\left(\frac{\phi_2}{f}\right)\right],
\end{eqnarray}
characterized by two mass scales $f$ and $\Lambda$, with $f\gg \Lambda$, and provided that $V_0\gg \Lambda^4$, the effects of this potential on the inflationary dynamics will be sub-dominant comparing to $V(\phi_1)$, however, as it will be shown, the new term will be relevant in the post-inflationary dynamics, since the light field will acquire a small mass term, stabilizing its dynamics at the end of reheating and providing the suppression of the isocurvature perturbation during reheating. Then, the total potential for studying the inflationary dynamics together with the transition into reheating, is considered to be:
\begin{eqnarray}
\label{total_PG}
V=V_1+V_2=V_0\left[\left(1+\left(\frac{\phi_1}{\tilde{\phi}_{0}}\right)^2\right)^{\frac{n}{2}}-1\right]+
\Lambda^4\left[1-\cos\left(\frac{\phi_2}{f}\right)\right].
\end{eqnarray}
In order to model reheating for this potential and determine the possible effects of the coupling between the inflaton fields and the radiation fluid, parameterized by $q_1$, and
the value of the field $\phi_1$ for which the concavity of potential changes, in our numerical implementation, we will study three different cases separately, namely i) $q_1=3.28$ and $\tilde{\phi}_0=\phi_{10}/50$, ii) $q_1=3.28$ and $\tilde{\phi}_0=\phi_{10}/90$, and finally, iii) $q_1=3.33$ and $\tilde{\phi}_0=\phi_{10}/50$. For all cases to be studied, we use the following set of values:
\begin{eqnarray}
&& n = \frac{1}{2} \textrm{, } \quad R_0 = \frac{2}{3} \textrm{, } \quad V_0 = A \times 10^{-11}\left(\frac{\tilde{\phi}_0}{\phi_{10}}\right)^n, \\
&& \Lambda = 2\times 10^{-2}V_0^{1/4} \textrm{, } \quad f=0.4M_p, \\
&& \phi_{10} = 11M_p \textrm{, } \quad \phi_{20}=-0.2R_0M_p,
\end{eqnarray}
where $\phi_{10}$ and $\phi_{20}$ denote the initial field values and $A$ is a dimensionless parameter. These choices allow us to compare our results with the previous work \cite{Achucarro:2016fby}. In addition, for each case, the $A$ parameter must be set in order to give the correct normalization of the curvature power spectra. Finally, the corresponding initial values for the dimensionless density parameter for radiation $\Omega_R$ are set to $9.74\times 10^{-15}$, $1.21\times 10^{-11}$, and $1.28\times 10^{-11}$, respectively.\\

\begin{figure}[ht]
  \centering
  \begin{subfigure}[b]{0.5\linewidth}
    \centering\includegraphics[width=300pt]{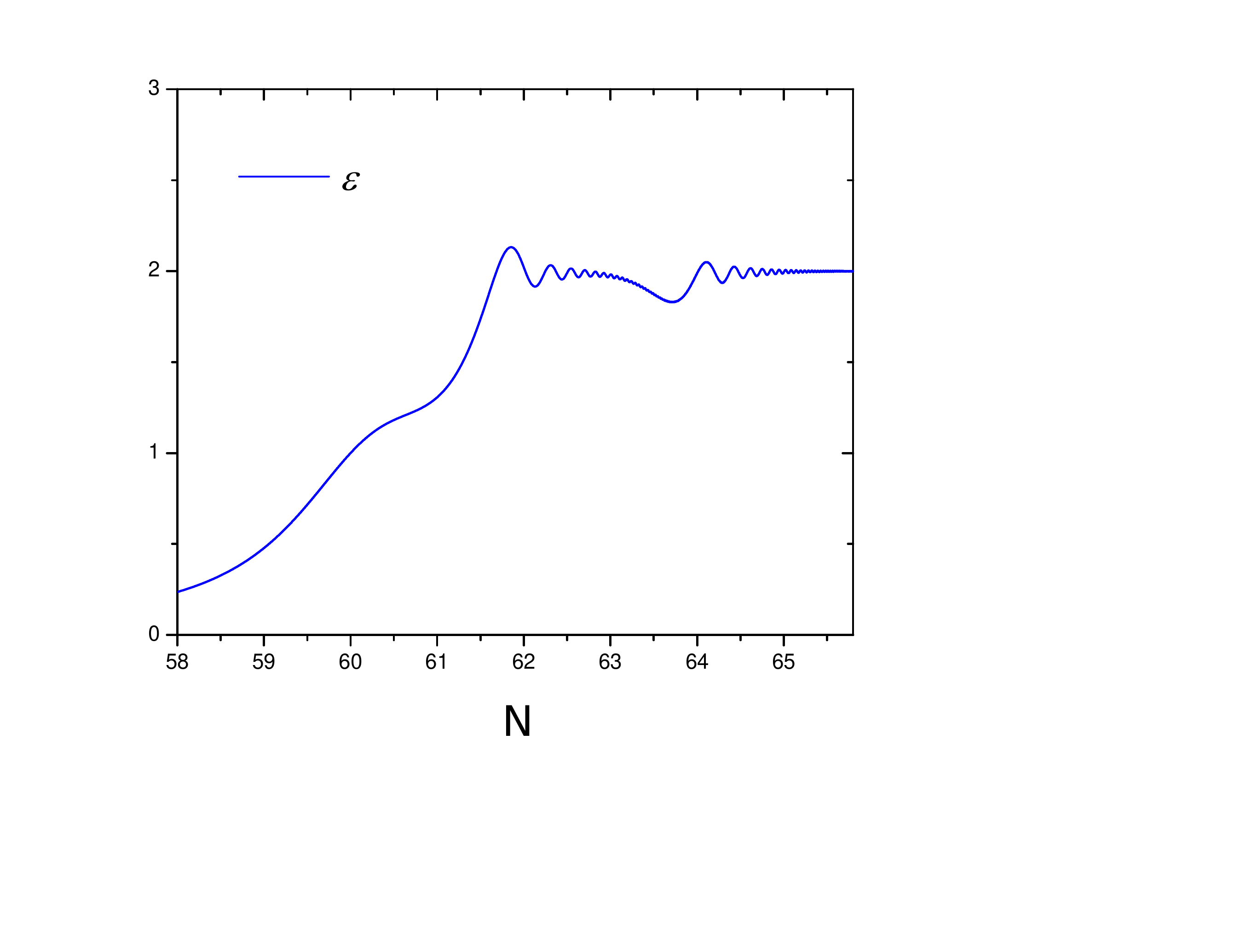}
    \vspace{-2 cm}
    \caption{\label{fig:fig4a}}
  \end{subfigure}
    \hspace{-1 cm}
  \begin{subfigure}[b]{0.5\linewidth}
    \centering\includegraphics[width=300pt]{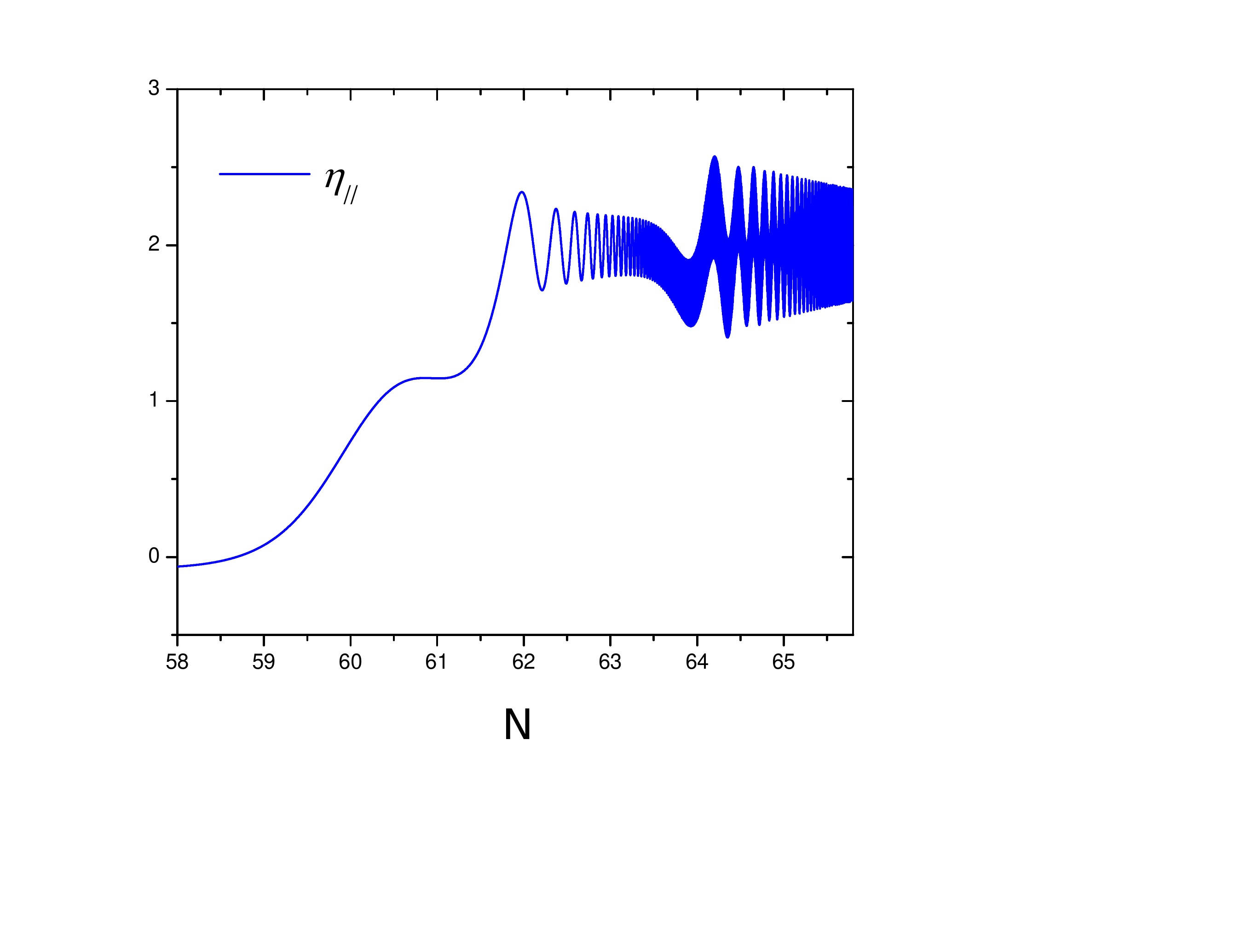}
     \vspace{-2 cm}
    \caption{\label{fig:fig4b}}
  \end{subfigure}
  \begin{subfigure}[b]{0.5\linewidth}
    \centering\includegraphics[width=300pt]{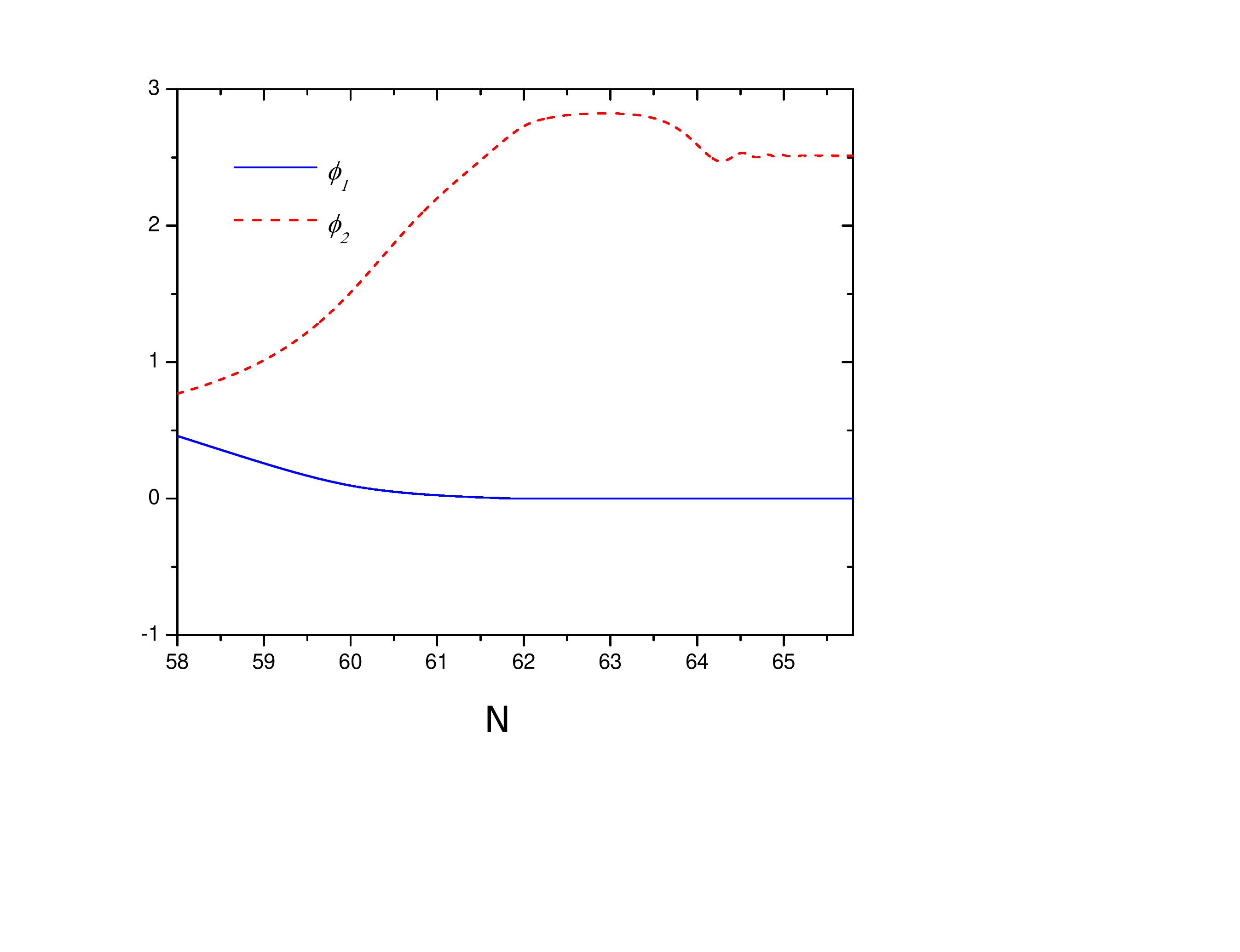}
    \vspace{-2 cm}
    \caption{\label{fig:fig4c}}
  \end{subfigure}
    \hspace{-1 cm}
  \begin{subfigure}[b]{0.5\linewidth}
    \centering\includegraphics[width=300pt]{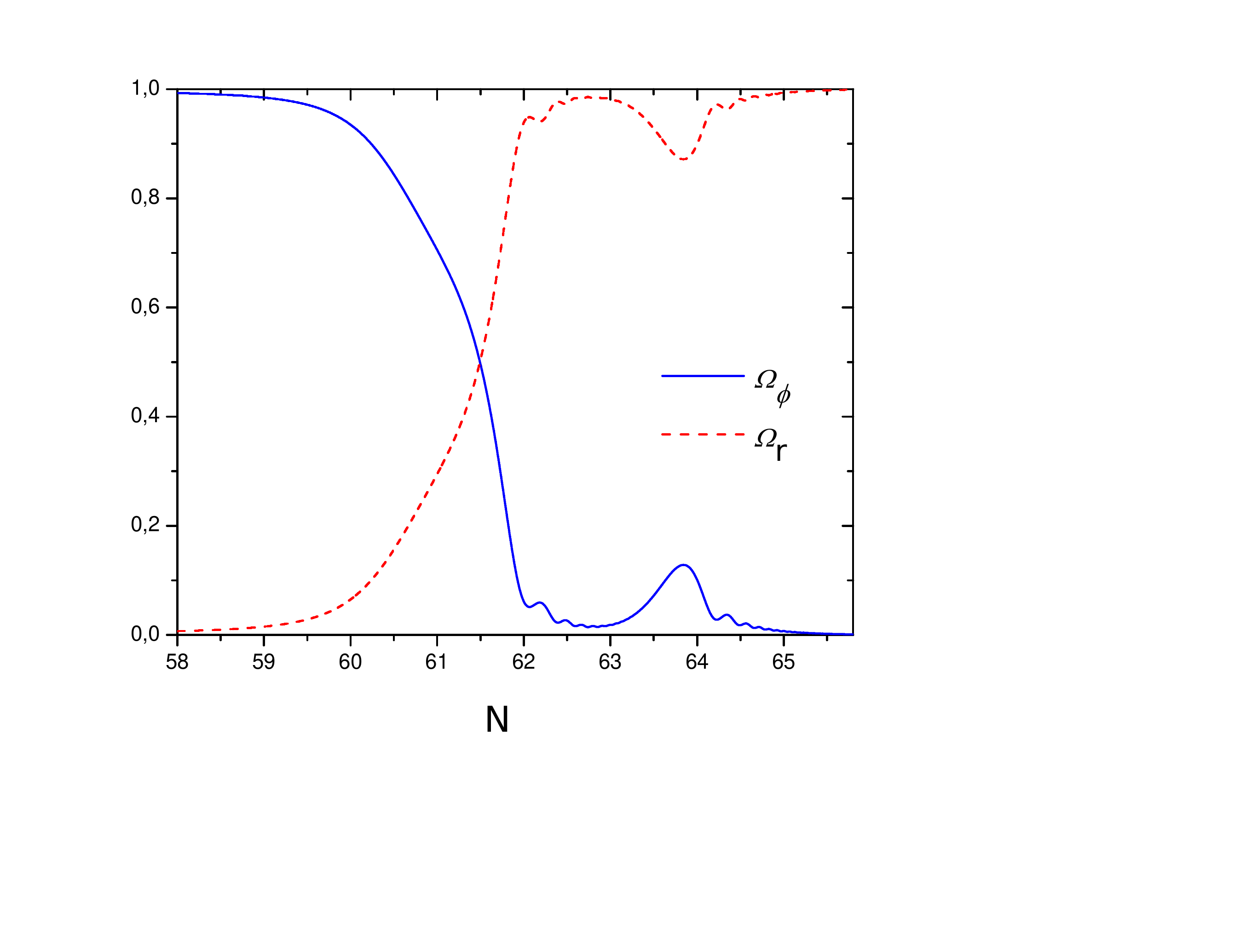}
     \vspace{-2 cm}
    \caption{\label{fig:fig4d}}
  \end{subfigure}
\caption{Transition from inflation to the radiation dominated era, through reheating, at background level for the case (i), corresponding to the potential~(\ref{total_PG}). In particular, Figures~\subref{fig:fig4a} and~\subref{fig:fig4b} show the evolution from the last stage of inflation to the radiation dominated epoch for the slow-roll parameters$\varepsilon$ and $\eta_{\parallel}$, respectively, which are plotted against the number of $e$-folds from the end of inflation. The fields $\phi_1$ (solid blue line) and $\phi_2$ (dashed red line) as function of the number of $e$-folds from the last $e$-folds of inflation to the end of reheating is depicted in Figure~\subref{fig:fig4c}. Figure~\subref{fig:fig4d} shows the evolution of fractional contributions of the energy density from the inflaton fields (solid blue line) and radiation (dashed red line).}
  \label{Fig4PG}
\end{figure}

\afterpage{\clearpage}

\subsubsection*{i) $q_1=3.28$ and $\tilde{\phi}_0=\phi_{10}/50$:}

\begin{figure}[ht]
  \centering
  \begin{subfigure}[b]{0.5\linewidth}
    \centering\includegraphics[width=300pt]{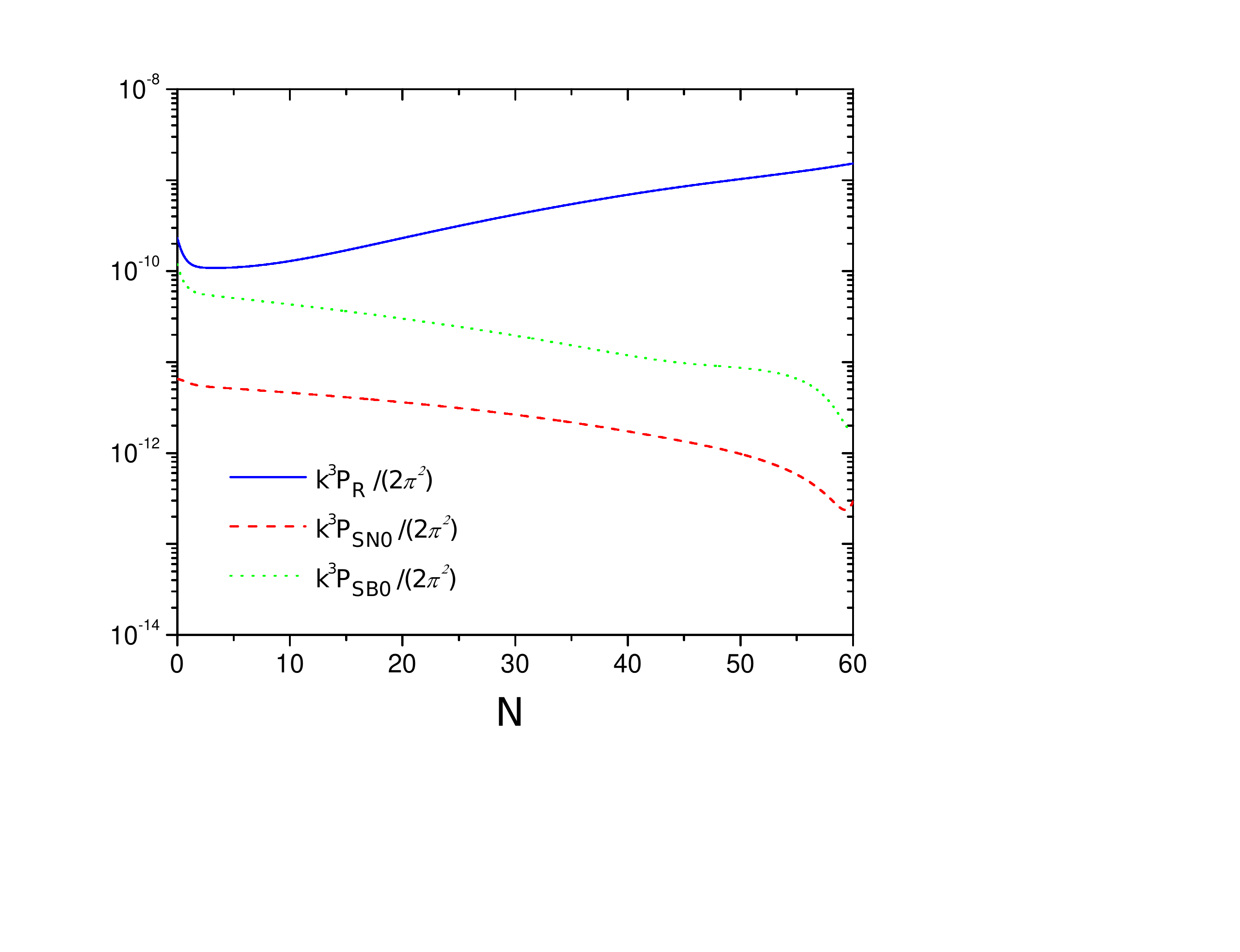}
    \vspace{-2 cm}
    \caption{\label{fig:fig5a}}
  \end{subfigure}
    \hspace{-1 cm}
  \begin{subfigure}[b]{0.5\linewidth}
    \centering\includegraphics[width=300pt]{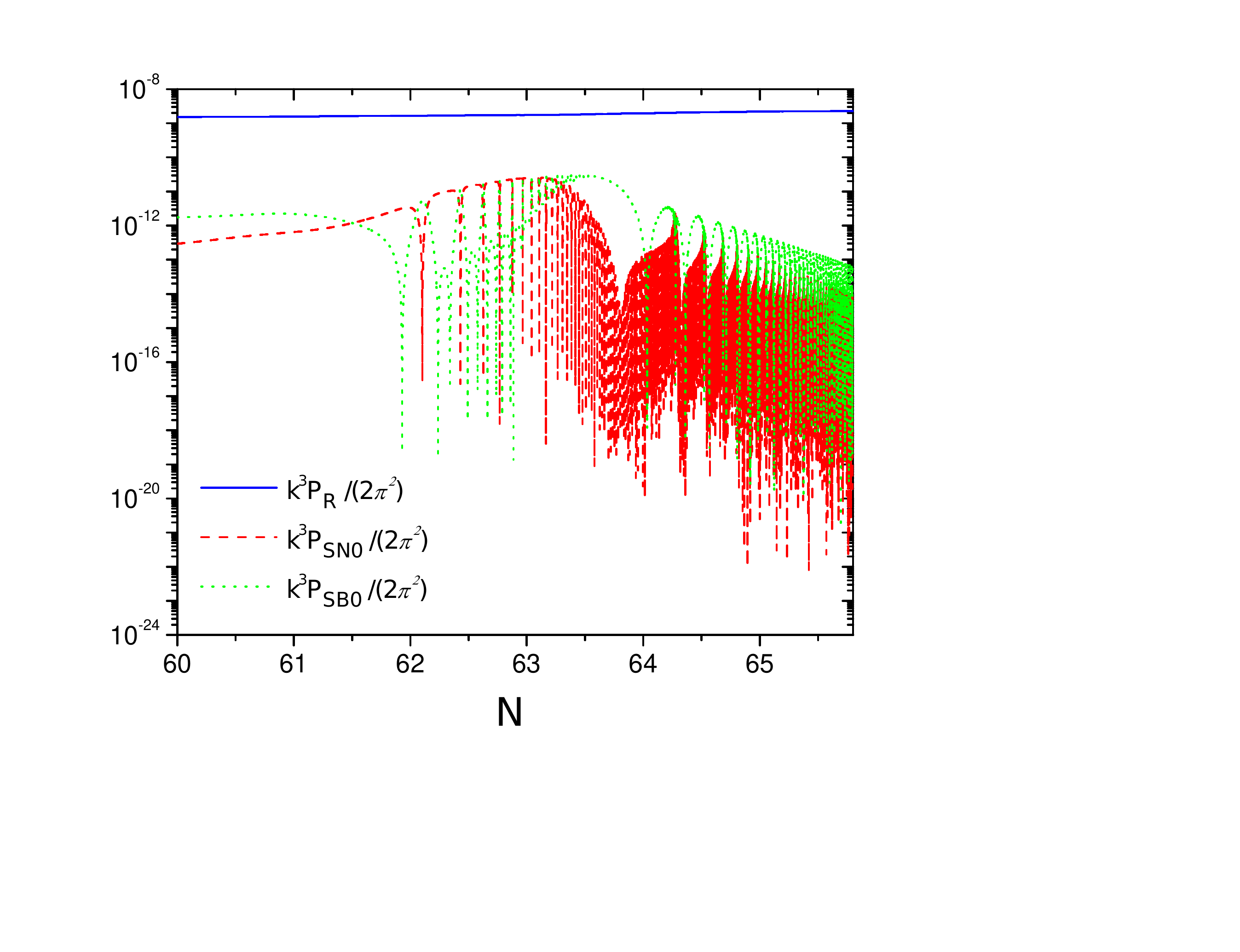}
     \vspace{-2 cm}
    \caption{\label{fig:fig5b}}
  \end{subfigure}
  \begin{subfigure}[b]{0.5\linewidth}
    \centering\includegraphics[width=300pt]{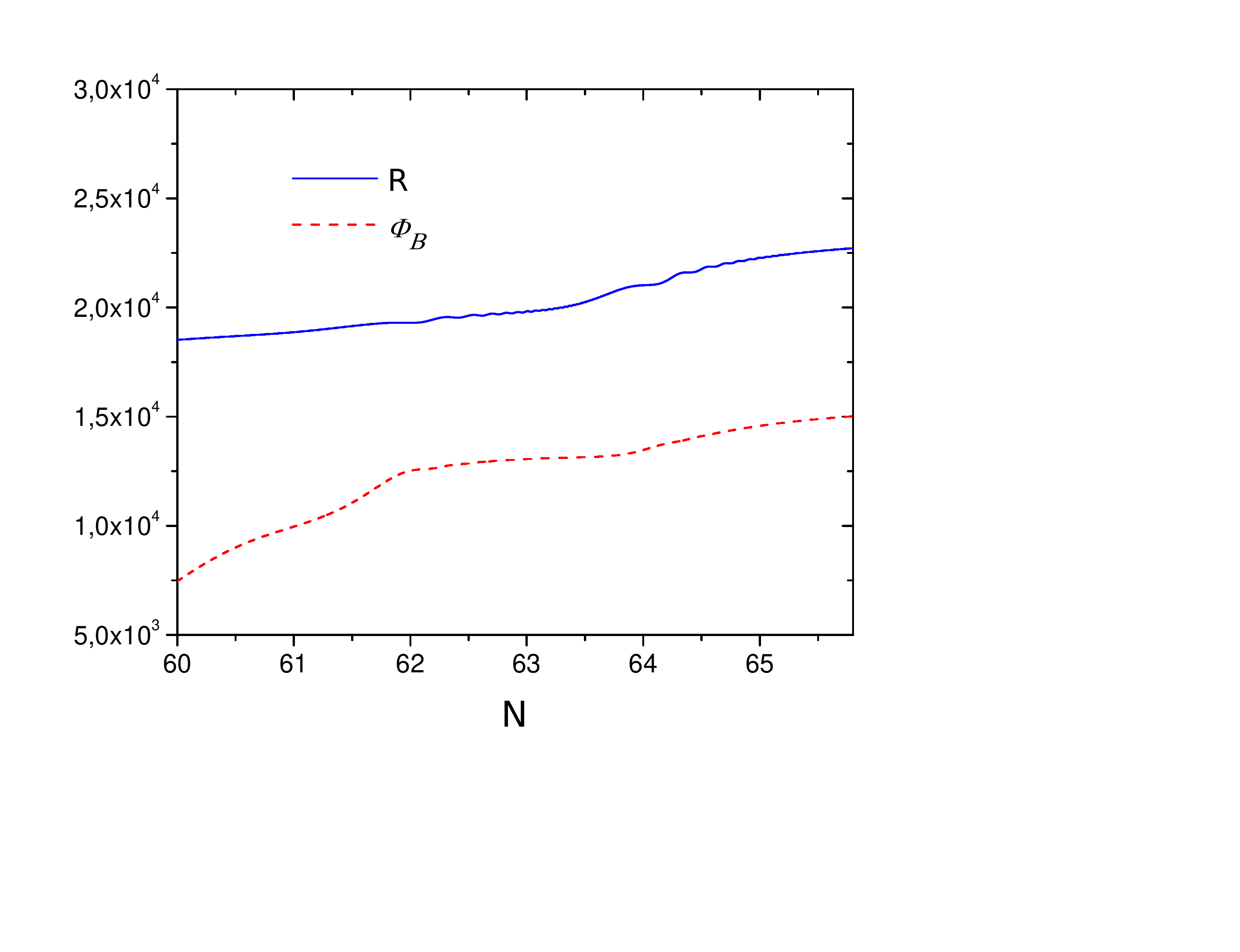}
    \vspace{-2 cm}
    \caption{\label{fig:fig5c}}
  \end{subfigure}
    \hspace{-1 cm}
  \begin{subfigure}[b]{0.5\linewidth}
    \centering\includegraphics[width=300pt]{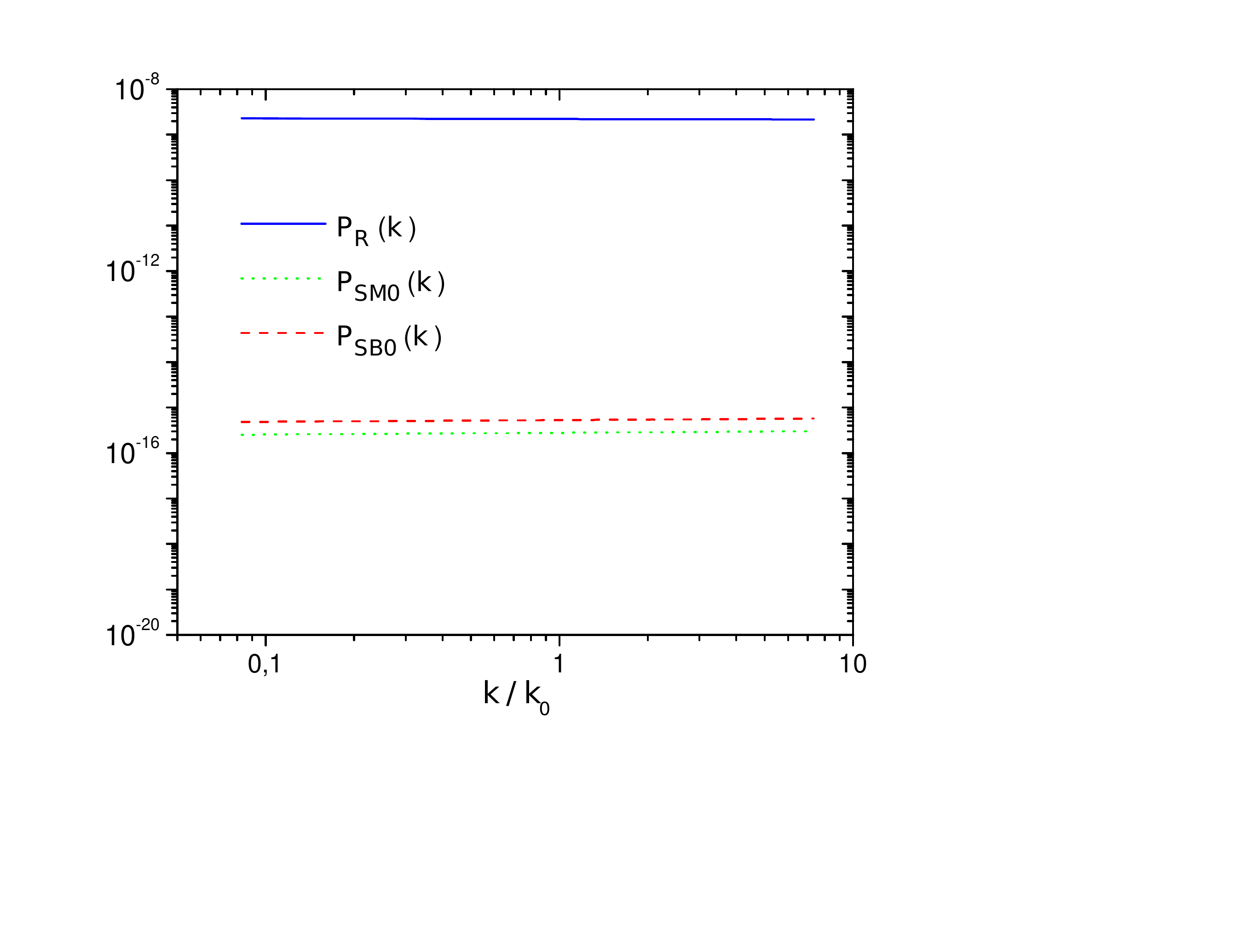}
     \vspace{-2 cm}
    \caption{\label{fig:fig5d}}
  \end{subfigure}
  \caption{Figures~\ref{fig:fig5a} and~\ref{fig:fig5b} show the comparison of the power spectra $\mathcal{P}_{\mathcal{R}}$ (solid blue line), $\mathcal{P}_{\mathcal{S}_{\mathcal{N}0}}$ (dashed red line), and $\mathcal{P}_{\mathcal{S}_{\mathcal{B}0}}$ (dotted green line) at the pivot scale $k_0=0.002$ Mpc$^{-1}$ during inflation and reheating, respectively, for the case (i). In Figure~\ref{fig:fig5c} we present the evolution of the amplitudes of the Bardeen potential $\Phi_\mathbf{B}$ and the total curvature perturbation during reheating, which also has been plotted as a function of the number of $e$-folds $N$. The power spectra $\mathcal{P}_{\mathcal{R}}$ (solid blue line), $\mathcal{P}_{\mathcal{S}_{\mathcal{N}0}}$ (dashed red line), and $\mathcal{P}_{\mathcal{S}_{\mathcal{B}0}}$ (dotted green line) in terms of the ratio $k/k_0$ at the end of reheating are plotted in Figure~\ref{fig:fig5d}.}
  \label{Fig5PG}
\end{figure}

The smooth transition from inflation to the radiation dominated-epoch is depicted in Figure \ref{Fig4PG}. Particularly, Figure \ref{fig:fig4a} shows the evolution of the $\varepsilon$ slow-roll parameter from the last e-folds of inflation to the radiation dominated-epoch. After oscillating during a few e-folds after the end of inflation, $\varepsilon$ reaches the constant value 2. Interestingly, at around 3 $e$-folds after inflation ends, $\varepsilon$ decreases and reaches a minimum value, and later it starts to increase and oscillate about 2. On the other hand, Figure \ref{fig:fig4b} presents the evolution of $\eta_{\parallel}$ during the last e-folds of inflation and the beginning of the radiation-dominated epoch. As the in previous model, $\eta_{\parallel}$ also has an oscillatory behavior about the value $2$, consistent with a radiation-dominated universe, however the amplitudes of oscillations of both $\varepsilon$ and $\eta_{\parallel}$ during reheating become smaller compared to the previous model. In addition, there is a feature between 3 and 4 $e$-folds after the end of inflation. This feature in the behavior of both $\varepsilon$ and $\eta_{\parallel}$ may be explained by studying the evolution of the fields during reheating, which is depicted in Figure \ref{fig:fig4c}. We observe that both field $\phi_1$ (solid blue line) and $\phi_2$ (dashed red line) oscillate about their minimum and the kinetic energy stored in the fields is transferred to the radiation fluid. As radiation becomes the dominant component, Hubble damping slows down the motion of $\phi_1$ and $\phi_2$. In particular, $\phi_1$ decays to its minimum before than $\phi_2$. This result suggest that most of the kinetic energy is stored in $\phi_2$, then the reheating phase for this case is driven by $\phi_2$. This becomes clearer from Figure \ref{fig:fig5d}, where we have plotted the fractional contributions of the energy densities of the inflaton fields and radiation. Before decaying at the minimum of its potential, $\phi_2$ reaches a maximum value during its oscillations to later invert its motion and finally decay. This behavior yields to a suddenly decrease of the amplitude of the oscillating $\varepsilon$ and $\eta_{\parallel}$, and a sudden increase of the fractional contribution of the inflaton field during a short period. The fractional contribution of radiation at the end of inflation becomes $\Omega_R \simeq 0.007$, then our choice for the values of parameters ensures that this contribution becomes negligible in comparison to the contribution of the inflaton fields. We can see that at the moment when $\phi_1$ reaches its minimum, most of the proportion of energy density in the fields comes from $\phi_2$ and, as the model reheats, the proportion of energy density in $\phi_2$ drops to zero and is converted into radiation. For this case, the numerical computation stops around $6$ $e$-folds of reheating, when $\Omega_R \simeq 0.993$ (dashed red line). Then, we set this time as the end of reheating.

\afterpage{\clearpage}

Regarding the dynamics of perturbations, we can see from Figure \ref{fig:fig5a} that, during inflation, the curvature power spectrum at pivot scale $k_0=0.002$ Mpc$^{-1}$ (solid blue line) increases monotonically, whereas the spectrum of the isocurvature perturbation $\mathcal{S}_{\mathcal{N}0}$ (red dashed line) is slowly decreasing and, as inflation ends, it becomes four order of magnitude smaller than the curvature perturbation. Since we are introducing a coupling between the two-field system and the radiation fluid from the beginning of the inflationary phase, the choice of parameters for the case $q_1=3.2$ and $\tilde{\phi}=\phi_{10}/50$ yields a value for the scalar spectral index of $n_{\mathcal{R}}=0.974$  for the pivot scale at the end of inflation, which deviates from the value obtained in~\cite{Achucarro:2016fby}, given by $0.967$. An explanation for this discrepancy may be found in the value we set for $\tilde{\phi}_0=\phi_{10}/50$. Since that $\tilde{\phi}$ represents the value for $\phi_1$ at which the potential changes its concavity in order to achieve the end of inflation, as $\tilde{\phi}_0$ is increased, $n_{\mathcal{R}}$ becomes greater than the maximum likelihood of Planck.

Regarding the analysis of post inflationary evolution of super-Hubble fluctuations, we recall that in the basis $\left(\mathcal{T}^A,\mathcal{N}_0^A,\mathcal{B}_0^A\right)$, $\mathcal{S}_{\mathcal{B}0}$ gives us an account on how the inflationary fields interact with radiation. On the other hand, $\mathcal{S}_{\mathcal{N}0}$ corresponds to the isocurvature fluctuation. In Figure \ref{fig:fig5b} we zoom into the evolution through the almost 6 $e$-folds of reheating. From Figure \ref{fig:fig5c} we can see that during the initial phase of reheating, the curvature perturbation $\mathcal{R}$ presents a small enhancement, however this feature cannot be seen clearly in Figure \ref{fig:fig5b}. Figure \ref{fig:fig5c} also shows the evolution of the Bardeen potential $\Phi_{\mathbf{B}}$ during reheating, and we notice that at the end of reheating, the curvature perturbation and the Bardeen potential satisfy the relation, $\Phi_{\mathbf{B}} \simeq 0.655 \mathcal{R}$, being close to the condition $\Phi_\mathbf{B}=\frac{2}{3}\mathcal{R}$, which is satisfied during the radiation-dominated phase. Back to Figure~\ref{fig:fig5b}, the isocurvature perturbation $\mathcal{S}_{\mathcal{N}0}$ begins to increase, presenting oscillatory patterns, but always being smaller than the curvature perturbation, which is still evolving. At around 3 $e$-folds after the end of inflation, it can be noticed that the isocurvature perturbation $\mathcal{S}_{\mathcal{N}0}$ begins to be suppressed and the curvature perturbation reaches an almost constant value when reheating ends. At this time, the curvature power spectra has the value $2.28\times 10^{-9}$, in agreement with Planck normalization of $\mathcal{P}_{\mathcal{R}}$, whereas the isocurvature perturbation $\mathcal{S}_{\mathcal{N}0}$ becomes seven orders of magnitude smaller than the curvature perturbation.

The comparison of the scale dependence of the several power spectra at the end of reheating is displayed in Figure \ref{fig:fig5d}, for scales within the range $0.1\leq k/k_0 \leq 10$. This plot illustrates the near scale invariance of the power spectra. In particular for the curvature power spectra (solid blue line), the scalar spectral index becomes $n_{\mathcal{R}}\simeq 0.989$, which is very close to scale invariance. Clearly this value presents a significant deviation from the maximum likelihood value from Planck 2015. On the other hand, for all the range $0.1\leq k/k_0 \leq 10$, the power spectrum of $\mathcal{S}_{\mathcal{N}0}$ is about 7 orders of magnitude smaller than the curvature power spectrum.

\begin{figure}[ht]
  \centering
  \begin{subfigure}[b]{0.5\linewidth}
    \centering\includegraphics[width=300pt]{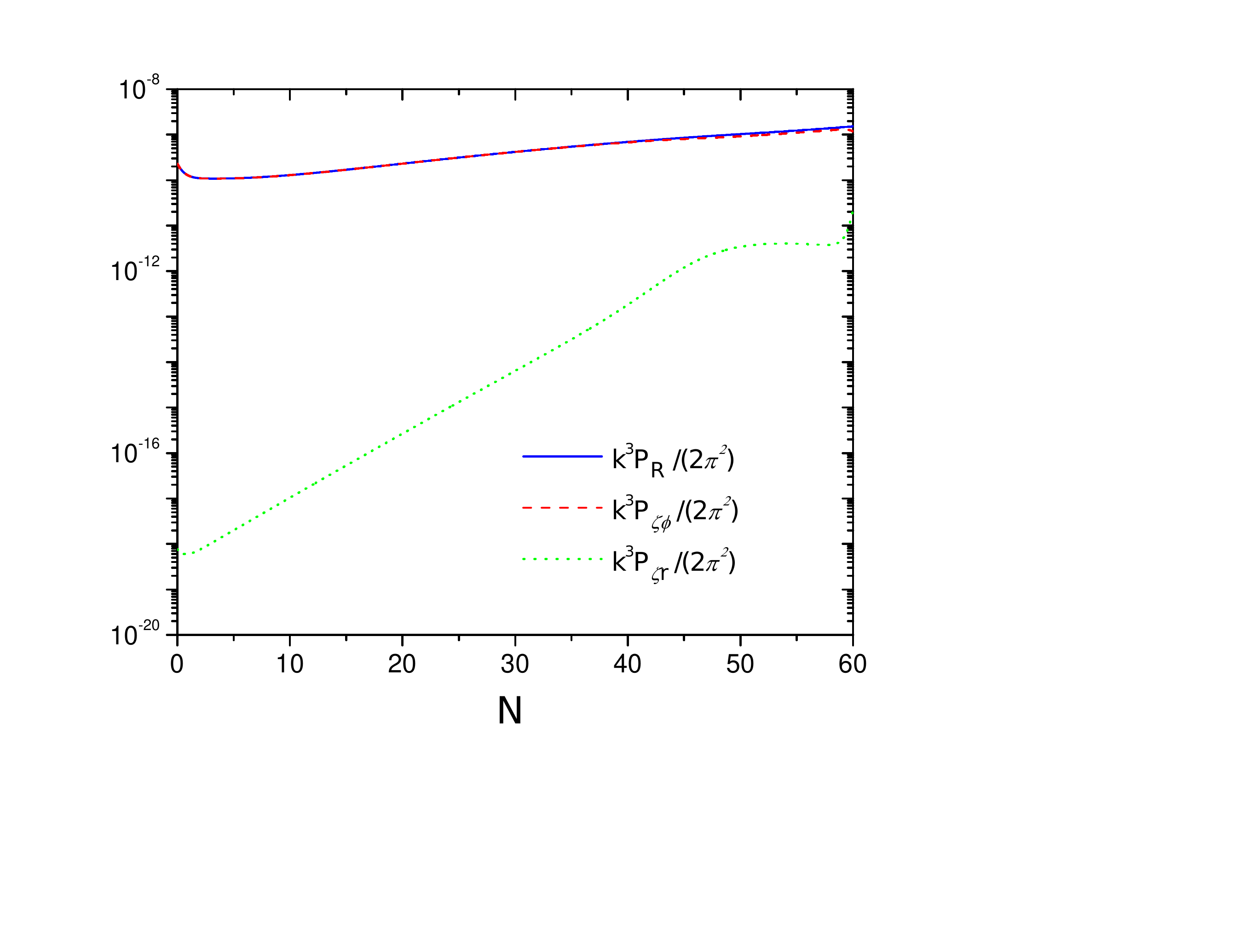}
    \vspace{-2 cm}
    \caption{\label{fig:fig6a}}
  \end{subfigure}
  \begin{subfigure}[b]{0.5\linewidth}
    \centering\includegraphics[width=300pt]{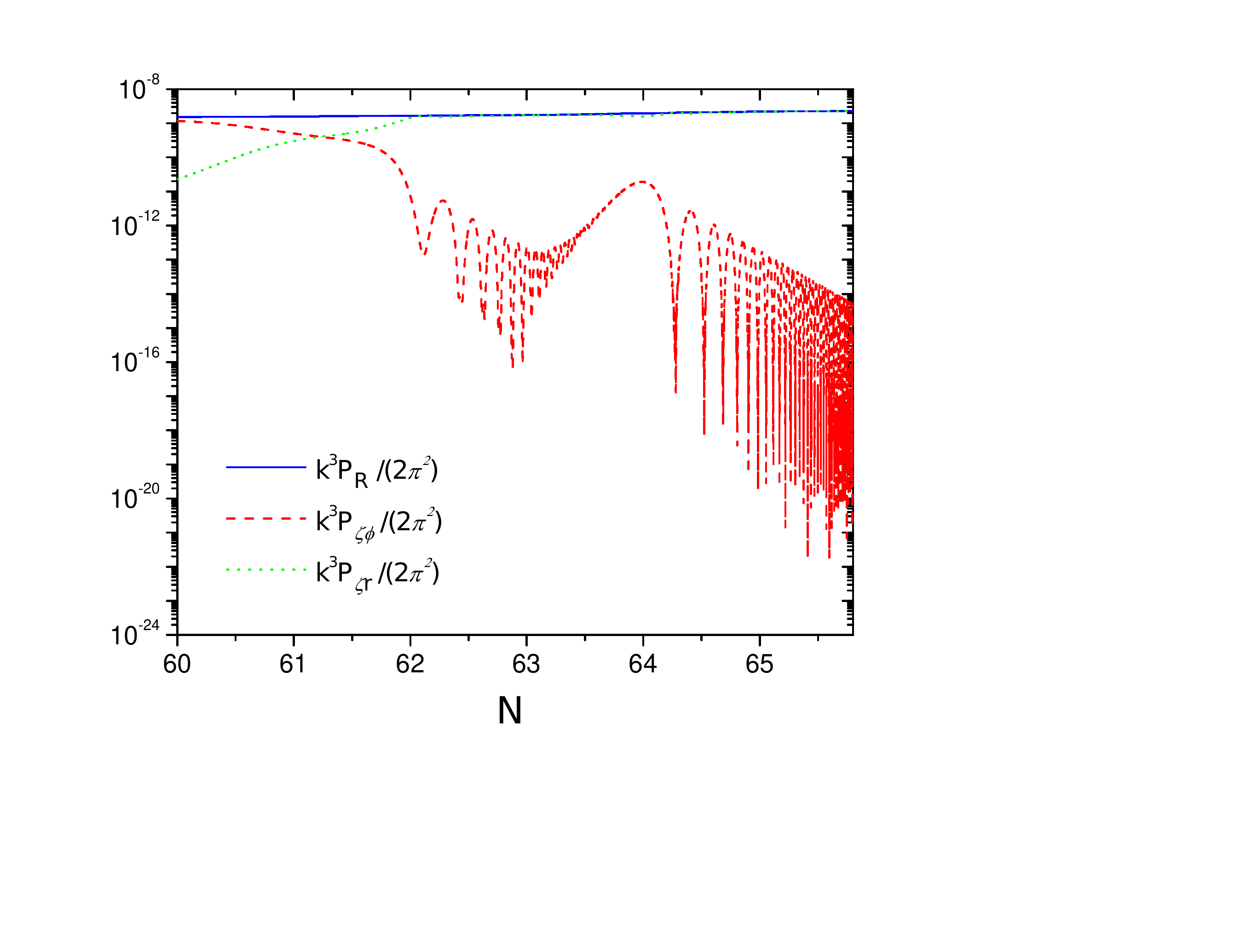}
     \vspace{-2 cm}
    \caption{\label{fig:fig6b}}
  \end{subfigure}
  \caption{Figures~\ref{fig:fig6a} and~\ref{fig:fig6b} show the evolution of curvature perturbations (total as solid blue line, dashed red and dotted green for the fields and fluid
 contribution, respectively) during inflation and reheating as a function of the number of $e$-folds, respectively, for the case (i).}
  \label{Fig6PG}
\end{figure}

Finally, by splitting the total curvature perturbation $\mathcal{R}$ as $\mathcal{R}=\zeta_{\phi}+\zeta_{R}$, as it was done for the product exponential potential, Figure \ref{fig:fig6a} shows that, during inflation, the main contribution to total curvature perturbation (solid blue line) comes from the individual curvature of inflaton fields $\zeta_{\phi}$ (dashed red line), and so its contribution due radiation fluid $\zeta_{R}$ (dotted green line) becomes negligible. However, as it can be noticed in Figure \ref{fig:fig6b}, 2 $e$-folds after inflation ends, $\zeta_{R}$ rapidly becomes the main contribution to curvature perturbation whereas the individual curvature of the inflatonary fields is rapidly suppressed although $\phi_2$ has not decayed totally in radiation. This result confirms our assumption that in order to have a smooth transition from inflation to the radiation-dominated epoch at background and perturbative level, the effects of radiation production must be taken into account in the computations.

\afterpage{\clearpage}

Then, we conclude that at the end of reheating for this case, the curvature perturbation becomes almost constant and the isocurvature mode becomes suppressed, achieving the adiabatic limit. Additionally, the initial conditions for the perturbations during the radiation-dominated phase are set at the end of the reheating phase. In addition, the curvature power spectrum is close to scale invariance, but it deviates from the maximum likelihood from current Planck data.\\

\subsubsection*{ii) $q_1=3.28$ and $\tilde{\phi}_0=\phi_{10}/90$:}

\begin{figure}[ht]
  \centering
  \begin{subfigure}[b]{0.5\linewidth}
    \centering\includegraphics[width=300pt]{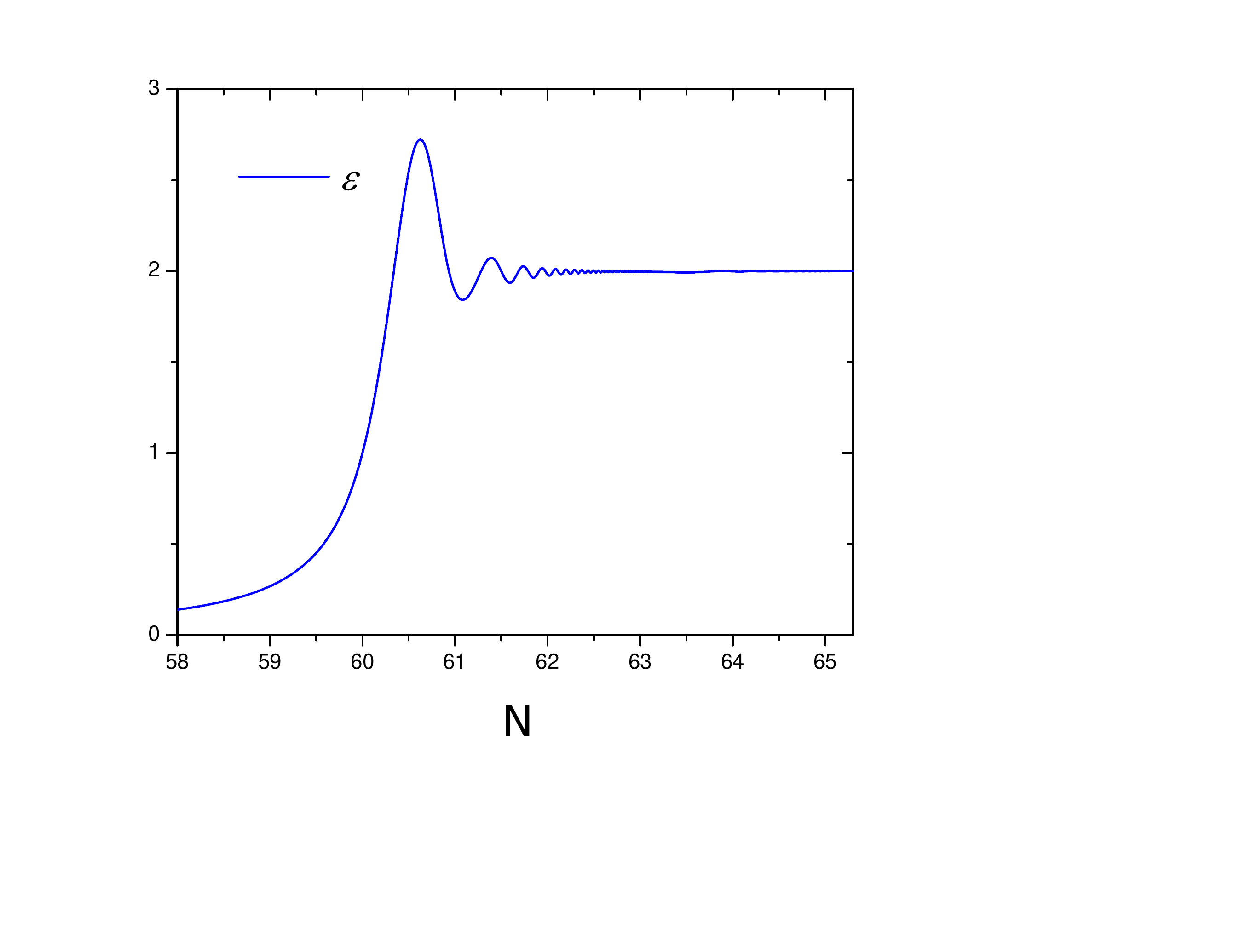}
    \vspace{-2 cm}
    \caption{\label{fig:fig7a}}
  \end{subfigure}
    \hspace{-1 cm}
  \begin{subfigure}[b]{0.5\linewidth}
    \centering\includegraphics[width=300pt]{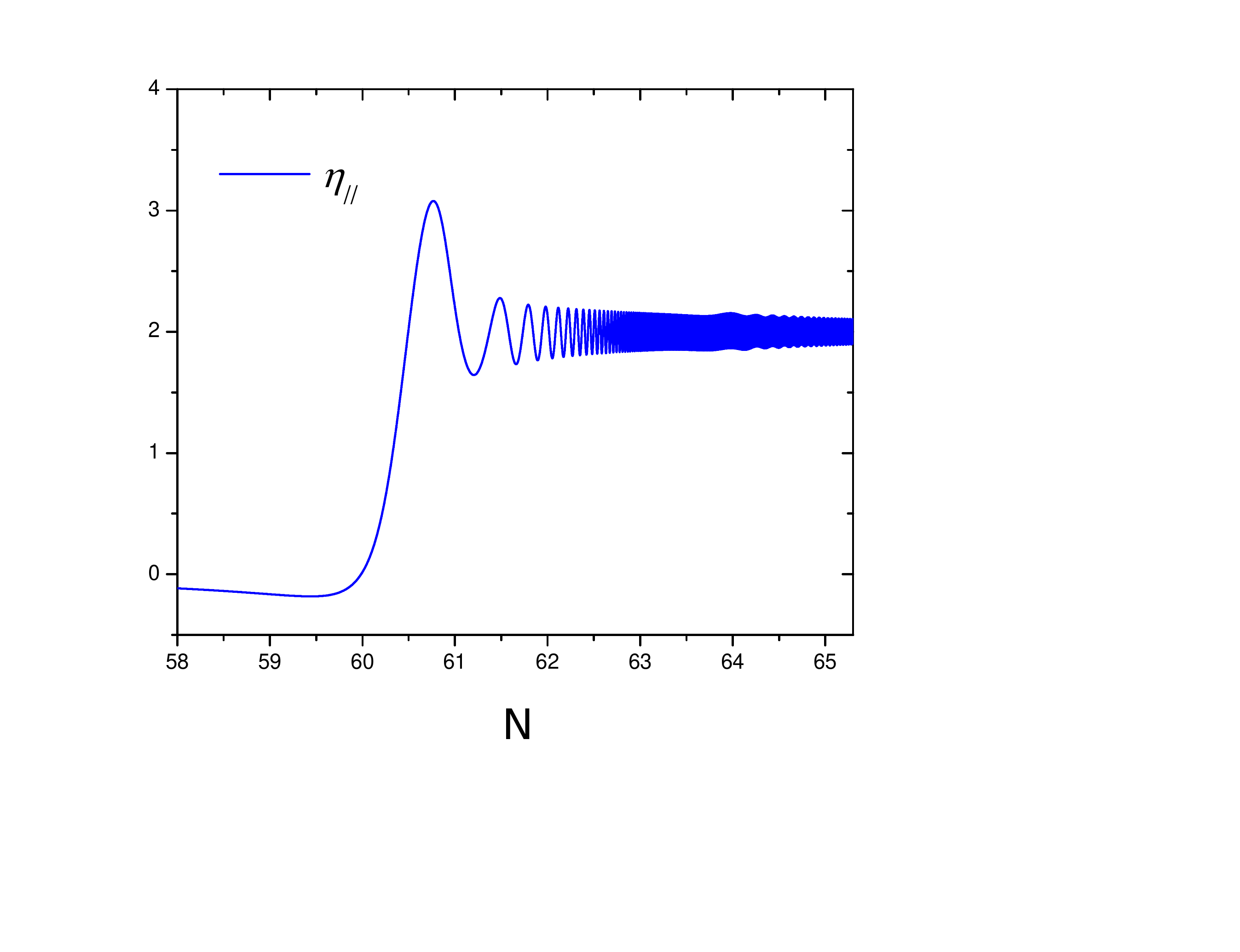}
     \vspace{-2 cm}
    \caption{\label{fig:fig7b}}
  \end{subfigure}
  \begin{subfigure}[b]{0.5\linewidth}
    \centering\includegraphics[width=300pt]{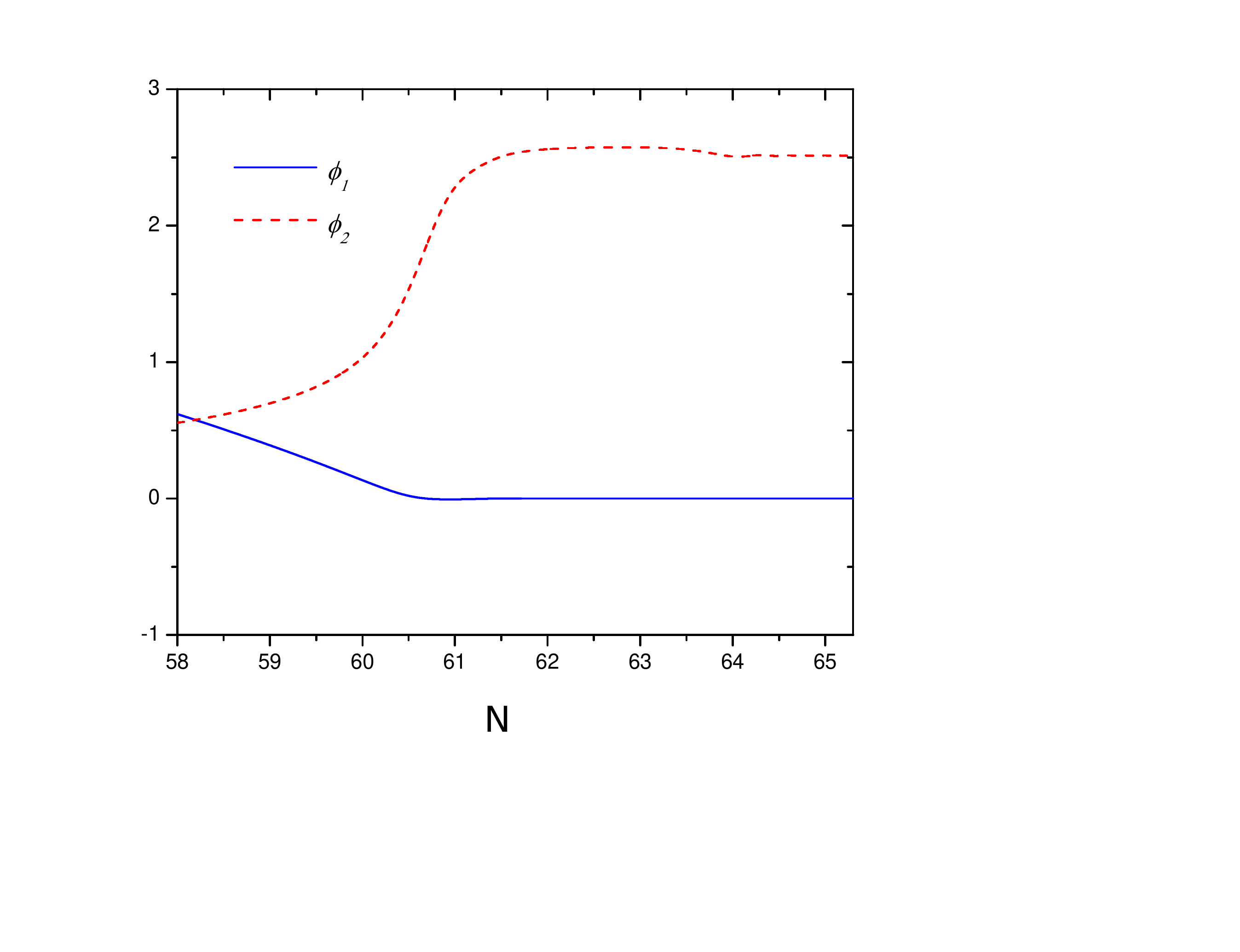}
    \vspace{-2 cm}
    \caption{\label{fig:fig7c}}
  \end{subfigure}
    \hspace{-1 cm}
  \begin{subfigure}[b]{0.5\linewidth}
    \centering\includegraphics[width=300pt]{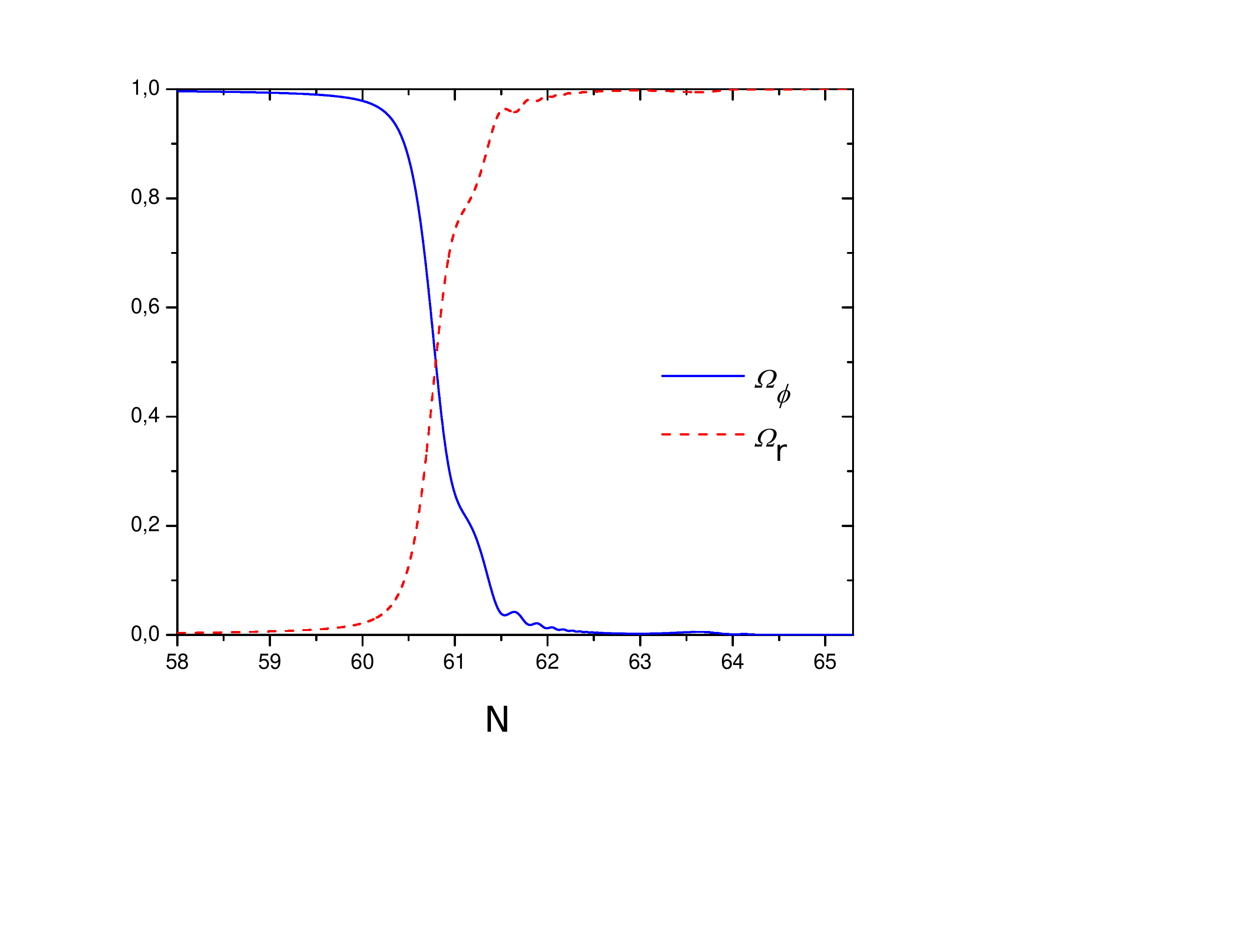}
     \vspace{-2 cm}
    \caption{\label{fig:fig7d}}
  \end{subfigure}
  \caption{Transition from inflation to the radiation dominated era, through reheating, at background level for the parameter choice (ii). In particular, Figures~\subref{fig:fig7a} and~\subref{fig:fig7b} show the evolution from the last stage of inflation to the radiation dominated epoch for the slow-roll parameters $\varepsilon$ and $\eta_{\parallel}$, respectively, which are plotted against the number of $e$-folds from the end of inflation. The fields $\phi_1$ (solid blue line) and $\phi_2$ (dashed red line) as functions of the number of $e$-folds from the last $e$-folds of inflation to the end of reheating is depicted in Figure \subref{fig:fig7c}. Figure \subref{fig:fig7d} shows the evolution of fractional contributions of the energy density from the inflaton fields (solid blue line) and radiation (dashed red line).}
  \label{Fig7PG}
\end{figure}

\afterpage{\clearpage}

In order to see the effects of modifying $\tilde{\phi}_0$ on the dynamics, the second case to be studied corresponds to $q_1=3.28$ and $\tilde{\phi}_0=\phi_{10}/90$, for which the  background evolution from inflation to the radiation-dominated epoch is displayed in Figure \ref{Fig7PG}. In particular, Figures \ref{fig:fig7a} and \ref{fig:fig7b} show the evolution of the slow-roll parameters $\varepsilon$ and $\eta_{\parallel}$ from the last $e$-folds of inflation up the radiation-dominated epoch, respectively. For this case, the behavior of both $\varepsilon$ and $\eta_{\parallel}$ is smoother than in the previous case, without any feature. As we can see, like the previous case $\phi_1$ decays before than $\phi_2$ does, and so most of the kinetic energy is stored in $\phi_2$. After $\phi_2$ reaches the minimum of its potential, the friction term produces a damping oscillation and finally decays into radiation. Regarding the behavior of the fractional energy densities, the contribution coming from radiation at the end of inflation becomes $\Omega_R\simeq 0.021$, which is smaller than the previous case. When the numerical computation stops, 5.3 $e$-folds after inflation ends, we find that $\Omega_R\simeq 0.9991$, signaling the end of reheating.

\begin{figure}[ht]
  \centering
  \begin{subfigure}[b]{0.5\linewidth}
    \centering\includegraphics[width=300pt]{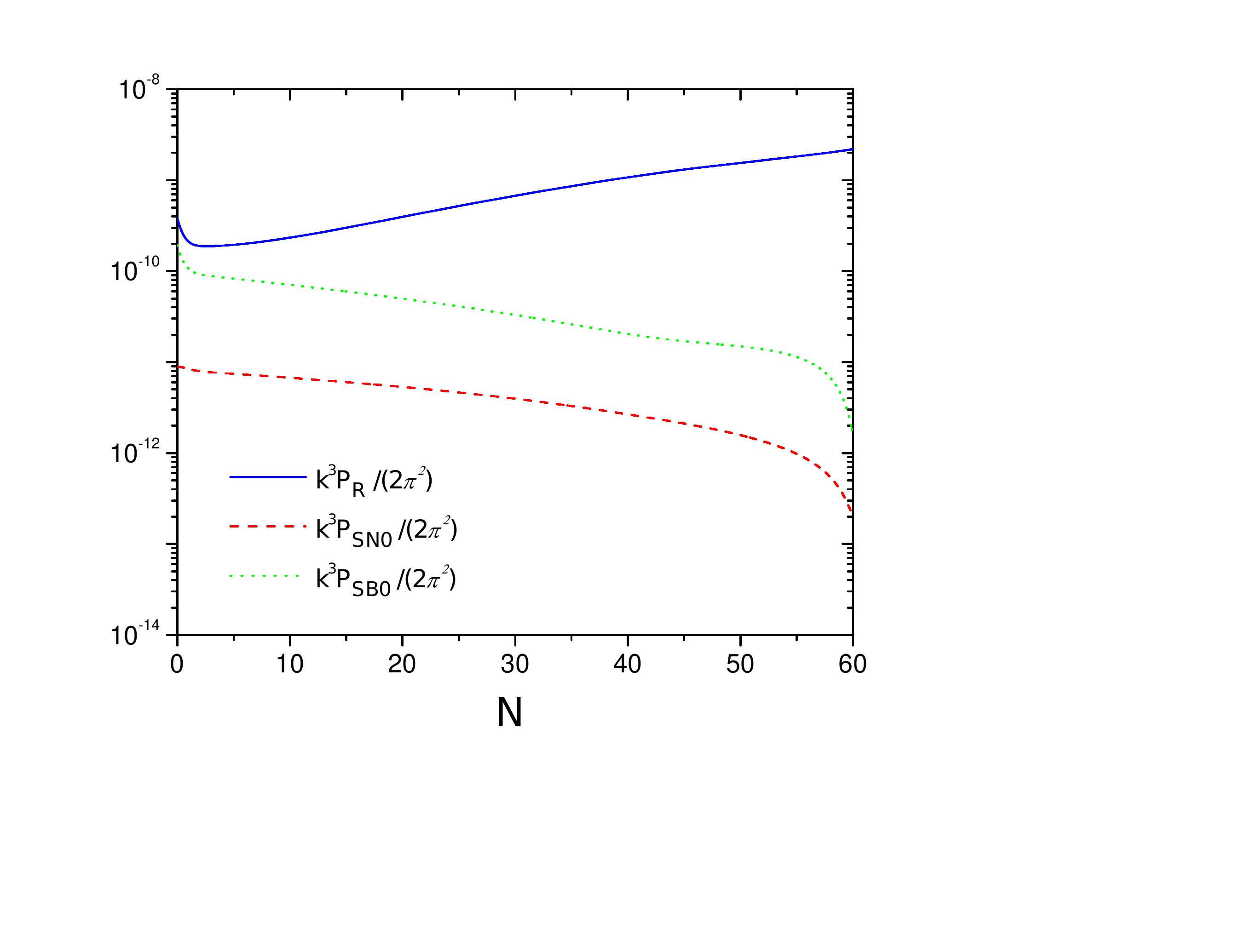}
    \vspace{-2 cm}
    \caption{\label{fig:fig8a}}
  \end{subfigure}
    \hspace{-1 cm}
  \begin{subfigure}[b]{0.5\linewidth}
    \centering\includegraphics[width=300pt]{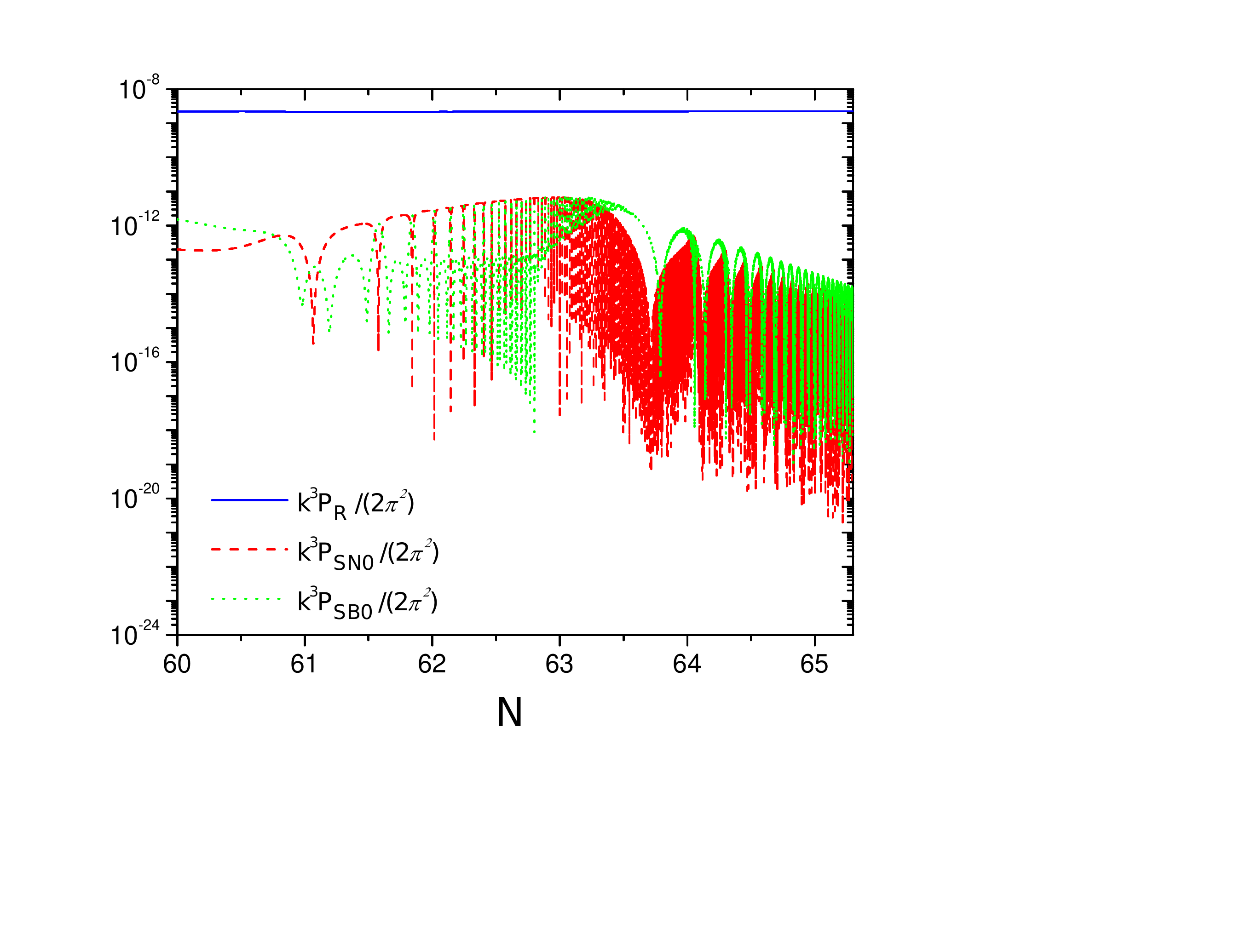}
     \vspace{-2 cm}
    \caption{\label{fig:fig8b}}
  \end{subfigure}
  \begin{subfigure}[b]{0.5\linewidth}
    \centering\includegraphics[width=300pt]{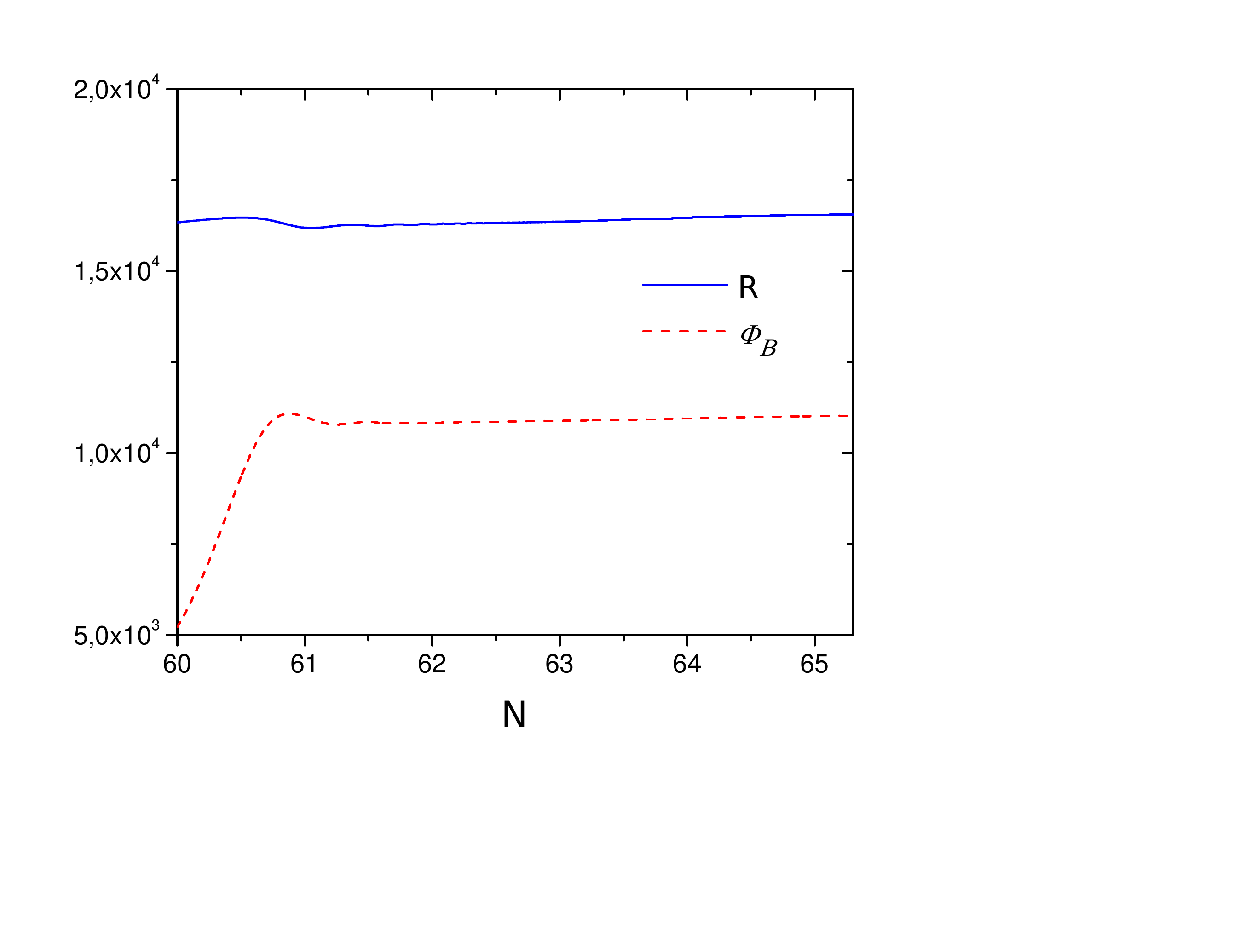}
    \vspace{-2 cm}
    \caption{\label{fig:fig8c}}
  \end{subfigure}
    \hspace{-1 cm}
  \begin{subfigure}[b]{0.5\linewidth}
    \centering\includegraphics[width=300pt]{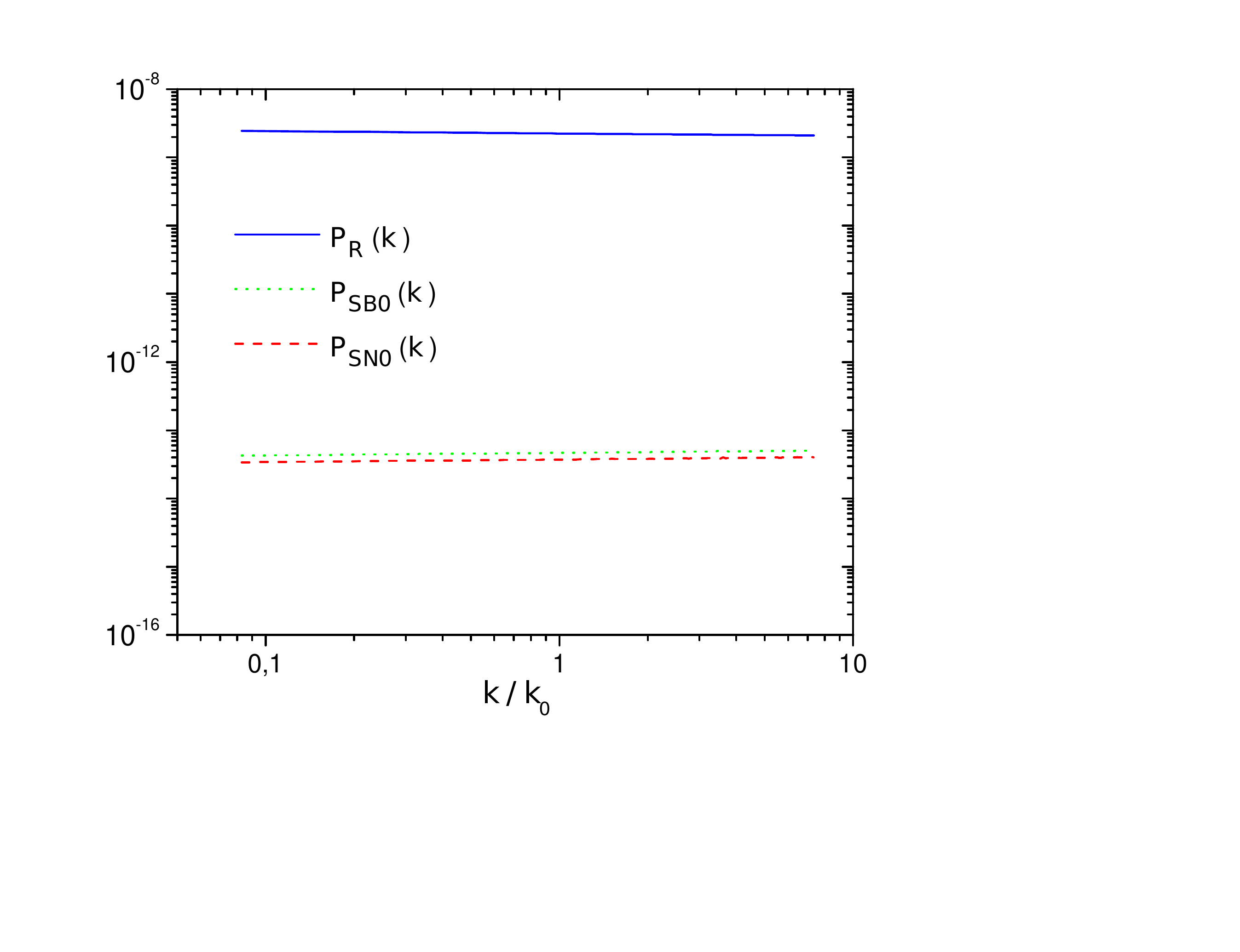}
     \vspace{-2 cm}
    \caption{\label{fig:fig8d}}
  \end{subfigure}
  \caption{Figures~\ref{fig:fig8a} and~\ref{fig:fig8b} show the comparison of the power spectra $\mathcal{P}_{\mathcal{R}}$ (solid blue line), $\mathcal{P}_{\mathcal{S}_{\mathcal{N}0}}$ (dashed red line), and $\mathcal{P}_{\mathcal{S}_{\mathcal{B}0}}$ (dotted green line) at the pivot scale $k_0=0.002$ Mpc$^{-1}$ during inflation and reheating, respectively, for the case (ii). In Figure~\ref{fig:fig8c} we present the evolution of the amplitudes of the Bardeen potential $\Phi_\mathbf{B}$ and total curvature perturbation during reheating, which also has been plotted as a function of the number of $e$-folds $N$. The power spectra $\mathcal{P}_{\mathcal{R}}$ (solid blue line), $\mathcal{P}_{\mathcal{S}_{\mathcal{N}0}}$ (dashed red line), and $\mathcal{P}_{\mathcal{S}_{\mathcal{B}0}}$ (dotted green line) in terms of the ratio $k/k_0$ at the end of reheating are plotted in Figure~\ref{fig:fig8d}.}
  \label{Fig8PG}
\end{figure}

\afterpage{\clearpage}

Figures~\ref{fig:fig8a} and~\ref{fig:fig8b} show the comparison of the power spectra $\mathcal{P}_{\mathcal{R}}$ (solid blue line), $\mathcal{P}_{\mathcal{S}_{\mathcal{N}0}}$ (dashed red line), and $\mathcal{P}_{\mathcal{S}_{\mathcal{B}0}}$ (dotted green line) at the pivot scale $k_0$ during inflation and reheating, respectively. From Figure \ref{fig:fig8a}, we observe that the behavior of the power spectra during inflation is practically indistinguishable from the previous case. However, the crucial difference between both cases comes from the tilt of the curvature power spectrum and the post inflationary evolution of $\mathcal{R}$. Regarding the value of the scalar spectral index at the the end of inflation, we found that $n_{\mathcal{R}}= 0.966$, which is closer than that obtained in \cite{Achucarro:2016fby}. This value for the tilt of the curvature power spectrum at the end of inflation can be achieved by decreasing the value of $\phi_1$ at which the potential changes its concavity. Figure \ref{fig:fig8c} shows the post-inflationary evolution of the curvature perturbation $\mathcal{R}$ and the Bardeen potential, $\Phi_{\mathbf{B}}$, which are plotted against the number of $e$-folds. The curvature perturbation is still evolving during the first 2 $e$-folds after inflation ends, and then becomes constant. A similar behavior is observed in $\Phi_{\mathbf{B}}$, which presents an enhancement during the first $e$-fold after the end of inflation, and then becomes constant until the end of reheating like as $\mathcal{R}$. In particular, at the end of reheating, these satisfy $\Phi_{\mathbf{B}}\simeq 0.666 \mathcal{R}$, being closer to the conditions which holds for radiation-dominated universe in comparison to previous case.

Back to Figure \ref{fig:fig8b}, during the initial phases of reheating, the isocurvature $\mathcal{S}_{\mathcal{N}0}$ is enhanced displaying a structure of spikes, but always being smaller than the curvature perturbation. At around 3 $e$-folds after inflation ends, it can be noticed that the isocurvature perturbation $\mathcal{S}_{\mathcal{N}0}$ reaches a maximum value and, shortly after, it begins to be suppressed and by the end of reheating, $\mathcal{P}_{\mathcal{S}_{\mathcal{N}0}}$ becomes five orders of magnitude smaller than $\mathcal{P}_{\mathcal{R}}$, whose value is $2.23\times 10^{-9}$.

The comparison of the scale dependence of the several power spectra at the end of reheating is displayed in Figure \ref{fig:fig8d}, for scales within the range $0.1\leq k/k_0 \leq 10$. For the curvature power spectra (solid blue line), the scalar spectral index now becomes $n_{\mathcal{R}}\simeq 0.965$, which corresponds to maximum likelihood value from Planck 2015. For all the scales within the range $0.1\leq k/k_0 \leq 10$, the power spectrum of $\mathcal{S}_{\mathcal{N}0}$ is about 5 orders of magnitude smaller than the curvature power spectra.

\begin{figure}[ht]
  \centering
  \begin{subfigure}[b]{0.5\linewidth}
    \centering\includegraphics[width=300pt]{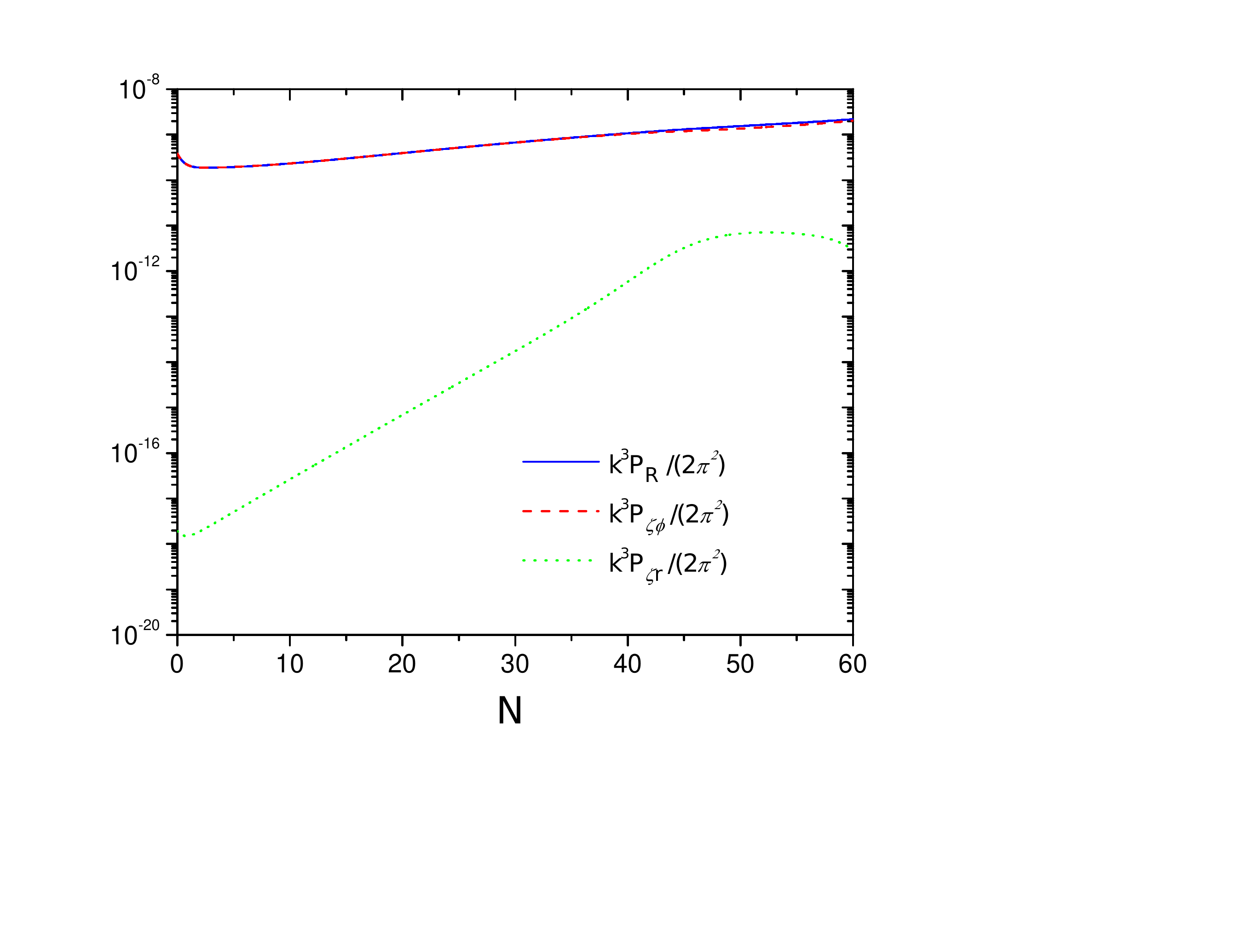}
    \vspace{-2 cm}
    \caption{\label{fig:fig9a}}
  \end{subfigure}
  \begin{subfigure}[b]{0.5\linewidth}
    \centering\includegraphics[width=300pt]{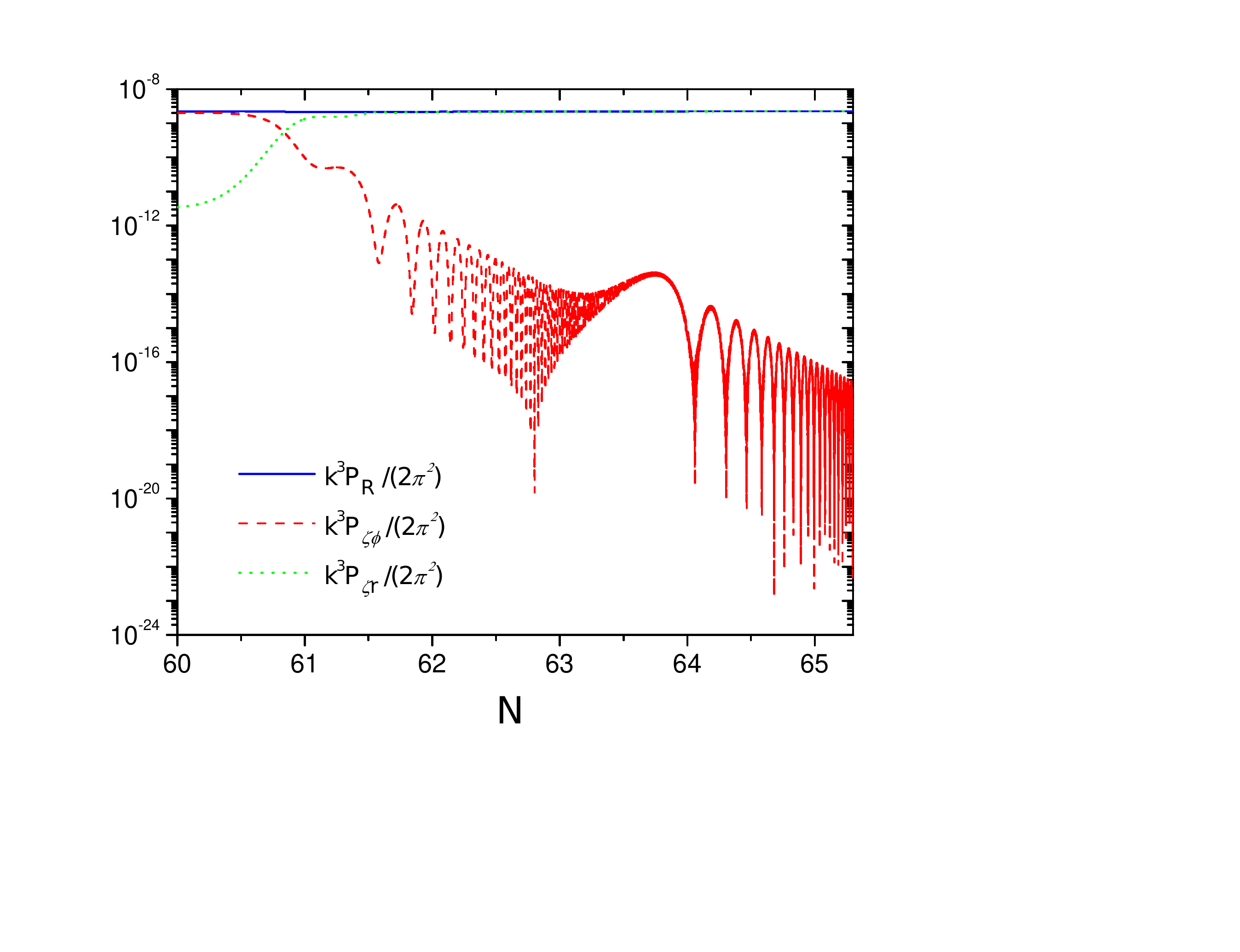}
     \vspace{-2 cm}
    \caption{\label{fig:fig9b}}
  \end{subfigure}
  \caption{Figures~\ref{fig:fig9a} and~\ref{fig:fig9b} show the evolution of curvature perturbations (total as solid blue line, dashed red and dotted green for the fields and fluid
 contribution, respectively) during inflation and reheating as a function of the number of $e$-folds, respectively, for the case (ii).}
  \label{Fig9PG}
\end{figure}

By splitting the total curvature perturbation $\mathcal{R}$ as $\mathcal{R}=\zeta_{\phi}+\zeta_{R}$, Figure \ref{fig:fig9a} shows that, during inflation, the evolution of $\mathcal{R}$, $\zeta_{\phi}$, and $\zeta_R$ becomes indistinguishable from the case (i). However, from Figure \ref{fig:fig9b}, it can be noticed that $\zeta_{R}$ becomes the main contribution to curvature perturbation around one $e$-fold after inflation ends, before than case (i) (see Figure \ref{fig:fig6b}). In a similar way, the individual curvature of the inflaton fields $\zeta_{\phi}$ is rapidly suppressed although $\phi_2$ has not decayed totally in radiation.

For this case we conclude that, for a value of $\tilde{\phi}_0$ closer to zero, the curvature perturbation does not evolve during the initial stages of reheating and the isocurvature mode becomes suppressed, achieving the adiabatic limit before the previous case. Additionally, the initial conditions for the perturbations during the radiation-dominated phase are also set at the end of reheating phase. Moreover, the scale dependence of the curvature power spectrum becomes compatible with maximum likelihood from current Planck data, since its value is smaller than in the previous case.\\

\subsubsection*{iii) $q_1=3.33$ and $\tilde{\phi}_0=\phi_{10}/50$:}

\begin{figure}[ht]
  \centering
  \begin{subfigure}[b]{0.5\linewidth}
    \centering\includegraphics[width=300pt]{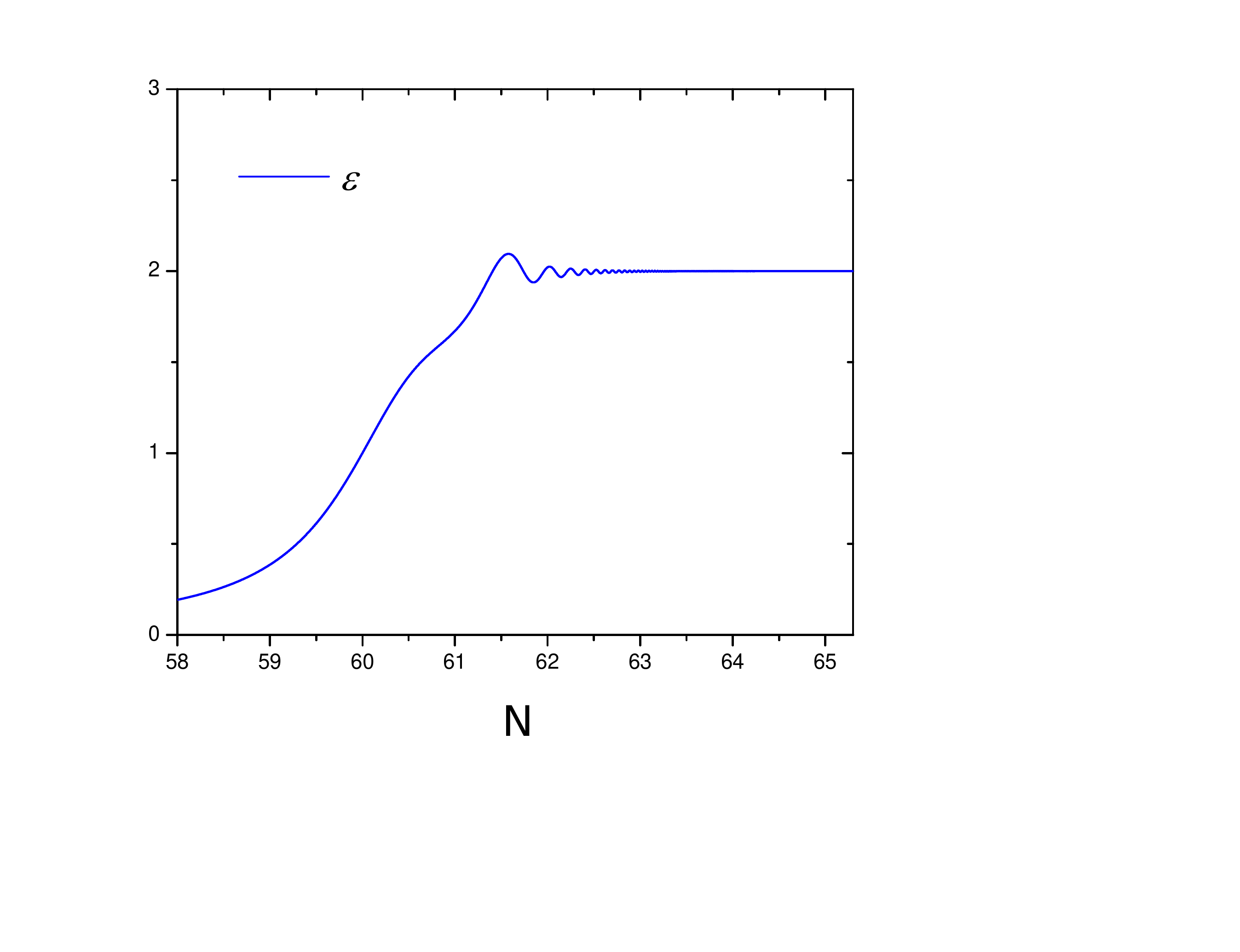}
    \vspace{-2 cm}
    \caption{\label{fig:fig10a}}
  \end{subfigure}
    \hspace{-1 cm}
  \begin{subfigure}[b]{0.5\linewidth}
    \centering\includegraphics[width=300pt]{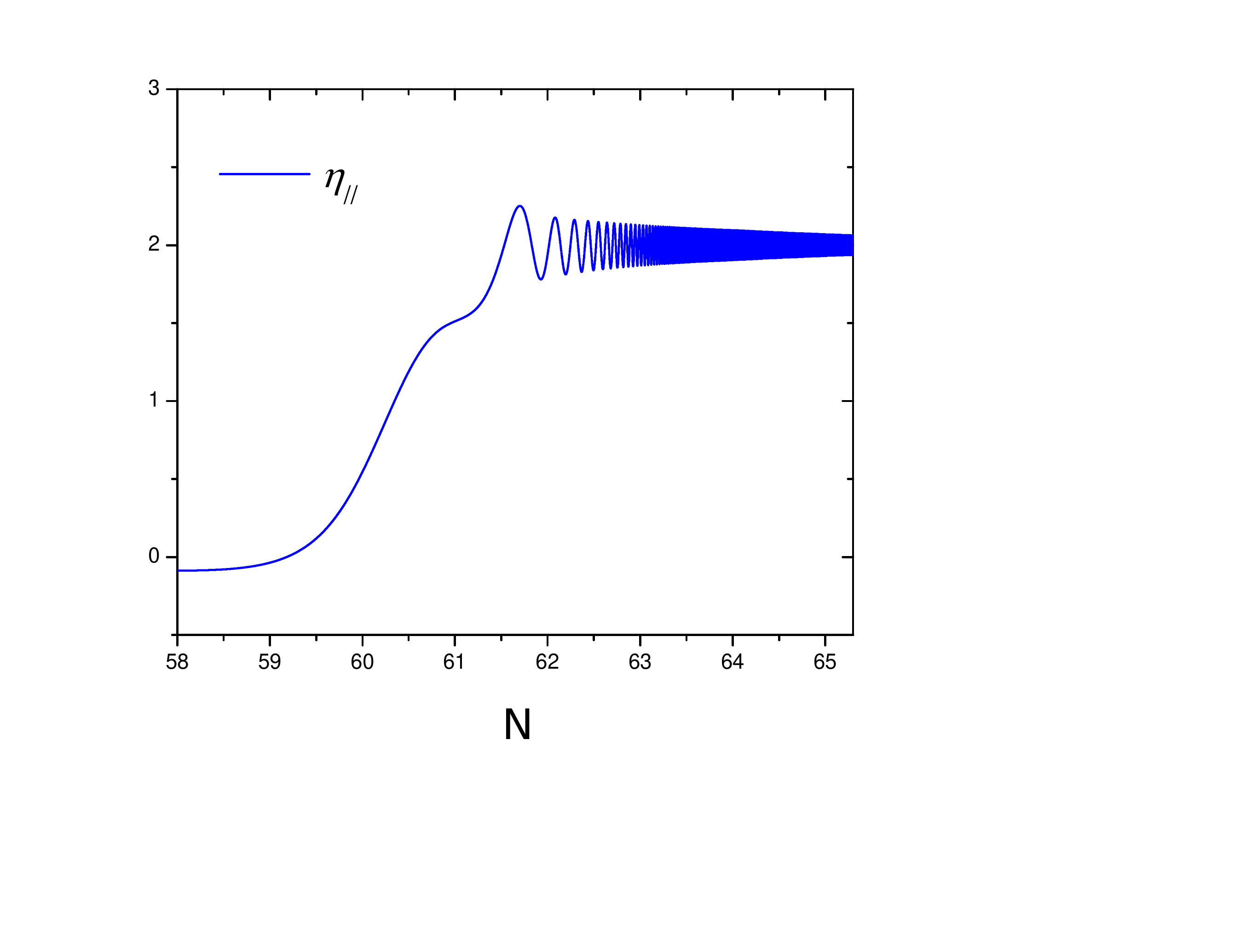}
     \vspace{-2 cm}
    \caption{\label{fig:fig10b}}
  \end{subfigure}
  \begin{subfigure}[b]{0.5\linewidth}
    \centering\includegraphics[width=300pt]{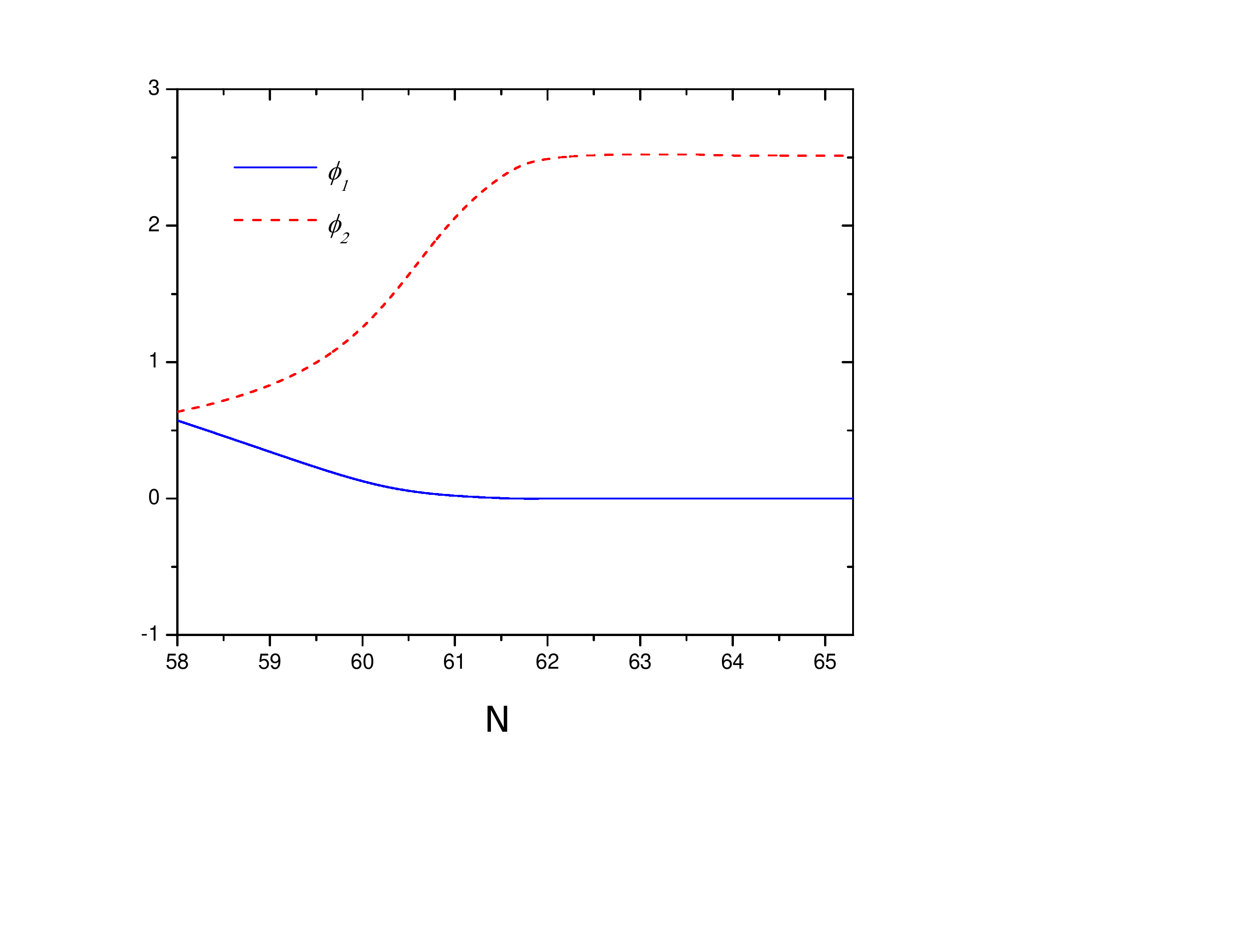}
    \vspace{-2 cm}
    \caption{\label{fig:fig10c}}
  \end{subfigure}
    \hspace{-1 cm}
  \begin{subfigure}[b]{0.5\linewidth}
    \centering\includegraphics[width=300pt]{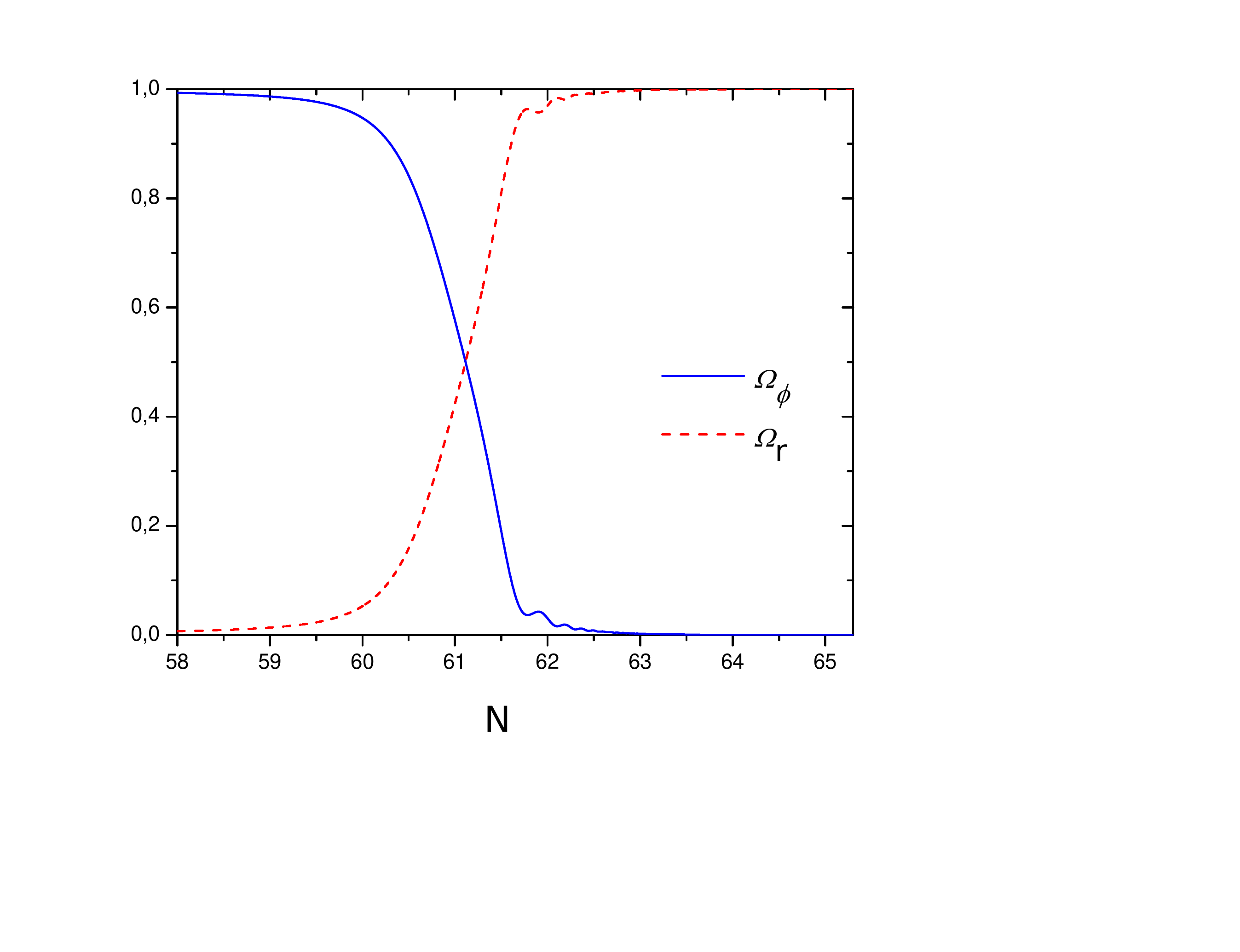}
     \vspace{-2 cm}
    \caption{\label{fig:fig10d}}
  \end{subfigure}
  \caption{Transition from inflation to the radiation dominated era, through reheating, at background level for the parameter choice (ii). Figures~\subref{fig:fig10a} and~\subref{fig:fig10b} show the evolution from the last stage of inflation to the radiation dominated epoch for the slow-roll parameters $\varepsilon$ and $\eta_{\parallel}$, respectively, which are plotted against the number of $e$-folds from the end of inflation. The fields $\phi_1$ (solid blue line) and $\phi_2$ (dashed red line) as function of the number of $e$-folds from the last $e$-folds of inflation to the end of reheating are depicted in Figure~\subref{fig:fig10c}. Figure~\subref{fig:fig10d} shows the evolution of the fractional contributions of the energy density from the inflaton fields (solid blue line) and radiation (dashed red line).}
  \label{Fig10PG}
\end{figure}

In order to see whether by increasing $q_1$ one modifies the results in comparison to the case (i), we set $q_1=3.33$ and $\tilde{\phi}_0=\phi_{10}/50$. As it can bee seen from Figure \ref{Fig10PG}, the transition from the last stage of inflation up to the radiation-dominated epoch at background level is smoother than in case (i). The reason for this behavior is that, in the present case, there is a stronger dissipation than in case (ii). Regarding the evolution of the fractional energy densities, the contribution coming from radiation at the end of inflation becomes $\Omega_R\simeq 0.052$, which is larger than the contribution of radiation compared to case (i), since $q_1=3.3$ leads to a stronger dissipation compared to $q_1=3.28$. For this case, the numerical computation stops around $5.3$ $e$-folds after inflation ends, and $\Omega_R\simeq 0.999$, signaling the end of reheating.

\afterpage{\clearpage}

\begin{figure}[ht]
  \centering
  \begin{subfigure}[b]{0.5\linewidth}
    \centering\includegraphics[width=300pt]{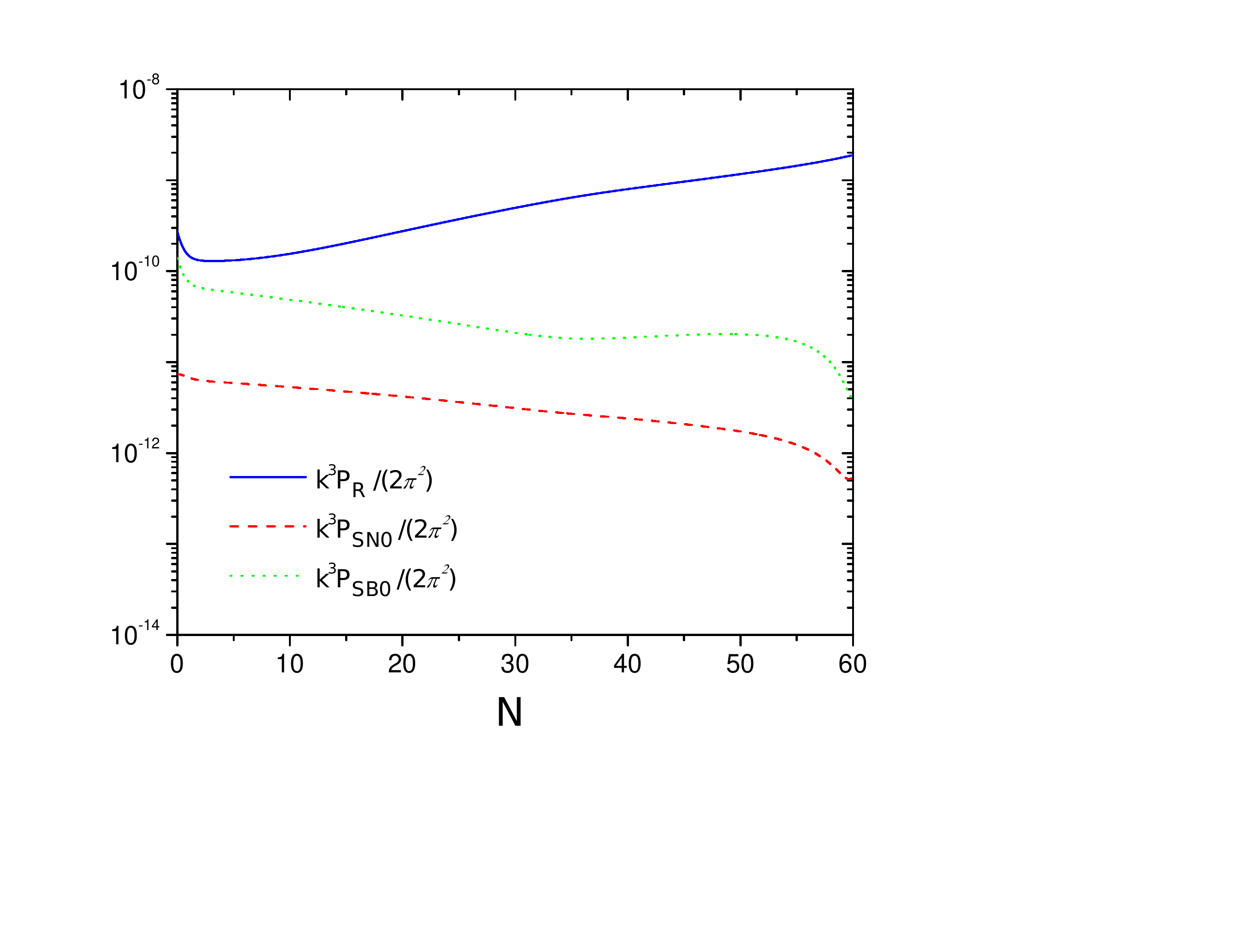}
    \vspace{-2 cm}
    \caption{\label{fig:fig11a}}
  \end{subfigure}
    \hspace{-1 cm}
  \begin{subfigure}[b]{0.5\linewidth}
    \centering\includegraphics[width=300pt]{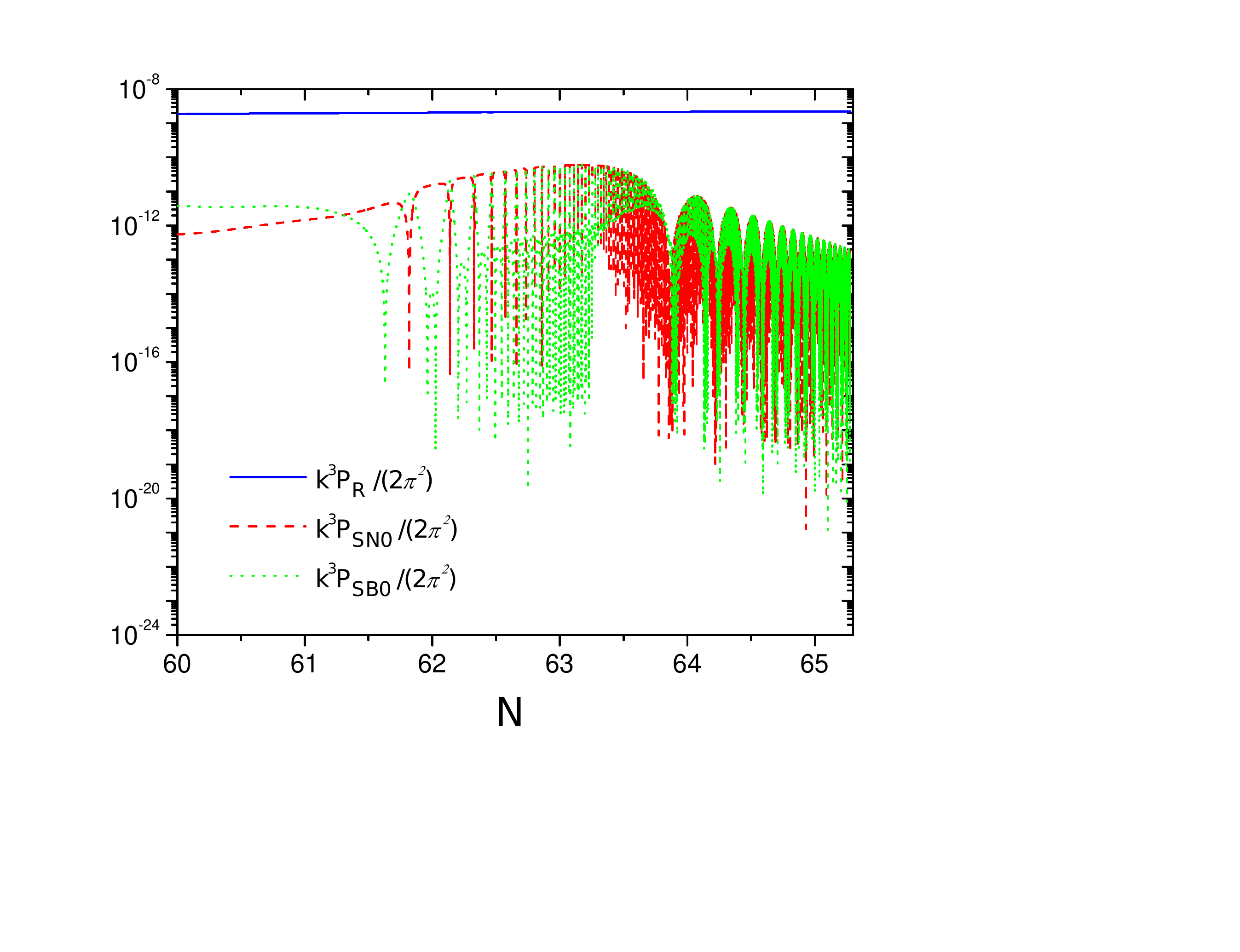}
     \vspace{-2 cm}
    \caption{\label{fig:fig11b}}
  \end{subfigure}
  \begin{subfigure}[b]{0.5\linewidth}
    \centering\includegraphics[width=300pt]{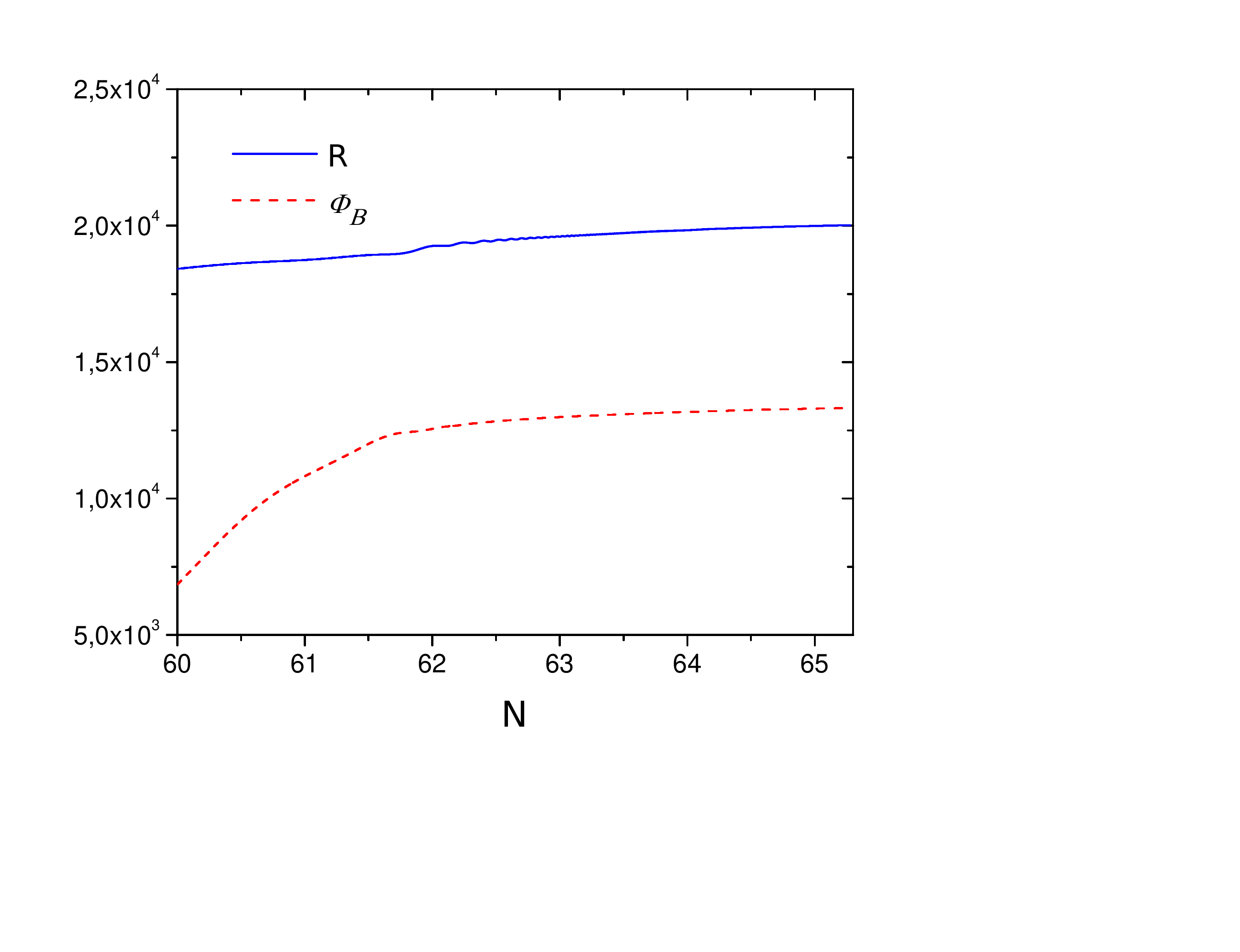}
    \vspace{-2 cm}
    \caption{\label{fig:fig11c}}
  \end{subfigure}
    \hspace{-1 cm}
  \begin{subfigure}[b]{0.5\linewidth}
    \centering\includegraphics[width=300pt]{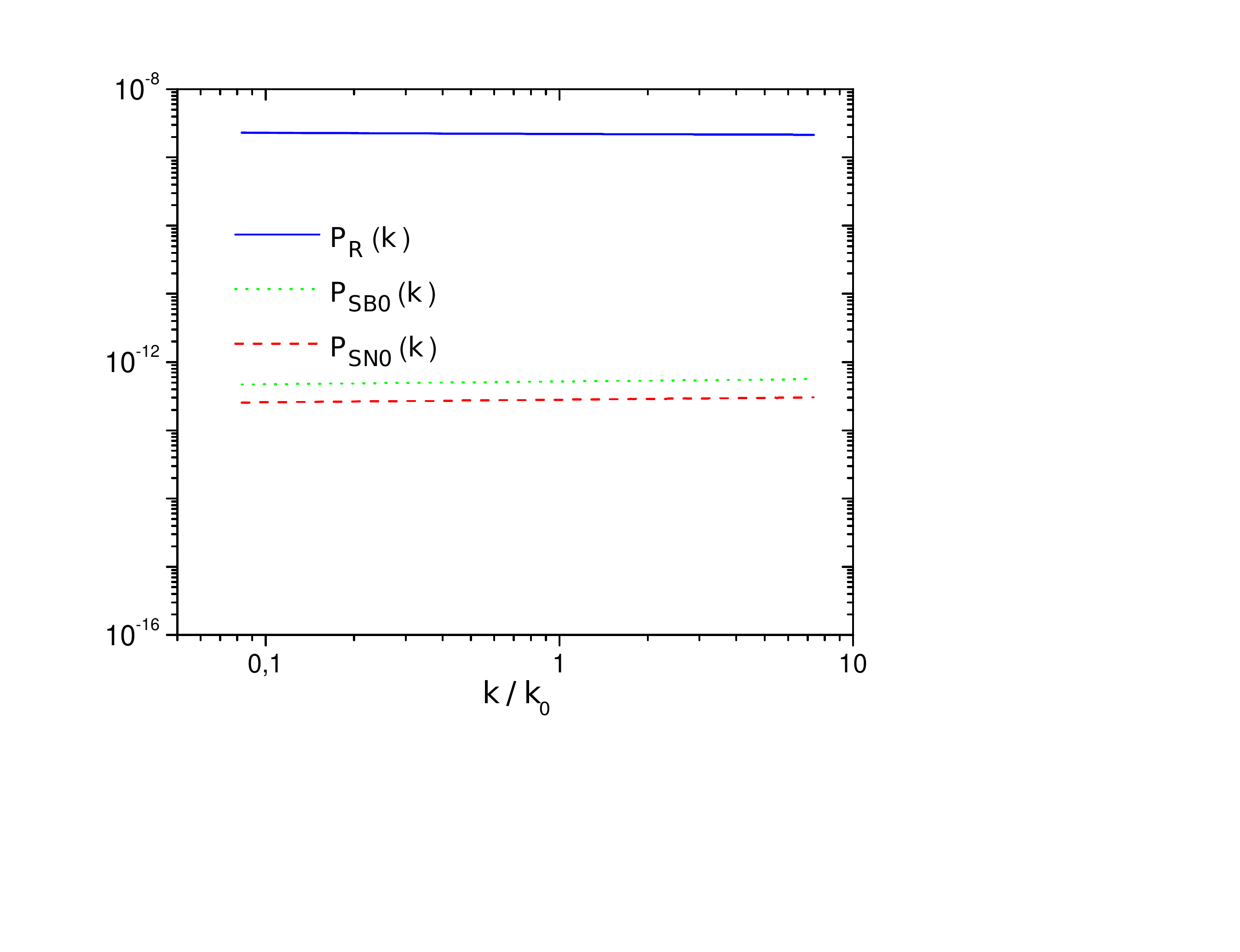}
     \vspace{-2 cm}
    \caption{\label{fig:fig11d}}
  \end{subfigure}
  \caption{Figures~\ref{fig:fig11a} and~\ref{fig:fig11b} show the comparison of the power spectra $\mathcal{P}_{\mathcal{R}}$ (solid blue line), $\mathcal{P}_{\mathcal{S}_{\mathcal{N}0}}$ (dashed red line), and $\mathcal{P}_{\mathcal{S}_{\mathcal{B}0}}$ (dotted green line) at the pivot scale $k_0=0.002$ Mpc$^{-1}$ during inflation and reheating, respectively, for the case (ii). In Figure~\ref{fig:fig11c} we present the evolution of the amplitudes of the Bardeen potential $\Phi_\mathbf{B}$ and the total curvature perturbation during reheating, which also has been plotted as a function of the number of $e$-folds $N$. The power spectra $\mathcal{P}_{\mathcal{R}}$ (solid blue line), $\mathcal{P}_{\mathcal{S}_{\mathcal{N}0}}$ (dashed red line), and $\mathcal{P}_{\mathcal{S}_{\mathcal{B}0}}$ (dotted green line) in terms of the ratio $k/k_0$ at the end of reheating are plotted in Figure \ref{fig:fig11d}.}
  \label{Fig11PG}
\end{figure}

From Figure \ref{fig:fig11a}, we observe that the behavior of the several power spectra during inflation is similar to that of case (i), being of the same order at the end of inflation. However, the tilt of the curvature power spectrum at the end of inflation becomes $n_{\mathcal{R}}=0.979$, which is smaller than that of case (i). Then, an enhancement of $q_1$ produces a lowering on the value of the scalar spectral index at the end of inflation. As it can be noticed from Figure \ref{fig:fig11c}, the curvature perturbation is stiil evolving during
first 2 $e$-folds of reheating, and then becomes constant up to the end
of reheating. This behavior is due the non-vanishing of the isocurvature mode, which after reaching a maximum value, displays a structure of spikes of decreasing amplitude, being suppressed until reheating ends, as it can be noticed from Figure \ref{fig:fig11b}. By the end of reheating, the curvature power spectrum becomes $2.2 \times 10^{-9}$, in agreement with Planck normalization of $\mathcal{P}_{\mathcal{R}}$, being grater by four orders of magnitude than the isocurvature perturbation $\mathcal{S}_{\mathcal{N}0}$. Moreover, at that time the ratio between $\mathcal{R}$ and $\Phi_{\mathbf{B}}$ becomes 0.665.

Figure \ref{fig:fig11d} displays the comparison of the scale dependence of the power spectrum at the end of reheating, for the range $0.1\leq k/k_0 \leq 10$. For the curvature power spectrum (solid blue line), the scalar spectral index now becomes $n_{\mathcal{R}}= 0.984$, which also deviates from the maximum likelihood value from Planck 2015 and is slower than in the previous cases. However, it is interesting to mention that this value becomes smaller than that obtained in case (i) at the end of reheating, which confirms the fact that, by increasing $q_1$, it produces a smaller tilt of the curvature power spectrum.

\afterpage{\clearpage}

Finally, regarding the evolution of the individual contributions of the total curvature coming from $\zeta_{\phi}$ and $\zeta_R$, we find that during and after inflation, we found that their behavior becomes indistinguishable than in the previous cases (figure not shown).

\begin{table}
  \centering
  \begin{tabular}{lllllll}
  \hline
  Model & $n_{\mathcal{R}}(t_{\textrm{end inf}})$ & $r(t_{\textrm{end inf}})$ & $\alpha_{\mathcal{R}}(t_{\textrm{end inf}})$ & $n_{\mathcal{R}}(t_{\textrm{end rh}})$ & $r(t_{\textrm{end rh}})$ & $\alpha_{\mathcal{R}}(t_{\textrm{end rh}})$ \\
  \hline
     PE   & 0.974 &  2.113$\times 10^{-2}$  & -4.272$\times 10^{-4}$ & 0.973 &  2.141$\times 10^{-2}$   & 2.116$\times 10^{-3}$ \\
  ULF (i)   & 0.974 & 6.548$\times 10^{-4}$&  1.680$\times 10^{-3}$ & 0.988 & 4.524$\times 10^{-4}$ & 2.040$\times 10^{-3}$ \\
  ULF (ii)  & 0.964 & 6.853$\times 10^{-4}$&  1.637$\times 10^{-3}$ & 0.965 & 6.717$\times 10^{-4}$ & 1.678$\times 10^{-3}$ \\
  ULF (iii) & 0.978 & 6.147$\times 10^{-4}$ & 1.622$\times 10^{-3}$ & 0.984& 5.243$\times 10^{-4}$& 1.769$\times 10^{-3}$ \\
  \hline
\end{tabular}
  \caption{In this table we summarize the values for the scalar spectral index ($n_{\mathcal{R}}$), the tensor-to-scalar ratio ($r$), and the running of the scalar spectral index ($\alpha_{\mathcal{R}}$), at the end of inflation ($t_{\textrm{end inf}}$) and reheating ($t_{\textrm{end rh}}$), respectively.}
  \label{Table1}
\end{table}

\afterpage{\clearpage}

Table \ref{Table1} summarizes the corresponding values of the scalar spectral index $n_{\mathcal{R}}$ at the end of inflation as well as at the end of reheating for the PE model and the three cases corresponding to the ULF model. In addition, we also have included the values of the tensor-to-scalar ratio $r$ and the running of the scalar spectral index $\alpha_{\mathcal{R}}$ both at the end of inflation and reheating.

\section{Conclusions} \label{Sec:conlusions}

We have studied the evolution of perturbations in two-field models of inflation in which the scalar fields remain coupled to a radiation fluid. We have payed special emphasis on those cases where the isocurvature mode remain nearly massless all the way from horizon crossing up to the reheating phase. To do so, we have introduced a formalism that allows one to treat the radiation fluid as an effective scalar field. This scalar field can be treated as the third partner of a three-field model, and so one can treat the perturbation system covariantly, in terms of a three dimensional target space.

By itself, the third scalar field $\phi^{(3)}$ has no physical meaning. On the other hand, the energy density $\rho_R$ introduced in Eq.~(\ref{Rad Cond}) does have a well defined meaning. This is in part behind the power of the method introduced here: One may deal with the full scalar field system, including the third fictitious scalar, as if it was a closed system. Another advantage of the present approach is that it allows us to deal with the initial conditions for the perturbations in a simple way, by imposing the Bunch-Davies conditions for the three scalar perturbations un sub-horizon scales.

We have checked well known results in the literature, and have examined how some observable quantities, such as the spectral index of the power spectrum of primordial curvature perturbations may change due to the coupling between the scalar fields and the thermal bath.  Such a coupling enables us to follow the post inflationary evolution, into reheating, for multi-field models having a non-vanishing amount of isocurvature when inflation ends. For the models examined here, we found that a non-vanishing coupling between the scalar fields and a radiation fluid (parametrized by $q_1$) forces the isocurvature modes to decay rapidly as the universe reheats. In general, when the energy stored in the scalar fields is completely transferred  to the radiation fluid we make contact with the radiation-dominated epoch of the hot big-bang scenario, and the power spectra of the isocurvature becomes completely negligible in comparison to the power spectrum of curvature. A more careful analysis of this process should consider the details of how the isocurvature mode decays into different species.

\subsection*{Acknowledgements}

The work of P.G. has been financed by CONICYT Programa de Postdoctorado FONDECYT $N^o$ 3150398 and Fondecyt Regular N$^o$ 1150390. G.A.P. acknowledges support from the Fondecyt Regular project N$^o$ 1171811. N.V. was supported by CONICYT Programa de Postdoctorado FONDECYT N$^o$ 3150490 and Fondecyt de Iniciaci\'on N$^o$ 1170162.

\begin{appendix}

\setcounter{equation}{0}
\renewcommand{\theequation}{\Alph{section}.\arabic{equation}}

\section{Fields Space.} \label{app:FieldsSpace}

In a field space with a metric $\mathbf{q}_{AB}$, we know that $(\mathcal{T}^A,\mathcal{N}^A,\mathcal{B}^A)$ make a complete orthogonal basis and:
\begin{eqnarray}
\mathbf{q}_{AB} = \mathcal{T}_A\mathcal{T}_B + \mathcal{N}_A\mathcal{N}_B + \mathcal{B}_A\mathcal{B}_B
\end{eqnarray}
and:
\begin{eqnarray}
\frac{D\mathbf{q}_{AB}}{dt} = 0 &\rightarrow& \frac{D\mathcal{T}^A}{dt}\mathcal{T}^B + \mathcal{T}^A\frac{D\mathcal{T}^B}{dt} + \frac{D\mathcal{N}^A}{dt}\mathcal{N}^B + \mathcal{N}^A\frac{D\mathcal{N}^B}{dt} + \frac{D\mathcal{B}^A}{dt}\mathcal{B}^B + \mathcal{B}^A\frac{D\mathcal{B}^B}{dt} = 0 \nonumber \\
&\rightarrow& - H\eta_{\perp}\mathcal{N}^A\mathcal{T}^B - H\eta_{\perp}\mathcal{T}^A\mathcal{N}^B + \frac{D\mathcal{N}^A}{dt}\mathcal{N}^B \nonumber \\
&& + \mathcal{N}^A\frac{D\mathcal{N}^B}{dt} + \frac{D\mathcal{B}^A}{dt}\mathcal{B}^B + \mathcal{B}^A\frac{D\mathcal{B}^B}{dt} = 0.
\end{eqnarray}
Then:
\begin{eqnarray}
\mathcal{T}^B\frac{D\mathbf{q}_{AB}}{dt} = 0 &\rightarrow& - H\eta_{\perp}\mathcal{N}^A + \mathcal{N}^A\mathcal{T}_B\frac{D\mathcal{N}^B}{dt} + \mathcal{B}^A\mathcal{T}_B\frac{D\mathcal{B}^B}{dt} = 0, \\[10pt]
\mathcal{N}^B\frac{D\mathbf{q}_{AB}}{dt} = 0 &\rightarrow& - H\eta_{\perp}\mathcal{T}^A + \frac{D\mathcal{N}^A}{dt} + \mathcal{N}^A\mathcal{N}_B\frac{D\mathcal{N}^B}{dt} + \mathcal{B}^A\mathcal{N}_B\frac{D\mathcal{B}^B}{dt} = 0, \\[10pt]
\mathcal{B}^B\frac{D\mathbf{q}_{AB}}{dt} = 0 &\rightarrow& \mathcal{N}^A\mathcal{B}_B\frac{D\mathcal{N}^B}{dt} + \frac{D\mathcal{B}^A}{dt} + \mathcal{B}^A\mathcal{B}_B\frac{D\mathcal{B}^B}{dt} = 0,
\end{eqnarray}
where we used Eq.~(\ref{bg eq 2}). Additionally, from the normalization of $\mathcal{N}^A$ and $\mathcal{B}^A$, we have $\frac{D\mathcal{N}^B}{dt}\mathcal{N}_B = \frac{D\mathcal{B}^B}{dt}\mathcal{B}_B = 0$. Then, defining $HC = \frac{D\mathcal{B}^B}{dt}\mathcal{N}_B = - \frac{D\mathcal{N}^B}{dt}\mathcal{B}_B$, we obtain from these equations:
\begin{eqnarray}
\frac{D\mathcal{N}^A}{dt} &=& H\eta_{\perp}\mathcal{T}^A - HC\mathcal{B}^A, \\
\frac{D\mathcal{B}^A}{dt} &=& HC\mathcal{N}^A.
\end{eqnarray}
They correspond to Eqs.~(\ref{bg eq 3}) and (\ref{bg eq 4}) respectively. These vectors contain 9 parameters, but they are constrained by $\mathcal{T}^A\mathcal{T}_A = \mathcal{N}^A\mathcal{N}_A = \mathcal{B}^A\mathcal{B}_A = 1$ and $\mathcal{T}^A\mathcal{N}_A = \mathcal{T}^A\mathcal{B}_A = \mathcal{N}^A\mathcal{B}_A = 0$. So, just three degrees of freedom will survive. During inflation, radiation must be negligible, so $\mathcal{T}^{(3)} = \frac{\phi_0^{(3)}}{\phi_0} \rightarrow 0$, connecting with the two fields case, where $\mathcal{T}^a\mathcal{T}_a = \mathcal{N}^a\mathcal{N}_a = 1$ and $\mathcal{T}^a\mathcal{N}_a = 0$, which means that $\mathcal{N}^{(3)} = 0$. Additionally, from $\mathcal{T}^a\mathcal{B}_a = \mathcal{N}^a\mathcal{B}_a = 0$, we obtain $\mathcal{B}^{(1)} = \mathcal{B}^{(2)} = 0$. Therefore, the normalization is reduced to $\mathcal{B}^A\mathcal{B}_A = \mathbf{q}_{(33)}\left(\mathcal{B}^{(3)}\right)^2 = 1$. To summarize, the initial conditions are given by:
\begin{eqnarray}
\label{Ini Cond}
\mathcal{B}^A =
\left(
  \begin{array}{c}
                 0              \\
                 0              \\
  \pm \frac{1}{\sqrt{\mathbf{q}_{(33)}}} \\
  \end{array}
\right).
\end{eqnarray}
On the other side, we can define another basis $(\mathcal{T}^A,\mathcal{N}_0^A,\mathcal{B}_0^A)$, given by:
\begin{eqnarray}
\mathcal{N}_0^A &\equiv& \cos(\beta)\mathcal{N}^A+\sin(\beta)\mathcal{B}^A, \\
\mathcal{B}_0^A &\equiv& -\sin(\beta)\mathcal{N}^A+\cos(\beta)\mathcal{B}^A,
\end{eqnarray}
where the equations of motion are:
\begin{eqnarray}
\label{EQ base0-1}
\frac{D\mathcal{T}^A}{dt} &=& - H\eta_{\perp}\left(\cos(\beta)\mathcal{N}_0^A - \sin(\beta)\mathcal{B}_0^A\right), \\
\label{EQ base0-2}
\frac{D\mathcal{N}_0^A}{dt} &=& H\eta_{\perp}\cos(\beta)\mathcal{T}^A + \left(\dot{\beta}-HC\right)\mathcal{B}_0^A,  \\
\label{EQ base0-3}
\frac{D\mathcal{B}_0^A}{dt} &=& - H\eta_{\perp}\sin(\beta)\mathcal{T}^A - \left(\dot{\beta}-HC\right)\mathcal{B}_0^A.
\end{eqnarray}
In this paper, we will use both bases to study the perturbative components (See Figure \ref{Fig:Appen B}). Also, we will study cases where $\mathbf{q}_{AB}$ is given by Eq.~ (\ref{Metric q}). So, we can parameterize our vectors as:
\begin{eqnarray}
\label{Parameterize}
\mathcal{T}^{(1)} &=& \frac{1}{\sqrt{\mathbf{q}_{(11)}}}\cos(\theta)\cos(\alpha), \nonumber \\
\mathcal{T}^{(2)} &=& - \frac{1}{\sqrt{\mathbf{q}_{(22)}}}\sin(\theta)\cos(\alpha), \nonumber \\
\mathcal{T}^{(3)} &=& \frac{1}{\sqrt{\mathbf{q}_{(33)}}}\sin(\alpha), \nonumber \\
\mathcal{N}^{(1)} &=& \frac{1}{\sqrt{\mathbf{q}_{(11)}}}\left(\sin(\theta)\cos(\beta) + \cos(\theta)\sin(\alpha)\sin(\beta)\right), \nonumber \\
\mathcal{N}^{(2)} &=& \frac{1}{\sqrt{\mathbf{q}_{(22)}}}\left(\cos(\theta)\cos(\beta) - \sin(\theta)\sin(\alpha)\sin(\beta)\right), \nonumber \\
\mathcal{N}^{(3)} &=& - \frac{1}{\sqrt{\mathbf{q}_{(33)}}}\cos(\alpha)\sin(\beta), \nonumber \\
\mathcal{B}^{(1)} &=& \frac{1}{\sqrt{\mathbf{q}_{(11)}}}\left(\sin(\theta)\sin(\beta) - \cos(\theta)\sin(\alpha)\cos(\beta)\right), \nonumber \\
\mathcal{B}^{(2)} &=& \frac{1}{\sqrt{\mathbf{q}_{(22)}}}\left(\cos(\theta)\sin(\beta) + \sin(\theta)\sin(\alpha)\cos(\beta)\right), \nonumber \\
\mathcal{B}^{(3)} &=& \frac{1}{\sqrt{\mathbf{q}_{(33)}}}\cos(\alpha)\cos(\beta)
\end{eqnarray}
and:
\begin{eqnarray}
\label{Parameterize0}
\mathcal{N}_0^{(1)} = \frac{1}{\sqrt{\mathbf{q}_{(11)}}}\sin(\theta) &\textrm{, }& \mathcal{B}_0^{(1)} = - \frac{1}{\sqrt{\mathbf{q}_{(11)}}}\cos(\theta)\sin(\alpha), \nonumber \\
\mathcal{N}_0^{(2)} = \frac{1}{\sqrt{\mathbf{q}_{(22)}}}\cos(\theta) &\textrm{, }& \mathcal{B}_0^{(2)} = \frac{1}{\sqrt{\mathbf{q}_{(22)}}}\sin(\theta)\sin(\alpha), \nonumber \\
\mathcal{N}_0^{(3)} = 0 &\textrm{, }& \mathcal{B}_0^{(3)} = \frac{1}{\sqrt{\mathbf{q}_{(33)}}}\cos(\alpha).
\end{eqnarray}
Applying Eq.~(\ref{Ini Cond}) in (\ref{Parameterize}), we note that the initial conditions are given by $\sin(\alpha) = \sin(\beta) = 0$. Actually, $\alpha$ defines the magnitude of radiation density as $\sin(\alpha) = \sqrt{\frac{2\Omega_R}{\varepsilon}}$, where we used Eq.~(\ref{Rad Cond}), $\rho_R^0 = 3H^2\Omega_R$ and $\dot{\phi}_0^2 = 2H^2\varepsilon$. Then, $\alpha$ goes from $0$, in the beginning of Inflation, to $\frac{\pi}{2}$ in the Radiation era. On the other side, $\beta$ is the angle to define the normal plane, such as $\mathcal{V}_\mathcal{B} + \mathcal{J}_\mathcal{B} = 0$ (See Figure \ref{Fig:Appen B}). This means:
\begin{figure}[ht]
\centering
\vspace{-2 cm}
\includegraphics[width=250pt]{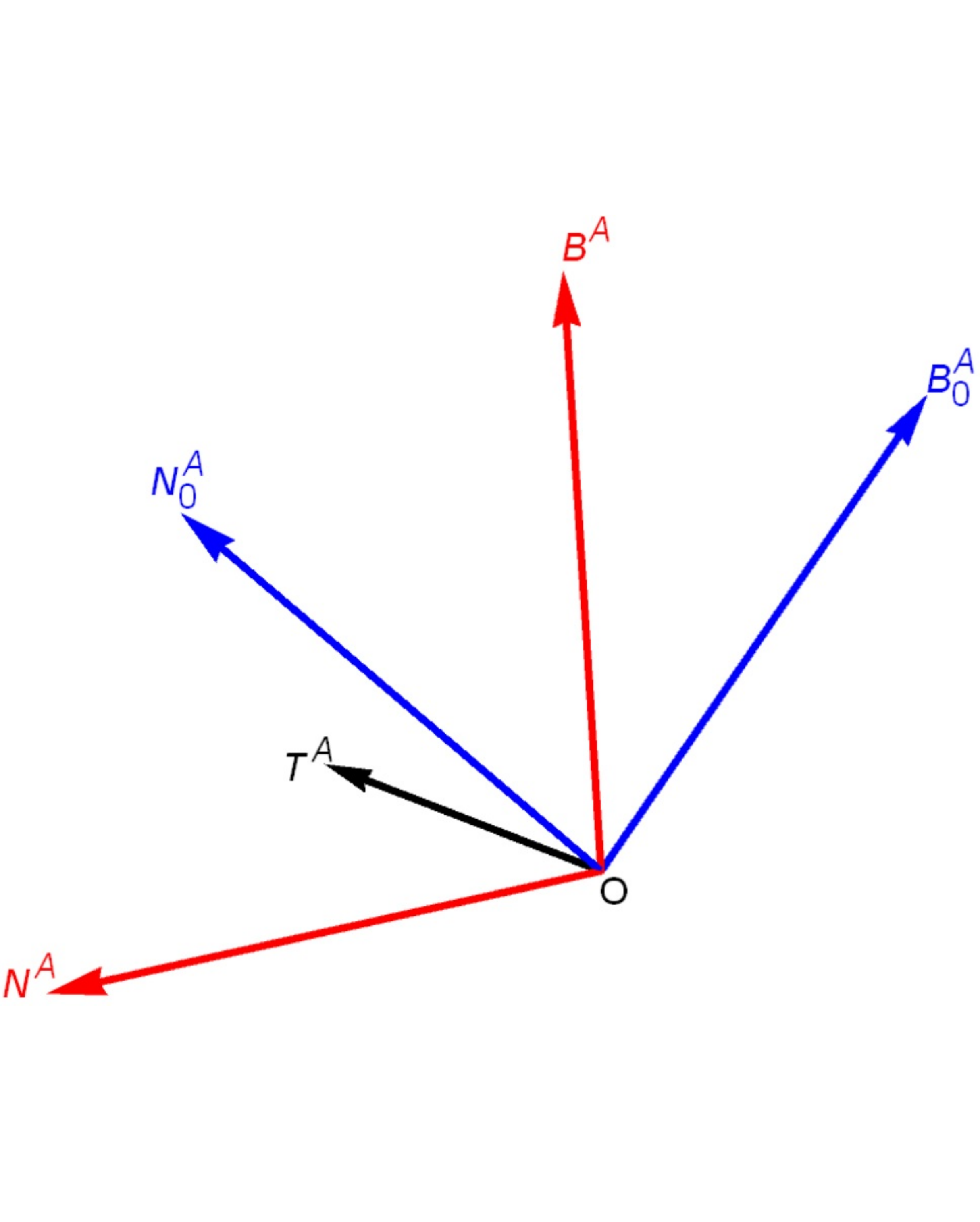}
\vspace{-1.5 cm}
\caption{Representation of the basis vectors in field space. In Black, the tangent vector $\mathcal{T}^A$. In Red, the normal vectors in the standard basis ($\mathcal{N}^A$,$\mathcal{B}^A$), given by Eq.~(\ref{Parameterize}). In Blue, the normal vectors in the \textit{zero} basis ($\mathcal{N}_0^A$,$\mathcal{B}_0^A$), given by Eq.~ (\ref{Parameterize0}). The $\beta$ angle corresponds to $\measuredangle$ $\mathcal{N}^A O \mathcal{N}_0^A$. This angle does not have a physical meaning; it is used to relate different bases. In the standard basis, $\beta$ is fixed in order to obtain $\mathcal{V}_\mathcal{B} + \mathcal{J}_\mathcal{B} = 0$.}
\label{Fig:Appen B}
\end{figure}
\begin{eqnarray}
&& \tan(\beta) = \nonumber \\
&& \frac{\sin(\alpha)\left(g\left(\theta\right) + \frac{1}{H^2\sqrt{2\varepsilon}}\left(\frac{J^0_{(1)}}{\sqrt{\mathbf{q}_{(11)}}}\cos(\theta) - \frac{J^0_{(2)}}{\sqrt{\mathbf{q}_{(22)}}}\sin(\theta)\right) + \left(1+\frac{Q}{3H^3\varepsilon\sin^2(\alpha)}\right)\cos(\alpha)\right)}{\left(f\left(\theta\right) + \frac{1}{H^2\sqrt{2\varepsilon}}\left(\frac{J^0_{(1)}}{\sqrt{\mathbf{q}_{(11)}}}\sin(\theta) + \frac{J^0_{(2)}}{\sqrt{\mathbf{q}_{(22)}}}\cos(\theta)\right)\right)},
\end{eqnarray}
with $Q = J^0_a\dot{\phi}_0^a$ and:
\begin{eqnarray}
\label{fyg1}
f\left(\theta\right) &=& \frac{1}{H^2\sqrt{2\varepsilon}}\left(\frac{V_{(1)}}{\sqrt{\mathbf{q}_{(11)}}}\sin(\theta) + \frac{V_{(2)}}{\sqrt{\mathbf{q}_{(22)}}}\cos(\theta)\right), \\
\label{fyg2}
g\left(\theta\right) &=& \frac{1}{H^2\sqrt{2\varepsilon}}\left(\frac{V_{(1)}}{\sqrt{\mathbf{q}_{(11)}}}\cos(\theta) - \frac{V_{(2)}}{\sqrt{\mathbf{q}_{(22)}}}\sin(\theta)\right).
\end{eqnarray}
Finally, from Eq.~(\ref{JA}) in Appendix \ref{app:Interaction}, we obtain that the $\beta$ angle is given by:
\begin{eqnarray}
\tan(\beta) = \frac{\sin(\alpha)\left(g\left(\theta\right) + \left(1+q_1\left(1-\frac{\cos^2(\alpha)}{3}\right)\right)\cos(\alpha)\right)}{f\left(\theta\right) + q_2},
\end{eqnarray}
where $Q = 2H^3\varepsilon q_1\sin^2(\alpha)\cos^2(\alpha)$.

\setcounter{equation}{0}
\renewcommand{\theequation}{\Alph{section}.\arabic{equation}}

\section{Interaction Analysis.} \label{app:Interaction}

In this Appendix, we will do a brief analysis about interactions, showing how we obtain the final expression used in this paper to describe the background and perturbative equations of motion. In Eq.~(\ref{action}), the interaction is given by $\mathcal{L}_{\textup{int}}$ and, in our prescription where we use a third scalar field to represent the radiation, it depend on $\phi^A$. Particulary, we expect that the interaction lagrangian is a function of $\phi^A$ and $\partial_{\mu}\phi^A$. Then, the interaction in a background level is given by:
\begin{eqnarray}
\label{JA bg}
\mathcal{J}^0_A &=& - \frac{\bar{\delta} \mathcal{L}_{\textup{int}}(\phi_0,\dot{\phi}_0)}{\bar{\delta} \phi_0^A} + \frac{D}{dt}\left(\frac{\delta \mathcal{L}_{\textup{int}}(\phi_0,\dot{\phi}_0)}{\delta \dot{\phi}_0^A}\right),
\end{eqnarray}
where $\frac{\bar{\delta}}{\bar{\delta} \phi_0^A}$ represents a covariant variation on the field space and $\frac{D}{dt}$ is the covariant time derivative. On the other side, we have that:
\begin{eqnarray}
\label{int test}
\dot{\mathcal{L}}_{\textup{int}}(\phi_0,\dot{\phi}_0) &=& \dot{\phi}_0^A\left(\frac{\bar{\delta} \mathcal{L}_{\textup{int}}}{\bar{\delta} \phi_0^A}\right) + \frac{D}{dt}\left(\dot{\phi}_0^A\right)\left(\frac{\delta \mathcal{L}_{\textup{int}}}{\delta \dot{\phi}_0^A}\right) = - \dot{\phi}_0^A\mathcal{J}^0_A + \frac{d}{dt}\left(\dot{\phi}_0^A\left(\frac{\delta \mathcal{L}_{\textup{int}}}{\delta \dot{\phi}_0^A}\right)\right) \nonumber \\
&=& \frac{d}{dt}\left(\dot{\phi}_0^A\left(\frac{\delta \mathcal{L}_{\textup{int}}}{\delta \dot{\phi}_0^A}\right)\right), \nonumber \\
\label{bgLint0}
&\rightarrow& \mathcal{L}_{\textup{int}}(\phi_0,\dot{\phi}_0) = \dot{\phi}_0^A\left(\frac{\delta \mathcal{L}_{\textup{int}}}{\delta \dot{\phi}_0^A}\right),
\end{eqnarray}
where we used that $\dot{\phi}_0^A = \dot{\phi}_0\mathcal{T}^A$ and $\mathcal{J}_\mathcal{T} \equiv \mathcal{T}^A\mathcal{J}^0_A = 0$. A quick analysis of the last expression says us $\mathcal{L}_{\textup{int}}$ is a first order expression on $\dot{\phi}^A$ in a background level. Actually in general, $\mathcal{L}_{\textup{int}}$ must be linear on $\partial_{\mu}\phi^A$ to obtain until a first order contribution in the equations of motion.

We can see that the Friedman equations (\ref{H2})-(\ref{dotH}) do not depend on the interactions. That is because they take into account the complete system, producing $\mathcal{J}_\mathcal{T} = 0$. This fact is related to (\ref{bgLint0}). Therefore, the interaction lagrangian and the source can be written in a background level as:
\begin{eqnarray}
\label{bgLint}
\mathcal{L}_{\textup{int}}(\phi_0,\dot{\phi}_0) &=& \lambda^0_A(\phi_0)\dot{\phi}_0^A, \\
\label{JA0}
\mathcal{J}^0_A &=& \left(D_B\lambda^0_A(\phi_0)-D_A\lambda^0_B(\phi_0)\right)\dot{\phi}_0^B,
\end{eqnarray}
where the covariant derivative $D_A$ appears because $\mathcal{J}^0_A$ must be define like a vector in the field space. Now, Eq.~(\ref{JA0}) means $\mathcal{J}_\mathcal{T} = 0$, however a more convenient expression to the interaction source can be defined using Eq.~(\ref{Parameterize0}) from Appendix \ref{app:FieldsSpace}. That is:
\begin{eqnarray}
\label{JA}
\mathcal{J}_0^A &=& H^2\sqrt{2\varepsilon}\left(q_2\mathcal{N}_0^A - q_1\sin(\alpha)\cos(\alpha)\mathcal{B}_0^A\right),
\end{eqnarray}
where $q_1$ and $q_2$ are arbitrary and dimensionless functions to represent different contributions from the interaction. They are related to $\lambda^0_A(\phi_0)$ in a particular but irrelevant way, because the form of $\mathcal{L}_{\textup{int}}$ at this moment is not important. For this reason, we will work with $q_1$ and $q_2$. Besides, they are defined in such a way that $\mathcal{J}_0^A$ is well-behaved for all values of $\alpha$. A third degree of freedom of $\lambda^0_A(\phi_0)$ is apparently irrelevant to the equations of motion. In Appendix \ref{app:FluidComp}, we will see that it is necessary to guarantee the equivalence between the radiation fluid and the additional scalar field. Now, using Eq.~(\ref{JA}), we can see that $Q = 2H^3\varepsilon q_1\sin^2(\alpha)\cos^2(\alpha)$ and (\ref{KG2})-(\ref{rad}) are:
\begin{eqnarray}
\frac{D\phi^{(1)\prime}_0}{dN} + \left(3-\varepsilon\right)\phi^{(1)\prime}_0 + \frac{V_{(1)}}{H^2\mathbf{q}_{(11)}} &=& - \frac{\sqrt{2\varepsilon}}{\sqrt{\mathbf{q}_{(11)}}}\left(q_2\sin(\theta)+q_1\sin^2(\alpha)\cos(\alpha)\cos(\theta)\right), \\
\frac{D\phi^{(2)\prime}_0}{dN} + \left(3-\varepsilon\right)\phi^{(2)\prime}_0 + \frac{V_{(2)}}{H^2\mathbf{q}_{(22)}} &=& - \frac{\sqrt{2\varepsilon}}{\sqrt{\mathbf{q}_{(22)}}}\left(q_2\cos(\theta)-q_1\sin^2(\alpha)\cos(\alpha)\sin(\theta)\right), \\
\Omega_R' + 2\left(2-\varepsilon\right)\Omega_R &=& \frac{2q_1}{3}\varepsilon\sin^2(\alpha)\cos^2(\alpha).
\end{eqnarray}
From these equations, we know that $q_1>0$ is related to decays from inflatons $(\phi_0^{(1)},\phi_0^{(2)})$ to radiation $(\Omega_R)$ and $q_2$ gives us the interaction between $\phi_0^{(1)}$ and $\phi_0^{(2)}$, depending on $\theta$.\\

On the other side, we need an expression to the perturbative interaction. We do not know enough about interactions during reheating in this level, but we can obtain a general expression to understand the behavior of our fields. Previously, we said that $\mathcal{L}_{\textup{int}}$ must be a first order function of $\partial_{\mu}\phi^A$, so the perturbative interaction, $\left(\Delta \mathcal{J}_A\right) = \mathcal{J}_A - \mathcal{J}^0_A$, must be linear too. Additionally, $\mathcal{L}_{\textup{int}}$ can be dependent on the metric, so some perturbative element in Eq.~(\ref{g}) must be taken into account. So, if we use (\ref{dphi}) and (\ref{J1I}) to define the invariant component in a way:
\begin{eqnarray}
\label{Inv Comp 1}
\left(\delta\phi^A\right) &=& \left(\delta\phi^A\right)^I - \frac{\dot{\phi}_0^A}{H}\Psi, \\
\label{Inv Comp 2}
\left(\Delta \mathcal{J}^A\right) &=& \left(\Delta \mathcal{J}^A\right)^I - \frac{1}{H}\frac{D}{dt}\left(\mathcal{J}_0^A\right)\Psi,
\end{eqnarray}
and Eqs.~(\ref{Einst EQ 1}) and (\ref{Einst EQ 4}) to reduce the metric contribution, then the most general expression to the invariant perturbative interaction can be written as:
\begin{eqnarray}
\label{Delta JA}
\left(\Delta \mathcal{J}_A\right)^I &=& \mathcal{I}_{AB}\left(\delta \phi^B\right)^I + \mathcal{J}_{AB}\frac{D}{dt}\left(\delta \phi^B\right)^I + \mathcal{K}_A\Phi_{\mathbf{B}},
\end{eqnarray}
with $\mathcal{I}_{AB}$ and $\mathcal{J}_{AB}$ are respectively symmetric and antisymmetric matrices. In Appendix \ref{app:FluidComp}, some components of this parameters are fixed to guarantee that the radiation fluid can be represented by an additional scalar field. It is proved that the tangent component of Eq.~(\ref{Delta JA}) is given by (\ref{DeltaJTI}). Using Eq.~(\ref{dphiInv}), it is reduced to:
\begin{eqnarray}
\label{DeltaJTI final}
\left(\Delta \mathcal{J}_\mathcal{T}\right)^I &=& \eta_{\perp}\mathcal{J}_\mathcal{N}\mathcal{R} + \dot{\phi}_0\tau_\mathcal{N}\dot{\mathcal{S}}_\mathcal{N} + \left(\frac{\dot{\mathcal{J}}_\mathcal{N}}{H}+C\mathcal{J}_\mathcal{B}+\dot{\phi}_0\left(\dot{\tau}_\mathcal{N}+H\left(3-2\eta_{\parallel}\right)\tau_\mathcal{N}\right)\right)\mathcal{S}_\mathcal{N} \nonumber \\
&& + \dot{\phi}_0\tau_\mathcal{B}\dot{\mathcal{S}}_\mathcal{B} + \left(\frac{\dot{\mathcal{J}}_\mathcal{B}}{H}-C\mathcal{J}_\mathcal{N}+\dot{\phi}_0\left(\dot{\tau}_\mathcal{B}+H\left(3-2\eta_{\parallel}\right)\tau_\mathcal{B}\right)\right)\mathcal{S}_\mathcal{B},
\end{eqnarray}
With this result, we can obtain the other components. They are:
\begin{eqnarray}
\label{DeltaJNI final}
\left(\Delta \mathcal{J}_\mathcal{N}\right)^I &=& - \dot{\phi}_0\tau_\mathcal{N}\dot{\mathcal{R}} + \dot{\phi}_0j_0\dot{\mathcal{S}}_\mathcal{B} + \left(\frac{\dot{\mathcal{J}}_\mathcal{N}}{H}+C\mathcal{J}_\mathcal{B}+\dot{\phi}_0\left(\dot{\tau}_\mathcal{N}+H\left(3-2\varepsilon\right)\tau_\mathcal{N}\right)+H\dot{\phi}_0C\tau_\mathcal{B}\right)\mathcal{R} \nonumber \\
&& + H\dot{\phi}_0\Lambda_{\mathcal{N}\mathcal{N}}\mathcal{S}_\mathcal{N} + H\dot{\phi}_0\left(\Lambda_{\mathcal{N}\mathcal{B}}+\left(\varepsilon-\eta_{\parallel}\right)j_0\right)\mathcal{S}_\mathcal{B} + H\dot{\phi}_0\kappa_\mathcal{N}\Phi_{\mathbf{B}}, \\
\label{DeltaJBI final}
\left(\Delta \mathcal{J}_\mathcal{B}\right)^I &=& - \dot{\phi}_0\tau_\mathcal{B}\dot{\mathcal{R}} - \dot{\phi}_0j_0\dot{\mathcal{S}}_\mathcal{N} \nonumber \\
&& + \left(\frac{\dot{\mathcal{J}}_\mathcal{B}}{H}-C\mathcal{J}_\mathcal{N}+\dot{\phi}_0\left(\dot{\tau}_\mathcal{B}+H\left(3-2\varepsilon\right)\tau_\mathcal{B}\right)+H\dot{\phi}_0\left(\eta_{\perp}j0-C\tau_\mathcal{N}\right)\right)\mathcal{R} \nonumber \\
&& + H\dot{\phi}_0\left(\Lambda_{\mathcal{N}\mathcal{B}}-\left(\varepsilon-\eta_{\parallel}\right)j_0-\eta_{\perp}\tau_\mathcal{B}\right)\mathcal{S}_\mathcal{N} + H\dot{\phi}_0\Lambda_{\mathcal{B}\mathcal{B}}\mathcal{S}_\mathcal{B} + H\dot{\phi}_0\kappa_\mathcal{B}\Phi_{\mathbf{B}},
\end{eqnarray}
where $\tau_\mathcal{N}$, $\tau_\mathcal{B}$ and:
\begin{eqnarray}
&& j_0 = \frac{\mathcal{J}_{\mathcal{N}\mathcal{B}}}{H}, \quad \Lambda_{\mathcal{N}\mathcal{N}} = \frac{I_{\mathcal{N}\mathcal{N}}}{H^2} - \eta_{\perp}\tau_\mathcal{N} - Cj_0, \\
&& \Lambda_{\mathcal{N}\mathcal{B}} = \frac{I_{\mathcal{N}\mathcal{B}}}{H^2}, \quad \Lambda_{\mathcal{B}\mathcal{B}} = \frac{I_{\mathcal{B}\mathcal{B}}}{H^2} - Cj_0, \\
&& \kappa_\mathcal{N} = \frac{\mathcal{K}_\mathcal{N}}{H\dot{\phi}_0}, \quad \kappa_\mathcal{B} = \frac{\mathcal{K}_\mathcal{B}}{H\dot{\phi}_0}
\end{eqnarray}
are the perturbative interaction parameters. These expressions will be used to obtain the perturbative equations in Section \ref{Sec: Perturbations}. From Eq.~(\ref{JA}), we can see that the interaction becomes completely determinate by two parameters in a background level. They are given by $q_1$ and $q_2$, represented by $\mathcal{J}_{\mathcal{N}}$ and $\mathcal{J}_{\mathcal{B}}$ in Eqs.~(\ref{DeltaJTI final})-(\ref{DeltaJBI final}). In the perturbative equations (\ref{KG EQ 1 Final})-(\ref{KG EQ 3 Final}), both parameters do not appear explicitly, so they are not relevant, however we have additional parameters to represent the interaction in a perturbative level.

In first place, we have $\tau_\mathcal{N}$ and $\tau_\mathcal{B}$ as the only not-fixed parameters in Appendix \ref{app:FluidComp} to represent the interaction. This means that they are not independent of the other six parameters. Additionally, they define the perturbative interaction in the tangent component or, in the same way, the effect of the interaction on the total fluid, affecting directly to the curvature $\mathcal{R}$. Then, the other parameters represent the internal effect. On one side, we have $j_0$, $\Lambda_{\mathcal{N}\mathcal{N}}$, $\Lambda_{\mathcal{N}\mathcal{B}}$ and $\Lambda_{\mathcal{B}\mathcal{B}}$ as the interaction parameters related to the isocurvature components. On the other side, we have $\kappa_\mathcal{N}$ and $\kappa_\mathcal{B}$, related to a geometric contribution in the interaction. At the beginning, we do not have information about these parameters, so they are completely arbitrary. For that, we need to know the interaction lagrangian. Each one of them could give us interesting properties, but in this paper we will fix them to zero in order to simplify the calculations. We present more details in Section \ref{Sec: Perturbations}.

\setcounter{equation}{0}
\renewcommand{\theequation}{\Alph{section}.\arabic{equation}}

\section{Fluid Components.} \label{app:FluidComp}

In this Appendix, we will describe the different components of the total fluid, particulary the perturbative contribution, in order to restrict the interaction contribution in our prescription where the radiation component is represented by an additional scalar field. In first place, we will find a general expression of density and pressure, and then fix the parameters. Finally, we will connect these results with $\left(\Delta J^A\right)$ defined in Appendix \ref{app:Interaction}.\\

The Energy-Momentum tensor is given by:
\begin{eqnarray}
T^{\mu \nu} = \frac{2}{\sqrt{-g}}\frac{\delta}{\delta g_{\mu\nu}}\left(\sqrt{-g}\mathcal{L}_M\right) = 2\frac{\delta \mathcal{L}_M}{\delta g_{\mu\nu}} + g^{\mu \nu}\mathcal{L}_M,
\end{eqnarray}
where $\mathcal{L}_M$ is the matter lagrangian. In our case, we have two inflatons and a radiation fluid, given by Eq.~(\ref{action}). However, in this paper, a third scalar field is used to represent the radiation fluid, moreover we introduce the coupling between inflatons and this additional field at the Lagrangian level following the procedure of Refs.\cite{Boehmer:2015sha,Boehmer:2015kta}. This means that the lagrangian and the energy-momentum tensor must be given respectively by:
\begin{eqnarray}
\label{LM EMT}
\mathcal{L}_M &=& -\frac{1}{2}g^{\mu \nu}\hat{\mathbf{q}}_{AB}\partial_{\mu}\phi^A\partial_{\nu}\phi^B-\mathcal{V}+\mathcal{L}_{\textup{int}}, \\
T_{\mu \nu} &=& \hat{\mathbf{q}}_{AB}\partial_{\mu}\phi^A\partial_{\nu}\phi^B + 2\hat{\mathcal{H}}_{\mu \nu} - g_{\mu \nu}\left(\frac{1}{2}g^{\alpha \beta}\hat{\mathbf{q}}_{AB}\partial_{\alpha}\phi^A\partial_{\beta}\phi^B+\mathcal{V}-\mathcal{L}_{\textup{int}}\right),
\end{eqnarray}
where $\hat{\mathbf{q}}_{AB}$ is a non-perturbative version of $\mathbf{q}_{AB}$ and $\hat{\mathcal{H}}^{\mu \nu} = \frac{\delta \mathcal{L}_{\textup{int}}}{\delta g_{\mu\nu}}$. So, using $T_{\mu\nu} = \left(\rho+p\right)u_{\mu}u_{\nu} + pg_{\mu\nu}$, we can deduce the total density, pressure and the 4-velocity from Eq.~(\ref{LM EMT}). They are:
\begin{eqnarray}
\label{tot dens}
\rho &=& \frac{\left(2u^{\alpha}u^{\beta}+g^{\alpha \beta}\right)}{2}\hat{\mathbf{q}}_{AB}\partial_{\alpha}\phi^A\partial_{\beta}\phi^B + \mathcal{V} - \mathcal{L}_{\textup{int}} + 2u^{\alpha}u^{\beta}\hat{\mathcal{H}}_{\alpha \beta}, \\
\label{tot press}
p &=& \frac{\left(2u^{\alpha}u^{\beta}-g^{\alpha \beta}\right)}{6}\hat{\mathbf{q}}_{AB}\partial_{\alpha}\phi^A\partial_{\beta}\phi^B - \mathcal{V} + \mathcal{L}_{\textup{int}} + \frac{2\left(u^{\alpha}u^{\beta}+g^{\alpha \beta}\right)}{3}\hat{\mathcal{H}}_{\alpha \beta}, \\
\label{tot q}
u_{\mu}u_{\nu} &=& \frac{\hat{\mathbf{q}}_{AB}\partial_{\mu}\phi^A\partial_{\nu}\phi^B + 2\hat{\mathcal{H}}_{\mu \nu} - \frac{g_{\mu \nu}}{3}\left(u^{\alpha}u^{\beta}+g^{\alpha \beta}\right)\left(\hat{\mathbf{q}}_{AB}\partial_{\alpha}\phi^A\partial_{\beta}\phi^B + 2\hat{\mathcal{H}}_{\alpha \beta}\right)}{\rho+p}.
\end{eqnarray}
From these equations, we can obtain the background components:
\begin{eqnarray}
\label{tot dens bg 0}
\rho^0 &=& \frac{1}{2}\dot{\phi}_0^2 + \mathcal{V}(\phi_0) - \lambda^0_A(\phi_0)\dot{\phi}_0^A + 2\mathcal{H}_{00}, \\
\label{tot press bg 0}
p^0 &=& \frac{1}{2}\dot{\phi}_0^2 - \mathcal{V}(\phi_0) + \lambda^0_A(\phi_0)\dot{\phi}_0^A,
\end{eqnarray}
where Eq.~(\ref{bgLint}) from Appendix \ref{app:Interaction} was used to represent $\mathcal{L}_{\textup{int}}$ in the background approximation and we will assume that $\hat{\mathcal{H}}_{\mu \nu} = \mathcal{H}_{\mu \nu} + \delta\mathcal{H}_{\mu \nu}$. These expressions are the components of the total fluid in the system, so Eqs.~(\ref{tot dens bg 0}) and (\ref{tot press bg 0}) have to satisfy (\ref{H2})-(\ref{dotH}), then:
\begin{eqnarray}
\label{cond 1}
\mathcal{L}_{\textup{int}}(\phi_0) \equiv \lambda^0_A(\phi_0)\dot{\phi}_0^A &\stackrel{On-Shell}{\hbox to 50pt{\rightarrowfill}}& 0, \\
\label{cond 2 0}
\mathcal{H}_{00} &\stackrel{On-Shell}{\hbox to 50pt{\rightarrowfill}}& 0.
\end{eqnarray}
Basically, (\ref{cond 1}) means that $\lambda^0_A(\phi_0)$ has two effective degrees of freedom and they are represented by $q_1$ and $q_2$ in Eq.~(\ref{JA}) from Appendix \ref{app:Interaction}. Additionally, we can use the fluid reference frame where:
\begin{eqnarray}
\label{tot q bg 0}
u^0_{\mu} &=&
\left(
  \begin{array}{cccc}
    1 & 0 & 0 & 0 \\
  \end{array}
\right).
\end{eqnarray}
In this case, (\ref{tot q}) says that $\mathcal{H}_{ij} = \mathcal{H}_{i0} = \mathcal{H}_{0i} = 0$. Therefore, the complete condition on $\mathcal{H}_{\mu \nu}$ is:
\begin{eqnarray}
\label{cond 2}
\mathcal{H}_{\mu \nu} &\stackrel{On-Shell}{\hbox to 50pt{\rightarrowfill}}& 0.
\end{eqnarray}
In any case, we will evaluate this condition to the end.\\

On the other side, we need the perturbative component of the fluid. At the first order, the interaction lagrangian can be write as:
\begin{eqnarray}
\label{Lint pert 0}
\delta\mathcal{L}_{\textup{int}} &=& \left(\frac{\bar{\delta} \mathcal{L}_{\textup{int}}(\phi_0,\dot{\phi}_0)}{\bar{\delta} \phi_0^A}\right)\left(\delta \phi^A\right) + \left(\frac{\delta \mathcal{L}_{\textup{int}}(\phi_0,\dot{\phi}_0)}{\delta \dot{\phi}_0^A}\right)\frac{D}{dt}\left(\delta \phi^A\right) + \mathcal{O}_A(\phi_0,\dot{\phi}_0)\left(\delta \phi^A\right) \nonumber \\
&& + \omega^0_A(\phi_0)\frac{D}{dt}\left(\delta \phi^A\right),
\end{eqnarray}
where $\delta\mathcal{L}_{\textup{int}} = \mathcal{L}_{\textup{int}}(\phi,\partial_{\mu}\phi) - \mathcal{L}_{\textup{int}}(\phi_0,\dot{\phi}_0)$ and we used the same variation presented in Eq.~(\ref{JA bg}) to covariantize the expression, and the last two terms are additional contributions in the lagrangian interaction in a perturbative level, given by $\mathcal{O}_A(\phi_0,\dot{\phi}_0)$ and $\omega^0_A(\phi_0)$. Using Eq.~(\ref{bgLint}) in (\ref{Lint pert 0}), we obtain:
\begin{eqnarray}
\label{Lint pert}
\delta\mathcal{L}_{\textup{int}} = \left(\dot{\phi}_0^B\left(D_A\lambda^0_B\right)+\mathcal{O}_A\right)\left(\delta \phi^A\right) + \left(\lambda^0_A+\omega^0_A\right)\frac{D}{dt}\left(\delta \phi^A\right).
\end{eqnarray}
With all these, the perturbative components are given by:
\begin{eqnarray}
\label{tot dens per 0}
\rho^1 &=& \left(\dot{\phi}^0_A-\lambda^0_A-\omega^0_A\right)\frac{D}{dt}\left(\delta \phi^A\right) - \dot{\phi}_0^2\Phi + \left(\mathcal{V}_A-\dot{\phi}_0^B\left(D_A\lambda^0_B\right)-\mathcal{O}_A\right)\left(\delta \phi^A\right) \nonumber \\
&& + 2\delta\mathcal{H}_{00} + 4u_0^{\alpha}u_1^{\beta}\mathcal{H}_{\alpha \beta}, \\
\label{tot press per 0}
p^1 &=& \left(\dot{\phi}^0_A+\lambda^0_A+\omega^0_A\right)\frac{D}{dt}\left(\delta \phi^A\right) - \dot{\phi}_0^2\Phi - \left(\mathcal{V}_A-\dot{\phi}_0^B\left(D_A\lambda^0_B\right)-\mathcal{O}_A\right)\left(\delta \phi^A\right) \nonumber \\
&& + \frac{2\left(2u_0^{\alpha}u_1^{\beta}+\delta g^{\alpha \beta}\right)}{3}\mathcal{H}_{\alpha \beta}, \\
\label{tot u per 1 0}
u^1_{(0)} &=& \Phi \\
\label{tot u per 2 0}
u^1_{(i)} &=& \frac{\left(\dot{\phi}_0\mathcal{T}_A\partial_i\left(\delta\phi^A\right)\right) + \delta\mathcal{H}_{0i}}{\dot{\phi}_0^2+2\mathcal{H}_{00}},
\end{eqnarray}
where we used Eq.~(\ref{g inv}) and $\hat{\mathbf{q}}_{AB}$ depends on $\phi^A$. In particular, the expression in (\ref{tot u per 2 0}) is completely inconvenient. $u^1_{i}$ is usually used to define the curvature, presented in Eq.~(\ref{Rinv}), so it must be directly related to the tangent component of $\left(\delta\phi^A\right)$, without interaction terms. So, we need the On-Shell condition:
\begin{eqnarray}
\label{cond 3}
\delta\mathcal{H}_{0i} &\stackrel{On-Shell}{\hbox to 50pt{\rightarrowfill}}& 0.
\end{eqnarray}
Additionally, we know that $\hat{\mathcal{H}}_{00}$ is linear in $\dot{\phi}^A$, then $\delta\mathcal{H}_{00}$ depends on $\frac{D}{dt}\left(\delta \phi^A\right)$ and $\left(\delta \phi^A\right)$. Besides, any term proportional to $\Phi$ in $\delta\mathcal{H}_{00}$  will be zero when we impose Eqs.~(\ref{cond 1}) and (\ref{cond 2}). With all these, the fluid components in our prescription using (\ref{cond 1}) and (\ref{cond 2}) are finally given by:
\begin{eqnarray}
\label{tot dens bg}
\rho^0 &=& \frac{1}{2}\dot{\phi}_0^2 + \mathcal{V}(\phi_0), \\
\label{tot press bg}
p^0 &=& \frac{1}{2}\dot{\phi}_0^2 - \mathcal{V}(\phi_0), \\
\label{tot q bg}
u^0_{\mu} &=&
\left(
  \begin{array}{cccc}
    1 & 0 & 0 & 0 \\
  \end{array}
\right) \\
\label{tot dens per}
\rho^1 &=& \left(\dot{\phi}^0_A+\chi^0_A-\lambda^0_A\right)\frac{D}{dt}\left(\delta \phi^A\right) - \dot{\phi}_0^2\Phi + \left(\mathcal{V}_A-\dot{\phi}_0^B\left(D_A\lambda^0_B\right)-\mathcal{U}_A\right)\left(\delta \phi^A\right), \\
\label{tot press per}
p^1 &=& \left(\dot{\phi}^0_A+\omega^0_A+\lambda^0_A\right)\frac{D}{dt}\left(\delta \phi^A\right) - \dot{\phi}_0^2\Phi - \left(\mathcal{V}_A-\dot{\phi}_0^B\left(D_A\lambda^0_B\right)-\mathcal{O}_A\right)\left(\delta \phi^A\right), \\
\label{tot u per 1}
u^1_{(0)} &=& \Phi \\
\label{tot u per 2}
u^1_{i} &\equiv& \partial_iu^1 \rightarrow u^1 = \frac{\mathcal{T}_A\left(\delta\phi^A\right)}{\dot{\phi}_0},
\end{eqnarray}
These expressions give us the more general relation between the scalar fields and the fluid. Now, we can fix some of these parameters to restrict the fluid components using particular rules. In first place, we can write Eqs.~(\ref{tot dens per})-(\ref{tot u per 2}) in term of invariant components using (\ref{Inv Comp 1})-(\ref{Inv Comp 2}). Now, Eq.~ (\ref{tot u per 1}) is just the Newtonian Potential, obeying Eq.~(\ref{bard1}), and Eq.~(\ref{tot u per 2}) is given by (\ref{Rinv}), where:
\begin{eqnarray}
\label{dphiInv}
\left(\delta\phi^A\right)^I &=& \frac{\dot{\phi}_0}{H}\left(\mathcal{R}\mathcal{T}^A + \mathcal{S}_\mathcal{N}\mathcal{N}^A + \mathcal{S}_\mathcal{B}\mathcal{B}^A\right).
\end{eqnarray}
On the other side, $\rho^1$ and $p^1$ are respectively given by Eqs.~(\ref{drhoinv})-(\ref{dpinv}) with:
\begin{eqnarray}
\label{tot dens per inv 0}
\rho^{1I} &=& \left(\dot{\phi}^0_A+\chi^0_A-\lambda^0_A\right)\frac{D}{dt}\left(\delta \phi^A\right)^I \nonumber \\
&& + \left(\mathcal{V}_A-\dot{\phi}_0^B\left(D_A\lambda^0_B\right)-\mathcal{U}_A- \frac{\left(\dot{\phi}_0^2+\chi^0_B\dot{\phi}_0^B\right)}{2H}\dot{\phi}^0_A\right)\left(\delta \phi^A\right)^I \nonumber \\
&& + \left(3H\chi^0_A\dot{\phi}_0^A+\chi^0_A\left(\mathcal{V}^A+\mathcal{J}_0^A\right)+\dot{\phi}_0^A\mathcal{U}_A\right)\left(\frac{\Phi_{\mathbf{B}}}{H} - f\right) + \chi^0_A\dot{\phi}_0^A\left(\Phi_{\mathbf{B}} + \dot{f}\right), \\
\label{tot press per inv 0}
p^{1I} &=& \left(\dot{\phi}^0_A+\omega^0_A+\lambda^0_A\right)\frac{D}{dt}\left(\delta \phi^A\right)^I \nonumber \\
&& - \left(\mathcal{V}_A-\dot{\phi}_0^B\left(D_A\lambda^0_B\right)-\mathcal{O}_A+ \frac{\left(\dot{\phi}_0^2+\omega^0_B\dot{\phi}_0^B\right)}{2H}\dot{\phi}^0_A\right)\left(\delta \phi^A\right)^I \nonumber \\
&& + \left(3H\omega^0_A\dot{\phi}_0^A+\omega^0_A\left(\mathcal{V}^A+\mathcal{J}_0^A\right)-\dot{\phi}_0^A\mathcal{O}_A\right)\left(\frac{\Phi_{\mathbf{B}}}{H} - f\right) + \omega^0_A\dot{\phi}_0^A\left(\Phi_{\mathbf{B}} + \dot{f}\right),
\end{eqnarray}
where $f=a^2\left(\dot{E}-\frac{B}{a}\right)$ and we used the background equations in Section \ref{SubSec: Radiation as an effective scalar}, Eqs.~(\ref{bard1}), (\ref{bard2}), (\ref{Einst EQ 1}) and (\ref{Einst EQ 4}). $f$ is a non-invariant element, so we need additional conditions on the interactions given by:
\begin{eqnarray}
\label{cond 4}
\chi^0_A\dot{\phi}_0^A &=& 0, \\
\label{cond 5}
\omega^0_A\dot{\phi}_0^A &=& 0, \\
\label{cond 6}
\dot{\phi}_0^A\mathcal{U}_A &=& - \chi^0_A\left(\mathcal{V}^A+\mathcal{J}_0^A\right) = - \chi^0_\mathcal{N}H\dot{\phi}_0\eta_{\perp}, \\
\label{cond 7}
\dot{\phi}_0^A\mathcal{O}_A &=& \omega^0_A\left(\mathcal{V}^A+\mathcal{J}_0^A\right) = \omega^0_\mathcal{N}H\dot{\phi}_0\eta_{\perp},
\end{eqnarray}
and Eqs.~(\ref{tot dens per inv 0})-(\ref{tot press per inv 0}) are reduced to:
\begin{eqnarray}
\label{tot dens per inv 1}
\rho^{1I} &=& \left(\dot{\phi}^0_A+\chi^0_A-\lambda^0_A\right)\frac{D}{dt}\left(\delta \phi^A\right)^I + \left(\mathcal{V}_A-\dot{\phi}_0^B\left(D_A\lambda^0_B\right)-\mathcal{U}_A- \frac{\dot{\phi}_0^2}{2H}\dot{\phi}^0_A\right)\left(\delta \phi^A\right)^I \nonumber \\
&=& \frac{\dot{\phi}_0^2}{H}\dot{\mathcal{R}} - 3\dot{\phi}_0^2\mathcal{R} + \frac{\left(\chi^0_\mathcal{N}-\lambda^0_\mathcal{N}\right)\dot{\phi}_0}{H}\dot{\mathcal{S}}_\mathcal{N} \nonumber \\
&& + \dot{\phi}_0\left(2\dot{\phi}_0\eta_{\perp}+\left(\chi^0_\mathcal{N}-\lambda^0_\mathcal{N}\right)\left(\varepsilon-\eta_{\parallel}\right)-\chi^0_\mathcal{B}C-\frac{\dot{\lambda}^0_\mathcal{N}+\mathcal{U}_\mathcal{N}}{H}\right)\mathcal{S}_\mathcal{N} \nonumber \\
&& + \frac{\left(\chi^0_\mathcal{B}-\lambda^0_\mathcal{B}\right)\dot{\phi}_0}{H}\dot{\mathcal{S}}_\mathcal{B} + \dot{\phi}_0\left(\left(\chi^0_\mathcal{B}-\lambda^0_\mathcal{B}\right)\left(\varepsilon-\eta_{\parallel}\right)+\chi^0_\mathcal{N}C-\frac{\dot{\lambda}^0_\mathcal{B}+\mathcal{U}_\mathcal{B}}{H}\right)\mathcal{S}_\mathcal{B}, \\
\label{tot press per inv 1}
p^{1I} &=& \left(\dot{\phi}^0_A+\omega^0_A+\lambda^0_A\right)\frac{D}{dt}\left(\delta \phi^A\right)^I - \left(\mathcal{V}_A-\dot{\phi}_0^B\left(D_A\lambda^0_B\right)-\mathcal{O}_A+ \frac{\dot{\phi}_0^2}{2H}\dot{\phi}^0_A\right)\left(\delta \phi^A\right)^I \nonumber \\
&=& \frac{\dot{\phi}_0^2}{H}\dot{\mathcal{R}} + \left(3-2\eta_{\parallel}\right)\dot{\phi}_0^2\mathcal{R} + \frac{\left(\omega^0_\mathcal{N}+\lambda^0_\mathcal{N}\right)\dot{\phi}_0}{H}\dot{\mathcal{S}}_\mathcal{N} \nonumber \\
&& + \dot{\phi}_0\left(\left(\omega^0_\mathcal{N}+\lambda^0_\mathcal{N}\right)\left(\varepsilon-\eta_{\parallel}\right)-\omega^0_\mathcal{B}C+\frac{\dot{\lambda}^0_\mathcal{N}+\mathcal{O}_\mathcal{N}}{H}\right)\mathcal{S}_\mathcal{N} \nonumber \\
&& + \frac{\left(\omega^0_\mathcal{B}+\lambda^0_\mathcal{B}\right)\dot{\phi}_0}{H}\dot{\mathcal{S}}_\mathcal{B} + \dot{\phi}_0\left(\left(\omega^0_\mathcal{B}+\lambda^0_\mathcal{B}\right)\left(\varepsilon-\eta_{\parallel}\right)+\omega^0_\mathcal{N}C+\frac{\dot{\lambda}^0_\mathcal{B}+\mathcal{O}_\mathcal{B}}{H}\right)\mathcal{S}_\mathcal{B},
\end{eqnarray}
where we used the background equations of motion and Eq.~(\ref{dphiInv}) in the last steps. On the other side, the perturbative Einstein equations say that $p^{1I}$ satisfy Eq.~ (\ref{Einst EQ 3}), so we need that:
\begin{eqnarray}
\label{cond 8}
\omega^0_A &=& - \lambda^0_A, \\
\label{cond 9}
\mathcal{O}_A &=& - \lambda^0_\mathcal{N}H\eta_{\perp}\mathcal{T}_A - \left(\dot{\lambda}^0_\mathcal{N}+HC\lambda^0_\mathcal{B}\right)\mathcal{N}_A - \left(\dot{\lambda}^0_\mathcal{B}-HC\lambda^0_\mathcal{N}\right)\mathcal{B}_A \nonumber \\
&=& - \frac{D \lambda^0_A}{dt}.
\end{eqnarray}
These conditions satisfy Eqs.~(\ref{cond 5}) and (\ref{cond 7}), and we can see from (\ref{Lint pert}) that the interaction lagrangian must be represented by:
\begin{eqnarray}
\label{Lint pert final}
\mathcal{L}_{\textup{int}}(\phi,\dot{\phi}) = \dot{\phi}_0^A\lambda^0_A(\phi_0) - \mathcal{J}^0_A(\phi_0,\dot{\phi}_0)\left(\delta \phi^A\right) + O\left(\left(\delta \phi\right)^2\right)
\end{eqnarray}
and $\hat{\mathcal{H}}_{\mu \nu} = \frac{\delta \mathcal{L}_{\textup{int}}}{\delta g_{\mu \nu}}$ gives us a first order contribution, all these in the on-shell approximation. Therefore, we can rewrite Eqs.~(\ref{tot dens per inv 1})-(\ref{tot press per inv 1}) as:
\begin{eqnarray}
\label{tot dens per inv 2}
\rho^{1I} &=& \left(\dot{\phi}^0_A+H\vartheta_A\right)\frac{D}{dt}\left(\delta \phi^A\right)^I + \left(\mathcal{V}_A+\mathcal{J}^0_A+H\dot{\phi}_0\tau_A- \frac{\dot{\phi}_0^2}{2H}\dot{\phi}^0_A\right)\left(\delta \phi^A\right)^I \nonumber \\
&=& \frac{\dot{\phi}_0^2}{H}\dot{\mathcal{R}} - 3\dot{\phi}_0^2\mathcal{R} + \vartheta_\mathcal{N}\dot{\phi}_0\dot{\mathcal{S}}_\mathcal{N} + \dot{\phi}_0\left(\dot{\phi}_0\left(2\eta_{\perp}+\tau_\mathcal{N}\right)+H\left(\left(\varepsilon-\eta_{\parallel}\right)\vartheta_\mathcal{N}-C\vartheta_\mathcal{B}\right)\right)\mathcal{S}_\mathcal{N} \nonumber \\
&& + \vartheta_\mathcal{B}\dot{\phi}_0\dot{\mathcal{S}}_\mathcal{B} + \dot{\phi}_0\left(\dot{\phi}_0\tau_\mathcal{B}+H\left(\left(\varepsilon-\eta_{\parallel}\right)\vartheta_\mathcal{B}+C\vartheta_\mathcal{N}\right)\right)\mathcal{S}_\mathcal{B}, \\
\label{tot press per inv 2}
p^{1I} &=& \dot{\phi}^0_A\frac{D}{dt}\left(\delta \phi^A\right)^I - \left(\mathcal{V}_A+\mathcal{J}^0_A+ \frac{\dot{\phi}_0^2}{2H}\dot{\phi}^0_A\right)\left(\delta \phi^A\right)^I = \frac{\dot{\phi}_0^2}{H}\dot{\mathcal{R}} + \left(3-2\eta_{\parallel}\right)\dot{\phi}_0^2\mathcal{R},
\end{eqnarray}
with:
\begin{eqnarray}
\label{redef 1}
\chi^0_A &=& \lambda^0_A + H\vartheta_A, \\
\label{redef 2}
\mathcal{U}_A &=& - \frac{D \lambda^0_A}{dt} - H\dot{\phi}_0\tau_A,
\end{eqnarray}
and the remaining parameters of the interaction satisfy:
\begin{eqnarray}
\label{cond fin 1}
\mathcal{J}_\mathcal{T} &=& 0, \\
\label{cond fin 2}
\vartheta_\mathcal{T} &=& 0, \\
\label{cond fin 3}
\tau_\mathcal{T} &=& \frac{H\eta_{\perp}\vartheta_\mathcal{N}}{\dot{\phi}_0}.
\end{eqnarray}
Finally, we need to verify that Eqs.~(\ref{tot dens per})-(\ref{tot u per 2}) obey the perturbative equation of motion of a fluid. From \cite{Malik:2008im}, we know that it is given by:
\begin{eqnarray}
\label{PerEqRad 1}
\dot{\rho}^1 + 3H\left(\rho^1+p^1\right) + \frac{\partial^2}{a^2}q^1 - \left(\rho^0+p^0\right)\left(3\dot{\Psi}-\partial^2\left(\dot{E}-\frac{B}{a}\right)\right) = 0,
\end{eqnarray}
where $q^1 = -(\rho^0+p^0)u^1$. Besides, from Eqs.~(\ref{tot dens per})-(\ref{tot u per 2}) and considering (\ref{cond 8})-(\ref{redef 2}), we have:
\begin{eqnarray}
\left(\dot{\phi}^0_A+H\vartheta_A\right)\frac{D}{dt}\left(\delta \phi^A\right) &=& \rho^1 + \dot{\phi}_0^2\Phi - \left(\mathcal{V}_A+\mathcal{J}^0_A+H\dot{\phi}_0\tau_A\right)\left(\delta \phi^A\right), \nonumber \\
\left(2\left(\mathcal{V}_A+\mathcal{J}^0_A\right)+H\dot{\phi}_0\tau_A\right)\left(\delta\phi^A\right) &=& \rho^1 - p^1 - H\vartheta_A\frac{D}{dt}\left(\delta \phi^A\right), \nonumber \\
\dot{\phi}_0^A\left(\delta\phi^A\right) &=& -q^1.
\end{eqnarray}
Now, in this paper is proposed that the perturbative equation of the scalar fields in our formalism is given by Eq.~(\ref{pert eq 5}). If we Project it on $\left(\dot{\phi}^0_A+H\vartheta_A\right)$, we obtain:
\begin{eqnarray}
\label{test}
\dot{\rho}^1 + 3H\left(\rho^1 + p^1\right) + \frac{\partial^2}{a^2}q^1 - \dot{\phi}_0^2\left(3\dot{\Psi} - \partial^2\left(\dot{E}-\frac{B}{a}\right)\right) + \left(\dot{\phi}_0^A+H\vartheta^A\right)\left(\Delta \mathcal{J}_A\right) && \nonumber \\
- \frac{\partial^2}{a^2}H\vartheta_A\left(\delta \phi^A\right) + 2H\vartheta^A\left(\mathcal{V}_A+\mathcal{J}^0_A\right)\Phi - H\left(\frac{D \vartheta_A}{dt}-H\varepsilon\vartheta_A+\dot{\phi}_0\tau_A\right)\frac{D}{dt}\left(\delta\phi^A\right) && \nonumber \\
- \left(\frac{D}{dt}\left(\mathcal{J}^0_A+H\dot{\phi}_0\tau_A\right)+3H^2\dot{\phi}_0\tau_A-H\vartheta^B\left(\mathcal{V}_{AB} - \dot{\phi}_0^2\mathbb{R}_{A\mathcal{T}\mathcal{T}B}\right)\right)\left(\delta \phi^A\right) &=& 0.
\end{eqnarray}
This means that Eqs.~(\ref{pert eq 5}) and (\ref{PerEqRad 1}) are equivalent if:
\begin{eqnarray}
\label{cond JA}
\left(\dot{\phi}_0^A+H\vartheta^A\right)\left(\Delta \mathcal{J}_A\right)^I &=& \frac{\partial^2}{a^2}H\vartheta_A\left(\delta \phi^A\right)^I + H\left(\frac{D \vartheta_A}{dt}-H\varepsilon\vartheta_A+\dot{\phi}_0\tau_A\right)\frac{D}{dt}\left(\delta\phi^A\right)^I \\
&& + \left(\frac{D}{dt}\left(\mathcal{J}^0_A\right)+H\dot{\phi}_0\frac{D}{dt}\left(\tau_A\right)+H^2\dot{\phi}_0\left(3-\varepsilon-\eta_{\parallel}\right)\tau_A \nonumber \right. \\
&& \left.-H\vartheta^B\left(\mathcal{V}_{AB} - \dot{\phi}_0^2\mathbb{R}_{A\mathcal{T}\mathcal{T}B}+\frac{\dot{\phi}^0_A\left(\mathcal{V}_B+\mathcal{J}^0_B\right)}{H}\right)\right)\left(\delta \phi^A\right)^I, \nonumber
\end{eqnarray}
where we used (\ref{bard1}), (\ref{bard2}), (\ref{Einst EQ 1}), (\ref{Einst EQ 4}) and (\ref{Inv Comp 1}). From Eq.~(\ref{Delta JA}) in Appendix \ref{app:Interaction}, we know that (\ref{cond JA}) implies:
\begin{eqnarray}
\label{cond JA 1}
\left(\dot{\phi}_0^B+H\vartheta^B\right)\mathcal{I}_{BA} &=& H\vartheta_A\frac{\partial^2}{a^2}+\frac{D}{dt}\left(\mathcal{J}^0_A\right)+H\dot{\phi}_0\frac{D}{dt}\left(\tau_A\right)+H^2\dot{\phi}_0\left(3-\varepsilon-\eta_{\parallel}\right)\tau_A  \nonumber \\
&& -H\vartheta^B\left(\mathcal{V}_{AB}- \dot{\phi}_0^2\mathbb{R}_{A\mathcal{T}\mathcal{T}B}+\frac{\dot{\phi}^0_A\left(\mathcal{V}_B+\mathcal{J}^0_B\right)}{H}\right), \\
\label{cond JA 2}
\left(\dot{\phi}_0^B+H\vartheta^B\right)\mathcal{J}_{BA} &=& H\left(\frac{D \vartheta_A}{dt}-H\varepsilon\vartheta_A+\dot{\phi}_0\tau_A\right), \\
\label{cond JA 3}
\left(\dot{\phi}_0^B+H\vartheta^B\right)\mathcal{K}_{B} &=& 0.
\end{eqnarray}
In this paper, we will discard the Laplacian term in Eq.~(\ref{cond JA 1}), because we do not expect that second order terms appear in the interaction contributions, however could be considered in a future work. Taking all these into consideration, we will use $\vartheta_A = 0$, so:
\begin{eqnarray}
\label{comp JT 1}
\mathcal{I}_{\mathcal{T}A} &=& \frac{1}{\dot{\phi}_0}\frac{D}{dt}\left(\mathcal{J}^0_A\right)+H\frac{D}{dt}\left(\tau_A\right)+H^2\left(3-\varepsilon-\eta_{\parallel}\right)\tau_A, \\
\label{cond JT 2}
\mathcal{J}_{\mathcal{T}A} &=& H\tau_A, \\
\label{cond JT 3}
\mathcal{K}_{\mathcal{T}} &=& 0,
\end{eqnarray}
and:
\begin{eqnarray}
\label{DeltaJTI}
\left(\Delta \mathcal{J}_\mathcal{T}\right)^I &=& H\tau_A\frac{D}{dt}\left(\delta\phi^A\right)^I + \left(\frac{1}{\dot{\phi}_0}\frac{D}{dt}\left(\mathcal{J}^0_A\right)+H\frac{D}{dt}\left(\tau_A\right)+H^2\left(3-\varepsilon-\eta_{\parallel}\right)\tau_A\right)\left(\delta \phi^A\right)^I,
\end{eqnarray}
with $\mathcal{J}_\mathcal{T} = 0$ and $\tau_\mathcal{T} = 0$. The final expression of $\left(\Delta \mathcal{J}_\mathcal{T}\right)^I$ and the other components of perturbative interaction used in this paper are presented in Eqs.~(\ref{DeltaJTI final})-(\ref{DeltaJBI final}) from Appendix \ref{app:Interaction}. Finally, in relation to the fluid components, they are given by:
\begin{eqnarray}
\label{bg comp}
\rho^0 &=& \frac{1}{2}\dot{\phi}_0^2 + \mathcal{V} \textit{, } p^0 = \frac{1}{2}\dot{\phi}_0^2 - \mathcal{V}, \\
\label{per comp 1}
\rho^{1I} &=& \frac{\dot{\phi}_0^2}{H}\dot{\mathcal{R}} - 3\dot{\phi}_0^2\mathcal{R} + \dot{\phi}_0^2\left(2\eta_{\perp}+\tau_\mathcal{N}\right)\mathcal{S}_\mathcal{N} + \dot{\phi}_0^2\tau_\mathcal{B}\mathcal{S}_\mathcal{B}, \\
\label{per comp 2}
p^{1I} &=& \frac{\dot{\phi}_0^2}{H}\dot{\mathcal{R}} + \left(3-2\eta_{\parallel}\right)\dot{\phi}_0^2\mathcal{R}.
\end{eqnarray}
This final result prove that a system with two inflatons and a radiation fluid can be represented by our prescription with three scalar fields, where the perturbative Klein-Gordon equation is given by Eq.~(\ref{pert eq 5}), and the interaction for the total fluid is given by $\tau_\mathcal{N}$ and $\tau_\mathcal{B}$. In fact, if we use Eqs.~(\ref{per comp 1})-(\ref{per comp 2}) in (\ref{PerEqRad 1}), we obtain (\ref{KG EQ 1 Final}). Besides, (\ref{per comp 1}) is used to obtain (\ref{Einst EQ 2 Final}) from (\ref{Einst EQ 2}) to complete the perturbative system equations.
\end{appendix}

\end{document}